\documentclass[aps,prx,reprint,preprintnumbers,twocolumn,superscriptaddress,nofootinbib,floatfix,shortbibliography]{revtex4-2}
\usepackage{amsmath}
\usepackage{amssymb}
\usepackage{dsfont}
\usepackage{graphicx}
\usepackage{hyperref}
\usepackage{float}
\usepackage{wasysym}
\usepackage{bbm}

\usepackage{pgffor}

\setcitestyle{super}

\newcommand{\ket}[1]{|#1 \rangle}

\usepackage[dvipsnames]{xcolor}
\usepackage[normalem]{ulem}
\usepackage{comment}

\usepackage{color}
\usepackage{hyperref}

\makeatother
\begin{document}

\title{Characterizing a non-equilibrium phase transition on a quantum computer}

\author{Eli Chertkov}
\email{eli.chertkov@quantinuum.com}
\affiliation{Quantinuum, 303 South Technology Court, Broomfield, Colorado 80021, USA}
\author{Zihan Cheng}
\affiliation{Department of Physics, University of Texas at Austin, Austin, TX 78712, USA}
\author{Andrew C. Potter}
\affiliation{Department of Physics, University of Texas at Austin, Austin, TX 78712, USA}
\affiliation{Department of Physics and Astronomy, and Stewart Blusson Quantum Matter Institute, University of British Columbia, Vancouver, BC, Canada V6T 1Z1}
\author{Sarang Gopalakrishnan}
\affiliation{Department of Electrical and Computer Engineering, Princeton University, Princeton, NJ 08544, USA}
\author{Thomas M. Gatterman}
\affiliation{Quantinuum, 303 South Technology Court, Broomfield, Colorado 80021, USA}
\author{Justin A. Gerber}
\affiliation{Quantinuum, 303 South Technology Court, Broomfield, Colorado 80021, USA}
\author{Kevin Gilmore}
\affiliation{Quantinuum, 303 South Technology Court, Broomfield, Colorado 80021, USA}
\author{Dan Gresh}
\affiliation{Quantinuum, 303 South Technology Court, Broomfield, Colorado 80021, USA}
\author{Alex Hall}
\affiliation{Quantinuum, 303 South Technology Court, Broomfield, Colorado 80021, USA}
\author{Aaron Hankin}
\affiliation{Quantinuum, 303 South Technology Court, Broomfield, Colorado 80021, USA}
\author{Mitchell Matheny}
\affiliation{Quantinuum, 303 South Technology Court, Broomfield, Colorado 80021, USA}
\author{Tanner Mengle}
\affiliation{Quantinuum, 303 South Technology Court, Broomfield, Colorado 80021, USA}
\author{David Hayes}
\affiliation{Quantinuum, 303 South Technology Court, Broomfield, Colorado 80021, USA}
\author{Brian Neyenhuis}
\affiliation{Quantinuum, 303 South Technology Court, Broomfield, Colorado 80021, USA}
\author{Russell Stutz}
\affiliation{Quantinuum, 303 South Technology Court, Broomfield, Colorado 80021, USA}
\author{Michael Foss-Feig}
\affiliation{Quantinuum, 303 South Technology Court, Broomfield, Colorado 80021, USA}

\begin{abstract}
At transitions between phases of matter, physical systems can exhibit universal behavior independent of their microscopic details. Probing such behavior in quantum many-body systems is a challenging and practically important problem that can be solved by quantum computers, potentially exponentially faster than by classical computers. In this work, we use the Quantinuum H1-1 quantum computer to realize a quantum extension of a simple classical disease spreading process that is known to exhibit a non-equilibrium phase transition between an active and absorbing state. Using techniques such as qubit-reuse and error avoidance based on real-time conditional logic (utilized extensively in quantum error correction), we are able to implement large instances of the model with $73$ sites and up to $72$ circuit layers, and quantitatively determine the model's critical properties. This work demonstrates how quantum computers capable of mid-circuit resets, measurements, and conditional logic enable the study of difficult problems in quantum many-body physics: the simulation of open quantum system dynamics and non-equilibrium phase transitions.
\end{abstract}

\maketitle

A remarkable feature of nature is that complicated systems governed by completely different microscopic rules, such as water or kitchen magnets, can behave essentially identically on large length and time scales when changing between two phases of matter, as long as they share a few basic properties such as symmetry and topology. This universal behavior near phase transitions makes it possible to predict many behaviors of complex materials by studying simple models. Especially in equilibrium, physicists have made great progress in understanding phase transitions by computing the properties of simplified models and validating the calculations with experiments. However, thermal equilibrium is an idealization that often does not hold; many features of the world around us---from geological formations, to disease spreading, to the flocking of birds---arise due to processes that are fundamentally non-equilibrium in nature. Non-equilibrium systems also exhibit universal behavior at phase transitions, but with a richer and generally less-well understood phenomenology than their equilibrium counterparts. The experimental realization of many platforms for simulating quantum dynamics has brought a particularly poorly understood question to the forefront in recent years: Can microscopic quantum features of non-equilibrium systems persist at macroscopic scales and impact universal properties of the dynamics?

Quantum computers may be helpful in addressing this question for two reasons. First, classically simulating open quantum systems can be more difficult than simulating unitary dynamics~\cite{Weimer2021} often requiring parametrically larger memory or simulation time. Second, since energy-conservation does not play a role in many non-equilibrium processes, the essential features can be captured by discrete quantum circuits without the large overhead for continuous-time Hamiltonian simulation techniques.

In this work, we study a dissipative quantum circuit model that generalizes a canonical \emph{classical} model, the contact process\cite{Harris1974}, known to exhibit a non-equilibrium phase transition in the directed percolation (DP) universality class\cite{Hinrichsen2000}.  
We are motivated in part by related continuous-time models\cite{Marcuzzi2016,Carollo2019,Gillman2019,Minjae2021} and quantum cellular automata models\cite{Lesanovsky2019,Gillman2020,Gillman2021a,Gillman2021b,Gillman2022} relevant to Rydberg atom quantum simulators, for which a combination of field-theoretic\cite{Marcuzzi2016} and numerical\cite{Carollo2019,Gillman2019,Minjae2021,Lesanovsky2019,Gillman2020,Gillman2021a,Gillman2021b,Gillman2022} analyses have been interpreted as a possible deviation from classical DP scaling. In large part due to the classical difficulty of obtaining precise results for sufficiently large system sizes and evolution times, a definitive answer to this question remains elusive. 
We note that even if the universal scaling properties of a phase transition in a quantum model turn out to be  classical, \emph{verifying} such behavior may be classically intractable. 

We consider a model that can be thought of either as a discrete-time version of the continuous-time model studied in Refs.~\onlinecite{Marcuzzi2016,Carollo2019,Gillman2019,Minjae2021}, or as a quantum generalization of the usual discrete-time classical contact process in which branching/coagulation processes are replaced by entanglement generating controlled rotation gates. Using tensor-network methods we attempt to identify and characterize the phase transition away from the classical point. We then use a trapped-ion quantum computer to study the model on both sides of the non-equilibrium phase transition as well as at the approximate critical point determined from classical numerics. Utilizing the mid-circuit measurement and reset capabilities available on Quantinuum's H1 series quantum computer, we are able to study this model for large times and system sizes (72 layers of two-qubit gates acting on a 73-site lattice). The high gate fidelities, together with error avoidance techniques enabled by real-time (e.g., circuit-level) conditional logic, allow us to verify power-law growth of observables near the critical point. Our classical and quantum simulations suggest that classical DP universality is robust to the introduction of quantum fluctuations, at least for the initial states considered in this work. While the universal properties of the continuous-time quantum contact process are under debate\cite{Marcuzzi2016,Carollo2019,Gillman2019}, similar results consistent with DP universal scaling for single-seed simulations were recently obtained\cite{Gillman2019}.

\begin{figure*}
    \centering
    \includegraphics[width=\textwidth]{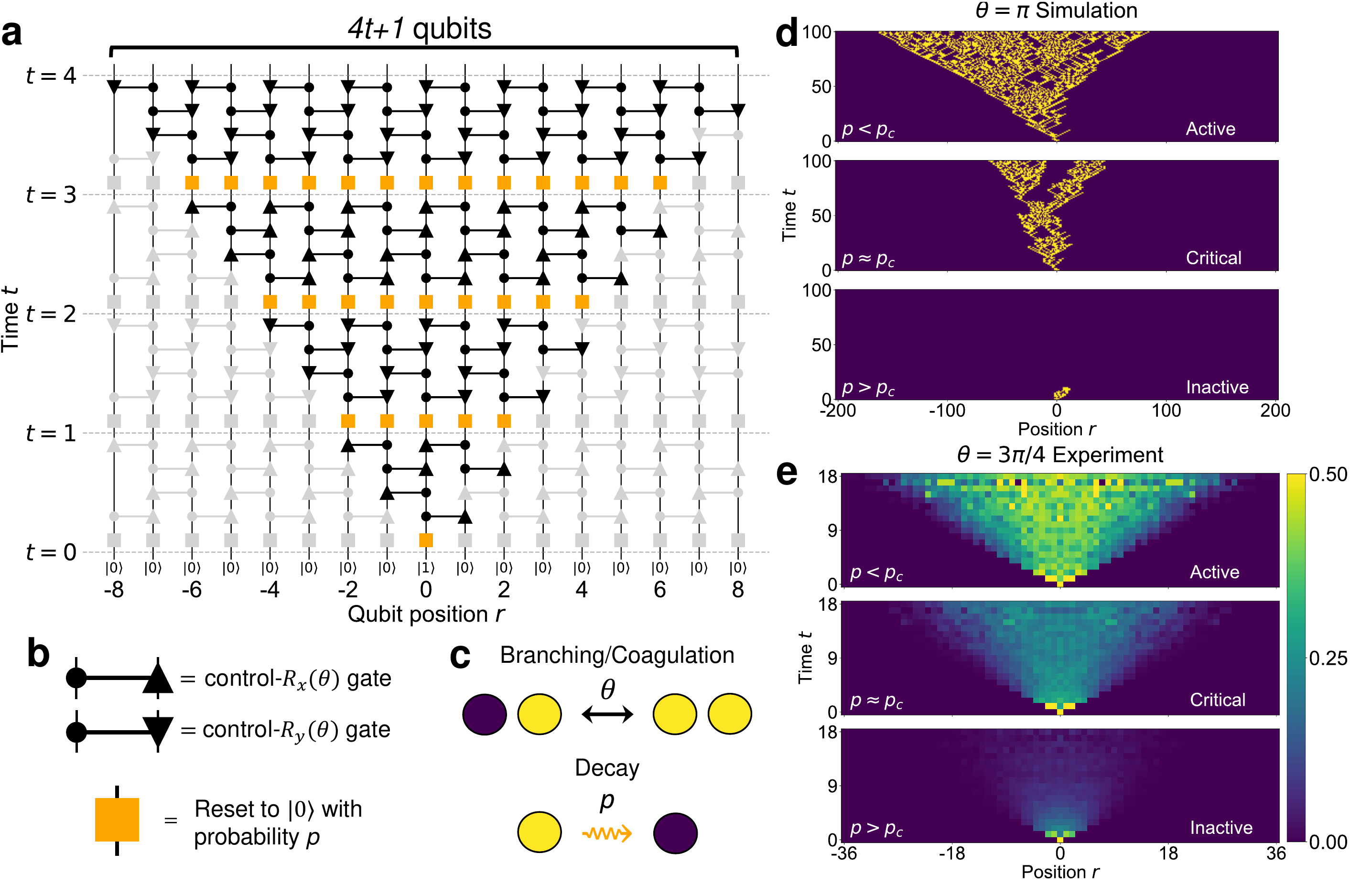}
    \caption{\textbf{A driven-dissipative quantum circuit.} \textbf{a} The one-dimensional Floquet quantum contact process evolving an initial state with a single active site at $r=0$ ($\ket{1}$ is active; $\ket{0}$ is inactive). The grayed-out gates and channels act as identity and do not need to be applied. \textbf{b} The two-qubit gates and quantum channels used in the model. \textbf{c} The two key processes in the model: quantum branching/coagulation of active sites produced by the controlled-rotation gates and decay of active sites produced by the probabilistic resets. Active sites ($\ket{1}$) are shown as yellow circles and inactive sites ($\ket{0}$) as dark blue circles.  \textbf{d} Sample trajectories at the classical point ($\theta = \pi$) of the model for $p=0.2 < p_c$, $p=0.3943\approx p_c$, and $p=0.6>p_c$. \textbf{e} The trajectory-averaged active-site density at $\theta=3\pi/4$ for $p=0.1 < p_c$, $p=0.3\approx p_c$, and $p=0.5>p_c$ obtained from experiments executed on the H1-1 quantum computer. Numerical estimates of $p_c$ are discussed in supplement.}
    \label{fig:fig1}
\end{figure*}

\textbf{A driven-dissipative quantum circuit. } In this work, we study a quantum circuit that we call the one-dimensional \emph{Floquet quantum contact process} (1D FQCP; see Fig.~\ref{fig:fig1}\textbf{a}--\textbf{c}). In each time step, the 1D FQCP circuit executes a layer of probabilistic resets (set each qubit to $\ket{0}$ with probability $p$) followed by four alternating layers of controlled-rotation gates: $CR_{x,y}(\theta) = e^{-\frac{i}{4}\theta (1+\hat{\sigma}^z)\otimes \hat{\sigma}^{x,y}}$ 
where $\hat{\sigma}^{x},\hat{\sigma}^{y},\hat{\sigma}^{z}$ are Pauli matrices.

The FQCP is a quantum circuit generalization of the contact process, a simple model for disease spreading by contact between infected individuals \cite{Hinrichsen2000,Marro1999,Odor2004} (other quantum generalizations can be found in Refs.~\cite{Marcuzzi2016,Carollo2019}). In the 1D FQCP, each site of a one-dimensional chain is labeled as ``active'' or ``inactive,'' corresponding to the single-qubit states $\ket{1}$ and $\ket{0}$, respectively. In the circuit, there is a competition between a unitary driving and non-unitary dissipative process. The driving (set by $\theta$) caused by the two-qubit gates enables active sites to spread to inactive neighbors through the processes: $\ket{10}\rightarrow \cos(\theta/2) \ket{10}+e^{i\phi}\sin(\theta/2)\ket{11}$ where $\phi\in \{0,\pi/2\}$ is a phase (see Fig.~\ref{fig:fig1}). For generic choices of $\theta,\phi$, such processes thermalize subsystems to the infinite-temperature maximally-mixed state $\hat{\rho} \propto \hat{\mathds{1}}$. 
We note that at $\theta=\pi$, the model maps bitstrings in the computational basis to other bitstrings without generating entanglement; at this special point our model is therefore classical, efficiently simulable, and guaranteed to have DP critical exponents (see Fig.~\ref{fig:fig1}\textbf{d} and supplement).  The dissipation (set by $p$) caused by the probabilistic reset channels drives the entire system to the product state $\hat{\rho} = |0\ldots0\rangle \langle 0\ldots 0 |$, called the absorbing state, which the system can reach but not exit. In chains of length $L$, the competition of unitary spreading and spontaneous decay separates two regimes: an inactive regime at high decay rate $p$ where the active sites decay exponentially in time, and an active regime where the spreading processes produce a finite density of active sites that survives to long times $\log t\sim L$. As $L\rightarrow \infty$, these regimes become sharply distinct phases separated by a critical decay rate $p_c$, at which the active site density grows super-diffusively forming a self-similar fractal active cluster.

To probe the transition, we scan $p$ for fixed $\theta$ and time-evolve an initial state $\hat{\rho}_0$, that has a single active ``seed" at position $r=0$ in a chain of $-L\leq r \leq L$ sites, for $t$ time steps and system size $L=2t$ (see Fig.~\ref{fig:fig1}\textbf{a}). Denoting the operator that indicates an active site at position $r$ by $\hat{n}_r = |1\rangle\langle 1|_r$ and $\mathcal{E}_t(\cdot)$ as $t$ steps of evolution, we examine the active-site density $\langle n(r,t) \rangle = \textrm{tr}(\mathcal{E}_t(\hat{\rho}_0) \hat{n}_r)$, total number of active sites $\langle N(t) \rangle = \sum_r \langle n(r,t) \rangle$, and mean-squared extent of the active cluster $\langle R^2(t) \rangle = \frac{1}{\langle N(t) \rangle}\sum_{r} r^2 \langle n(r,t) \rangle$.

Generically, it is postulated that models with absorbing states and no other explicit symmetries undergo non-equilibrium phase transitions in the directed percolation (DP) universality class \cite{Hinrichsen2000}. At the critical point $p=p_c$ of a DP transition, observables asymptotically scale as power-laws: $\langle N(t) \rangle \sim t^{\Theta}$ and $\langle R^2(t) \rangle \sim t^{2/z}$, where the universal exponents $\Theta$ and $z$ have been determined from numerical studies ($\Theta=0.313686(8)$ and $z=1.580745(10)$ in 1D) \cite{Jensen1999,Hinrichsen2000}. Likewise, in 1D, the active-site density profile at late times and $p=p_c$ obeys an asymptotic scaling form $\langle n(r,t) \rangle \sim t^{\Theta - 1/z}f(r/t^{1/z})$, where $f(x)$ is a universal scaling function \cite{Hinrichsen2006}.

\begin{figure*}
    \centering
    \includegraphics[width=\textwidth]{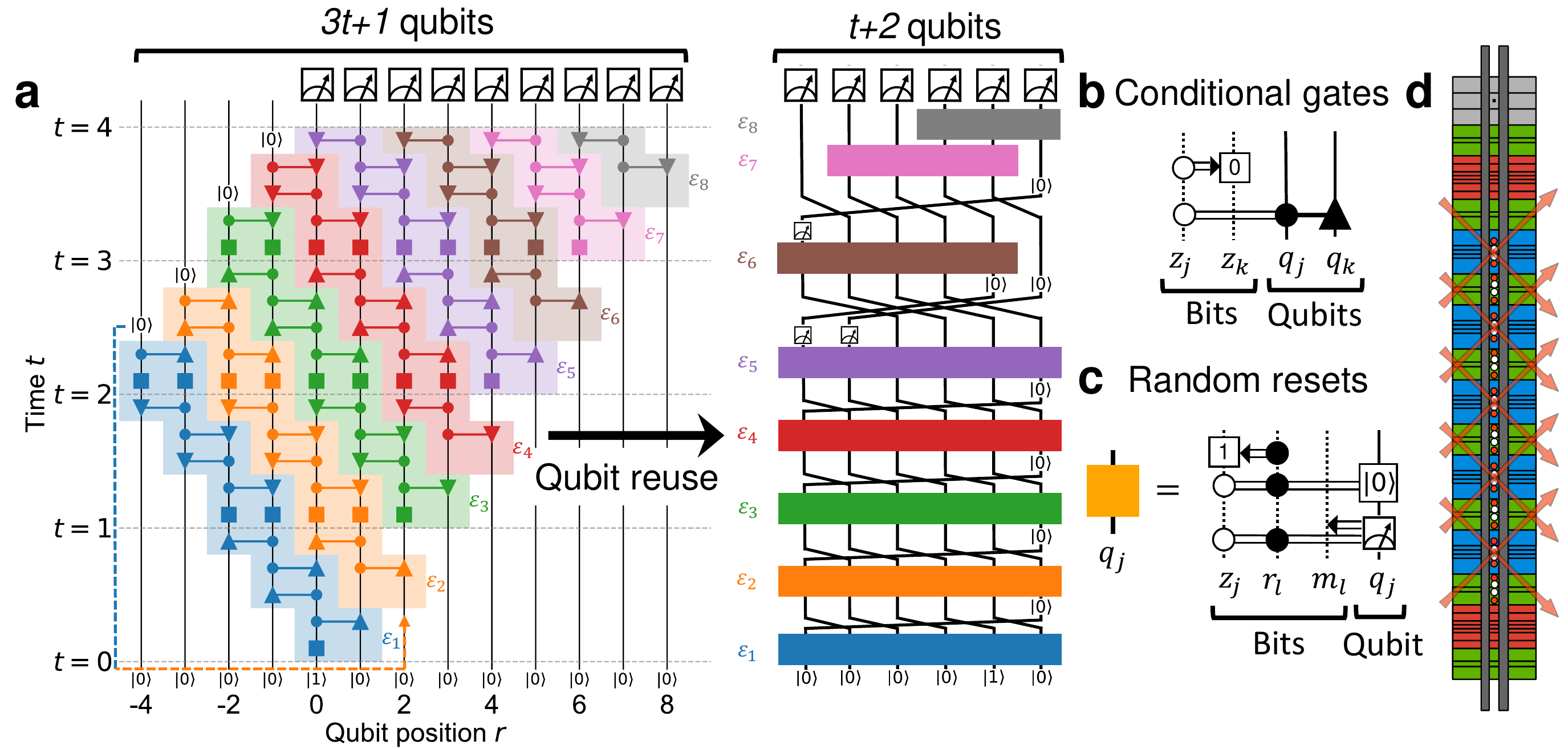}
    \caption{\textbf{Qubit-reuse and real-time conditional logic.} \textbf{a} In our experiments, we measure qubits at positions $r\geq 0$ at the final time, which contain only the depicted gates and channels in their causal cones. By executing the operations in the highlighted ``slices'' $\varepsilon_1,\varepsilon_2,\ldots$ and resetting and re-using qubits between slices, we can reduce the number of qubits needed for the simulation from $3t+1$ to $t+2$. For example, in this diagram, we first execute the gates and channels in $\varepsilon_1$, then reset the qubit at $r=-4$, and reuse it as the input qubit at $r=2$ for $\varepsilon_2$. During the circuit, we use three sets of classical bits $z_j$, $r_l$, and $m_l$ to log which qubits are known to be $\ket{0}$ from mid-circuit resets, to randomly determine which resets to apply, and to record mid-circuit measurement results, respectively. \textbf{b} To reduce the effect of two-qubit gate errors, we choose in real-time not to apply a gate controlled on qubit $q_j$ if we know that $q_j$ is $\ket{0}$ from prior resets. \textbf{c} We implement each probabilistic reset channel by applying a reset conditioned on the value of the random bit $r_l$ and the logging bit $z_j$. \textbf{d} The H1-1 trapped-ion quantum computer used in this work holds 20 qubit ions ($^{171}$Yb$^+$, red dots) and 20 sympathetic cooling ions ($^{138}$Ba$^+$, white dots). Parallel gating, measurement and reset operations are done in the five zones marked with crossing laser beams. H1-1 can perform: two-qubit gates (arbitrary-angle $e^{-i\phi\hat{\sigma}^z\otimes \hat{\sigma}^z / 2}$ gates) with about $3 \times 10^{-3}$ typical gate infidelity between any pair of qubits by rearranging ions via transport operations; single-qubit gates with $5 \times 10^{-5}$ infidelity; state-preparation and measurement with $3 \times 10^{-3}$ infidelity; mid-circuit measurements and resets with crosstalk below $1 \times 10^{-4}$; and real-time conditional logic (see Refs.~\cite{H11specsheet}~and~\cite{RyanAnderson2022} for more details).} \label{fig:fig2}
\end{figure*}

\textbf{Implementation on a trapped-ion quantum computer. } We implement the FQCP model on Quantinuum's H1-1 trapped-ion quantum processor \cite{RyanAnderson2022, pino2020} (see Fig.~\ref{fig:fig2}\textbf{d}). Using recently developed qubit-reuse and quantum tensor network techniques involving mid-circuit measurement and resets \cite{kim2017,fossfeig2020,Barratt2021,Chertkov2022,Niu2021,Lin2021,Zhang2022,Dborin2022,DeCross2022}, we implement up to $t=18$ steps of time evolution using $20$ qubits, a significant savings over the $4t+1=73$ qubits needed without qubit-reuse. High-fidelity and low-cross-talk mid-circuit resets are particularly crucial for this study as the FQCP's probabilistic reset channels would have required hundreds of additional ancilla qubits to implement without mid-circuit resets.

As shown in Fig.~\ref{fig:fig1}\textbf{a}, for the single active seed initial state many controlled gates and channels in the model act as identity, and therefore do not need to be applied, reducing the total number of noisy operations executed in experiments. Moreover, for this initial state, the averaged dynamics is reflection symmetric about $r=0$, so that $\langle n(r,t)\rangle = \langle n(-r,t)\rangle$, which allows us to collect data for all $r$ by measuring only at $r\geq 0$, further reducing the required number of qubits and gates. Fig.~\ref{fig:fig2}\textbf{a} shows the gates and channels causally connected to qubits measured at $r\geq 0, t>0$ and highlights how only $t+2$ physical qubits are needed to execute the model, which would have required $4t+1$ qubits without exploiting reflection symmetry ($3t+1$ using the symmetry).

We utilize an ``error avoidance'' technique that allows us to significantly reduce the effects of two-qubit errors in our circuits. During each experiment, we maintain a real-time log of qubits known to be currently in the $\ket{0}$ state -- from prior mid-circuit resets -- and use that log to conditionally avoid the application of any two-qubit gate for which the control qubit is known to be in $\ket{0}$, thereby avoiding errors accompanying gates that do not induce any dynamics (see Fig.~\ref{fig:fig2}\textbf{b},\textbf{c}).

\textbf{Results. } We experimentally perform dynamics simulations of the FQCP model for times $t=2,4,\ldots,18$ on both sides of the phase transition and near the critical point. From extensive numerical simulations using matrix product operator (MPO) techniques \cite{Bonnes2014,Weimer2021} (see supplement), we find evidence of a directed percolation phase transition for both classical ($\theta=\pi$, $p_c\approx 0.39$) and quantum ($\theta=3\pi/4$, $p_c\approx 0.3$) instances of the model. Informed by these classical numerics, we perform experiments on the quantum version of the model choosing $\theta=3\pi/4$ and targeting (a) the active phase ($p=0.1$), (b) the critical regime ($p=0.3$), and (c) the inactive phase ($p=0.5$). We employ zero-noise extrapolation \cite{Temme2017,Li2017} error mitigation at $p=0.3$ (see supplement).
We emphasize that, while our experiments are not in the quantum advantage regime since they involve only 20 qubits that can be classically simulated, the dynamics of the FQCP model is challenging to compute classically in practice, requiring large-scale MPO simulations at times late enough to see clear signatures of criticality.

The spreading of active sites as a function of space $r$ and time $t$ measured in our three sets of experiments are shown in the heatmaps of Fig.~\ref{fig:fig1}\textbf{e}. From these heatmaps, we can qualitatively observe the expected physical behavior: in the active phase ($p < p_c$) the cluster seeded from the single active site grows ballistically in size, forming a clear cone shape; at the critical point ($p\approx p_c$) the cluster also grows but sub-ballistically; and in the inactive phase ($p > p_c$) the cluster shrinks as the quantum state is pulled into the inactive absorbing state. 

\begin{figure}
    \centering
    \includegraphics[width=0.5\textwidth]{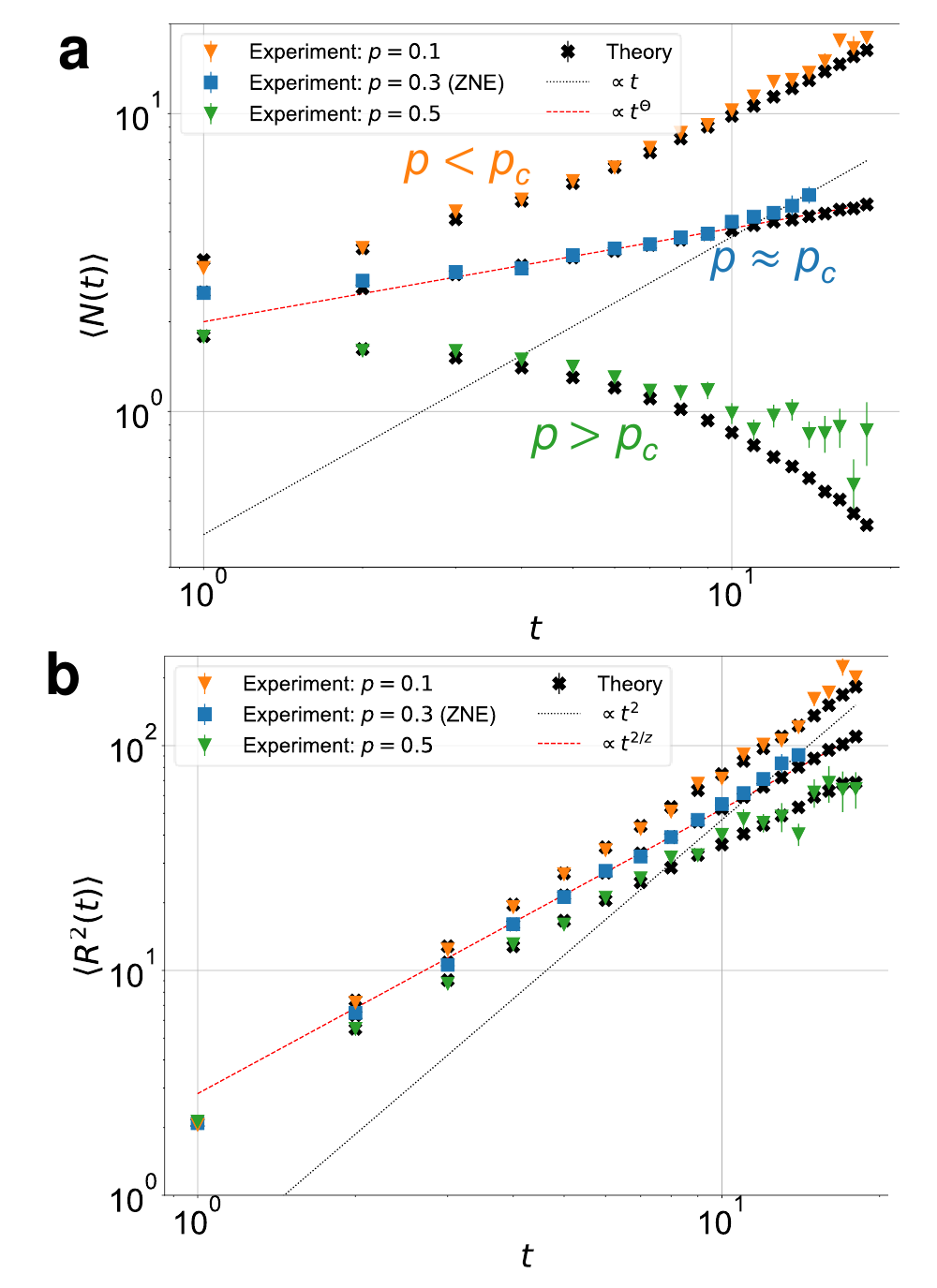}
    \caption{\textbf{Experimental results from quantum computer.} \textbf{a} The number of active sites $\langle N(t)\rangle$ and \textbf{b} the mean-squared extent of the active cluster $\langle R^2(t) \rangle$ versus time on a log-log scale for $p=0.1,0.3,0.5$, as measured in experiment (colored markers) and obtained from noise-less 10,000-shot quantum trajectory simulations (black crosses). As a guide to the eye, curves showing ballistic (black dotted line) and directed percolation power-law (dashed red line) growth of active sites are also depicted. The $p=0.3$ data has zero-noise extrapolation (ZNE) applied.}
    \label{fig:fig3}
\end{figure}

The total number of active sites measured experimentally, compared to the expected theoretical results obtained from noiseless quantum trajectory simulations, are shown in Fig.~\ref{fig:fig3}\textbf{a}. For reference, ballistic ($\propto t$) and directed percolation power-law ($\propto t^\Theta$) curves are included. These plots clearly show how two-qubit gate errors seed new active sites, which leads to a late-time ballistic propagation of errors for $p \approx p_c$. Despite the gate errors, our data at $p=0.1,0.5$ are quantitatively accurate to late times. After zero-noise extrapolation, the $p=0.3$ data is quantitatively accurate up to $t\approx 12$, up to which time it shows good agreement with DP critical scaling. Similarly, Fig.~\ref{fig:fig3}\textbf{b} shows the experimentally measured mean-squared extent of the active cluster compared to the theoretical results, which also shows a deviation from DP scaling after $t \approx 12$ time steps near the critical point. Error bars are standard errors of the mean obtained from boot-strap resampling with 100 resamples.

\begin{figure}[!t]
    \centering
    \includegraphics[width=0.5\textwidth]{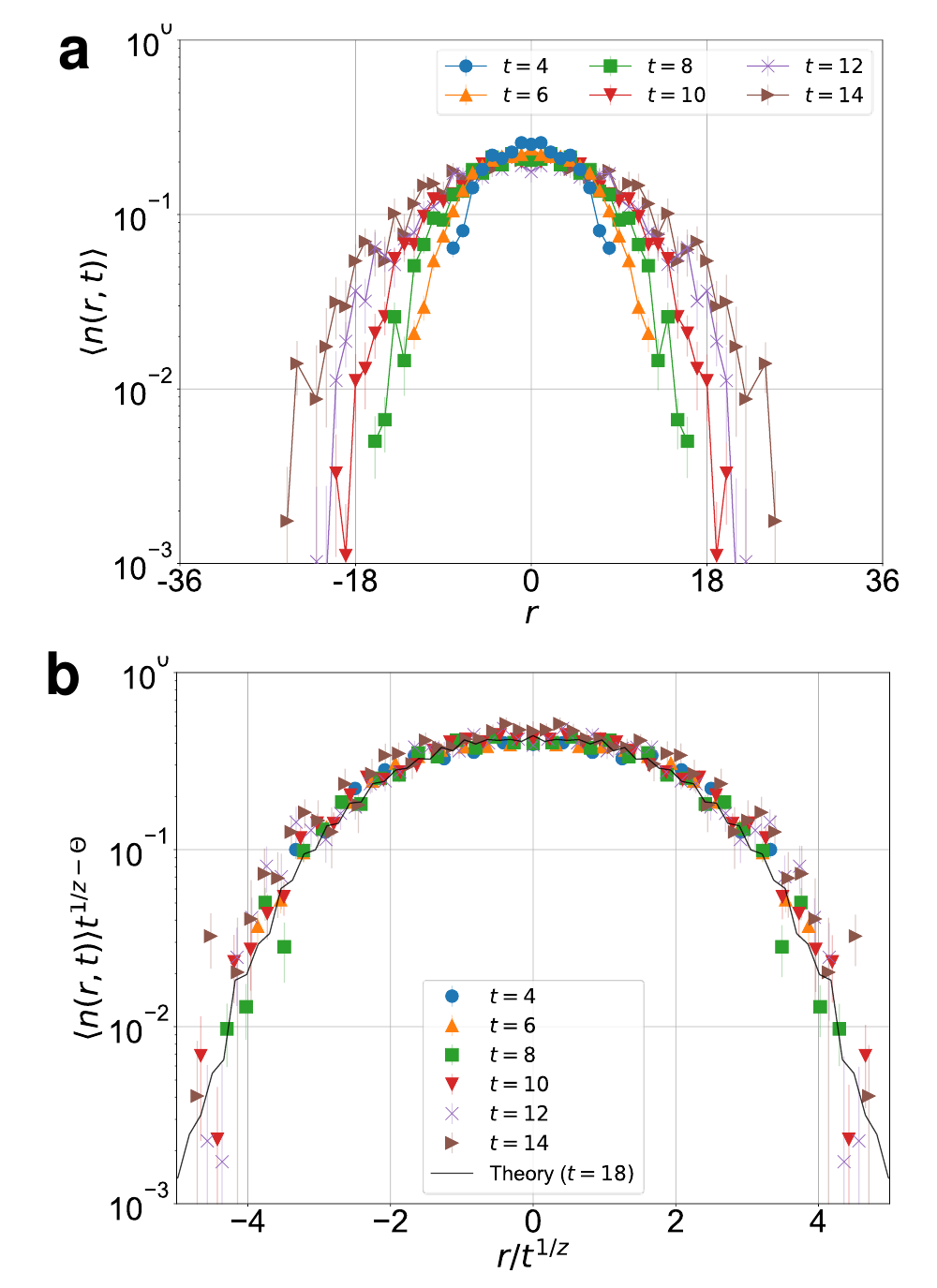}
    \caption{\textbf{Scaling collapse on quantum data.} \textbf{a} The experimentally measured active-site density $\langle n(r,t) \rangle$ versus position $r$ for fixed times $t$ near the critical point ($p=0.3\approx p_c$), mitigated by zero-noise extrapolation. \textbf{b} Scaling collapse of the experimental data using known directed percolation exponents $\Theta\approx 0.31$ and $z\approx 1.58$. The solid black curve shows the scaling collapsed curve for 10,000-shot noise-less quantum trajectory simulations performed at $t=18$.}
    \label{fig:fig4}
\end{figure}

A scaling collapse of the experimental data and theoretical simulations at the critical point is shown in Fig.~\ref{fig:fig4}\textbf{b}, with $y=\langle n(r,t)\rangle t^{1/z-\Theta}$ plotted versus $x=r/t^{1/z}$ and with $\Theta,z$ set to the known DP values. For large $t$ at the critical point, these rescaled points are expected to lie on a universal curve $y=f(x)$. Even for the limited $t$ available, we are able to see a reasonable collapse of the data to a universal curve. The largest deviations from the collapse are seen in the $t\gtrsim 12$ experimental data, which show clear deviations from the early-time data, consistent with a change in scaling behavior due to TQ gate errors.

\textbf{Discussion.} We have demonstrated that near-term quantum computers are capable of studying interesting non-equilibrium dissipative quantum phenomena, by examining the phase transition of a quantum circuit generalization of the contact process. The tools we have utilized to make this quantum simulation experimentally feasible -- such as qubit-reuse and error avoidance using real-time conditional logic -- are generally applicable and will be helpful primitives for future studies.

For the particular observables and initial states studied, both tensor-network numerics and direct simulation on a quantum computer indicate that our model exhibits directed percolation critical scaling at the phase transition. In related continuous-time models\cite{Marcuzzi2016,Carollo2019,Gillman2019}, scaling consistent with directed percolation has been observed in some contexts\cite{Gillman2019} and not in others\cite{Carollo2019}. In future work, it would be useful to better understand the origin of this discrepancy and whether it also exists in the discrete-time models considered in this work. Such a study would greatly benefit from the development of additional tools for simulating open quantum systems, since existing tools \cite{Verstraete2004,Bonnes2014,Cui2015,Mascarenhas2015,Werner2016,White2018,Jaschke2018,Cheng2021,Weimer2021,Cheng2022} are currently underdeveloped compared to closed system simulation methods.

In general, open quantum systems, which can have entropy increasing \emph{and decreasing} processes, provide a rich platform for realizing interesting dynamics and steady state physics and, in practice, are difficult to classically simulate\cite{Weimer2021}. One promising future direction would be to study open system dynamics in higher spatial dimensions on a quantum computer, since classical simulations of such dynamics are particularly challenging. It would also be important to explore different types of non-equilibrium phase transitions than the one considered in this work, particularly those that are robust to hardware errors (in our case, the directed percolation transition is sensitive to bit flips caused by gate errors). One could also investigate the non-equilibrium behavior of quantum circuits with classical feedback mechanisms \cite{Buchhold2022,Iadecola2022}, e.g., similar to those employed in quantum error correction, and see if such dynamics can give rise to interesting physics or error-robust critical phenomena.

\section{Acknowledgements}

This work was made possible by a large group of people, and the authors would like to thank the entire Quantinuum team for their many contributions. The experiments reported in this manuscript were performed on the Quantinuum system model H1-1 quantum computer\cite{H11}, which is powered by Honeywell ion traps. Numerical calculations were performed using the ITensor library \cite{itensor}. We thank Sebastian Diehl, Michael Buchold, Kevin Hemery, Henrik Dreyer, Reza Haghshenas, Natalie Brown, Ciaran Ryan-Anderson, Matthew DeCross, Karl Mayer, Christopher Langlett, Michael Lubasch, Michael Wall, Philip Daniel Blocher, Ivan Deutsch, Mohsin Iqbal, and Vedika Khemani for helpful discussions.
This research was supported in part by the National Science Foundation under Grant No. NSF PHY-1748958. ACP was supported by DOE  DE-SC0022102, and the Alfred P. Sloan Foundation through a Sloan Research Fellowship. ACP and SG performed this work in part at the Aspen Center for Physics which is supported by NSF grant PHY-1607611. This research used resources of the Oak Ridge Leadership Computing Facility, which is a DOE Office of Science User Facility supported under Contract DE-AC05-00OR22725.

\bibliography{refs}

\begin{thebibliography}{43}%
\makeatletter
\providecommand \@ifxundefined [1]{%
 \@ifx{#1\undefined}
}%
\providecommand \@ifnum [1]{%
 \ifnum #1\expandafter \@firstoftwo
 \else \expandafter \@secondoftwo
 \fi
}%
\providecommand \@ifx [1]{%
 \ifx #1\expandafter \@firstoftwo
 \else \expandafter \@secondoftwo
 \fi
}%
\providecommand \natexlab [1]{#1}%
\providecommand \enquote  [1]{``#1''}%
\providecommand \bibnamefont  [1]{#1}%
\providecommand \bibfnamefont [1]{#1}%
\providecommand \citenamefont [1]{#1}%
\providecommand \href@noop [0]{\@secondoftwo}%
\providecommand \href [0]{\begingroup \@sanitize@url \@href}%
\providecommand \@href[1]{\@@startlink{#1}\@@href}%
\providecommand \@@href[1]{\endgroup#1\@@endlink}%
\providecommand \@sanitize@url [0]{\catcode `\\12\catcode `\$12\catcode
  `\&12\catcode `\#12\catcode `\^12\catcode `\_12\catcode `\%12\relax}%
\providecommand \@@startlink[1]{}%
\providecommand \@@endlink[0]{}%
\providecommand \url  [0]{\begingroup\@sanitize@url \@url }%
\providecommand \@url [1]{\endgroup\@href {#1}{\urlprefix }}%
\providecommand \urlprefix  [0]{URL }%
\providecommand \Eprint [0]{\href }%
\providecommand \doibase [0]{https://doi.org/}%
\providecommand \selectlanguage [0]{\@gobble}%
\providecommand \bibinfo  [0]{\@secondoftwo}%
\providecommand \bibfield  [0]{\@secondoftwo}%
\providecommand \translation [1]{[#1]}%
\providecommand \BibitemOpen [0]{}%
\providecommand \bibitemStop [0]{}%
\providecommand \bibitemNoStop [0]{.\EOS\space}%
\providecommand \EOS [0]{\spacefactor3000\relax}%
\providecommand \BibitemShut  [1]{\csname bibitem#1\endcsname}%
\let\auto@bib@innerbib\@empty
\bibitem [{\citenamefont {Weimer}\ \emph {et~al.}(2021)\citenamefont {Weimer},
  \citenamefont {Kshetrimayum},\ and\ \citenamefont {Or\'us}}]{Weimer2021}%
  \BibitemOpen
  \bibfield  {author} {\bibinfo {author} {\bibfnamefont {H.}~\bibnamefont
  {Weimer}}, \bibinfo {author} {\bibfnamefont {A.}~\bibnamefont
  {Kshetrimayum}},\ and\ \bibinfo {author} {\bibfnamefont {R.}~\bibnamefont
  {Or\'us}},\ }\bibfield  {title} {\bibinfo {title} {Simulation methods for
  open quantum many-body systems},\ }\href
  {https://doi.org/10.1103/RevModPhys.93.015008} {\bibfield  {journal}
  {\bibinfo  {journal} {Rev. Mod. Phys.}\ }\textbf {\bibinfo {volume} {93}},\
  \bibinfo {pages} {015008} (\bibinfo {year} {2021})}\BibitemShut {NoStop}%
\bibitem [{\citenamefont {Harris}(1974)}]{Harris1974}%
  \BibitemOpen
  \bibfield  {author} {\bibinfo {author} {\bibfnamefont {T.~E.}\ \bibnamefont
  {Harris}},\ }\bibfield  {title} {\bibinfo {title} {Contact interactions on a
  lattice},\ }\href {http://www.jstor.org/stable/2959099} {\bibfield  {journal}
  {\bibinfo  {journal} {Ann. Prob.}\ }\textbf {\bibinfo {volume} {2}},\
  \bibinfo {pages} {969} (\bibinfo {year} {1974})}\BibitemShut {NoStop}%
\bibitem [{\citenamefont {Hinrichsen}(2000)}]{Hinrichsen2000}%
  \BibitemOpen
  \bibfield  {author} {\bibinfo {author} {\bibfnamefont {H.}~\bibnamefont
  {Hinrichsen}},\ }\bibfield  {title} {\bibinfo {title} {Non-equilibrium
  critical phenomena and phase transitions into absorbing states},\ }\href
  {https://doi.org/10.1080/00018730050198152} {\bibfield  {journal} {\bibinfo
  {journal} {Adv. Phys.}\ }\textbf {\bibinfo {volume} {49}},\ \bibinfo {pages}
  {815} (\bibinfo {year} {2000})}\BibitemShut {NoStop}%
\bibitem [{\citenamefont {Marcuzzi}\ \emph {et~al.}(2016)\citenamefont
  {Marcuzzi}, \citenamefont {Buchhold}, \citenamefont {Diehl},\ and\
  \citenamefont {Lesanovsky}}]{Marcuzzi2016}%
  \BibitemOpen
  \bibfield  {author} {\bibinfo {author} {\bibfnamefont {M.}~\bibnamefont
  {Marcuzzi}}, \bibinfo {author} {\bibfnamefont {M.}~\bibnamefont {Buchhold}},
  \bibinfo {author} {\bibfnamefont {S.}~\bibnamefont {Diehl}},\ and\ \bibinfo
  {author} {\bibfnamefont {I.}~\bibnamefont {Lesanovsky}},\ }\bibfield  {title}
  {\bibinfo {title} {Absorbing state phase transition with competing quantum
  and classical fluctuations},\ }\href
  {https://doi.org/10.1103/PhysRevLett.116.245701} {\bibfield  {journal}
  {\bibinfo  {journal} {Phys. Rev. Lett.}\ }\textbf {\bibinfo {volume} {116}},\
  \bibinfo {pages} {245701} (\bibinfo {year} {2016})}\BibitemShut {NoStop}%
\bibitem [{\citenamefont {Carollo}\ \emph {et~al.}(2019)\citenamefont
  {Carollo}, \citenamefont {Gillman}, \citenamefont {Weimer},\ and\
  \citenamefont {Lesanovsky}}]{Carollo2019}%
  \BibitemOpen
  \bibfield  {author} {\bibinfo {author} {\bibfnamefont {F.}~\bibnamefont
  {Carollo}}, \bibinfo {author} {\bibfnamefont {E.}~\bibnamefont {Gillman}},
  \bibinfo {author} {\bibfnamefont {H.}~\bibnamefont {Weimer}},\ and\ \bibinfo
  {author} {\bibfnamefont {I.}~\bibnamefont {Lesanovsky}},\ }\bibfield  {title}
  {\bibinfo {title} {Critical behavior of the quantum contact process in one
  dimension},\ }\href {https://doi.org/10.1103/PhysRevLett.123.100604}
  {\bibfield  {journal} {\bibinfo  {journal} {Phys. Rev. Lett.}\ }\textbf
  {\bibinfo {volume} {123}},\ \bibinfo {pages} {100604} (\bibinfo {year}
  {2019})}\BibitemShut {NoStop}%
\bibitem [{\citenamefont {Gillman}\ \emph {et~al.}(2019)\citenamefont
  {Gillman}, \citenamefont {Carollo},\ and\ \citenamefont
  {Lesanovsky}}]{Gillman2019}%
  \BibitemOpen
  \bibfield  {author} {\bibinfo {author} {\bibfnamefont {E.}~\bibnamefont
  {Gillman}}, \bibinfo {author} {\bibfnamefont {F.}~\bibnamefont {Carollo}},\
  and\ \bibinfo {author} {\bibfnamefont {I.}~\bibnamefont {Lesanovsky}},\
  }\bibfield  {title} {\bibinfo {title} {Numerical simulation of critical
  dissipative non-equilibrium quantum systems with an absorbing state},\ }\href
  {https://doi.org/10.1088/1367-2630/ab43b0} {\bibfield  {journal} {\bibinfo
  {journal} {New J. Phys.}\ }\textbf {\bibinfo {volume} {21}},\ \bibinfo
  {pages} {093064} (\bibinfo {year} {2019})}\BibitemShut {NoStop}%
\bibitem [{\citenamefont {Jo}\ \emph {et~al.}(2021)\citenamefont {Jo},
  \citenamefont {Lee}, \citenamefont {Choi},\ and\ \citenamefont
  {Kahng}}]{Minjae2021}%
  \BibitemOpen
  \bibfield  {author} {\bibinfo {author} {\bibfnamefont {M.}~\bibnamefont
  {Jo}}, \bibinfo {author} {\bibfnamefont {J.}~\bibnamefont {Lee}}, \bibinfo
  {author} {\bibfnamefont {K.}~\bibnamefont {Choi}},\ and\ \bibinfo {author}
  {\bibfnamefont {B.}~\bibnamefont {Kahng}},\ }\bibfield  {title} {\bibinfo
  {title} {Absorbing phase transition with a continuously varying exponent in a
  quantum contact process: A neural network approach},\ }\href
  {https://doi.org/10.1103/PhysRevResearch.3.013238} {\bibfield  {journal}
  {\bibinfo  {journal} {Phys. Rev. Research}\ }\textbf {\bibinfo {volume}
  {3}},\ \bibinfo {pages} {013238} (\bibinfo {year} {2021})}\BibitemShut
  {NoStop}%
\bibitem [{\citenamefont {Lesanovsky}\ \emph {et~al.}(2019)\citenamefont
  {Lesanovsky}, \citenamefont {Macieszczak},\ and\ \citenamefont
  {Garrahan}}]{Lesanovsky2019}%
  \BibitemOpen
  \bibfield  {author} {\bibinfo {author} {\bibfnamefont {I.}~\bibnamefont
  {Lesanovsky}}, \bibinfo {author} {\bibfnamefont {K.}~\bibnamefont
  {Macieszczak}},\ and\ \bibinfo {author} {\bibfnamefont {J.~P.}\ \bibnamefont
  {Garrahan}},\ }\bibfield  {title} {\bibinfo {title} {Non-equilibrium
  absorbing state phase transitions in discrete-time quantum cellular automaton
  dynamics on spin lattices},\ }\href
  {https://doi.org/10.1088/2058-9565/aaf831} {\bibfield  {journal} {\bibinfo
  {journal} {Quantum Sci. Technol.}\ }\textbf {\bibinfo {volume} {4}},\
  \bibinfo {pages} {02LT02} (\bibinfo {year} {2019})}\BibitemShut {NoStop}%
\bibitem [{\citenamefont {Gillman}\ \emph {et~al.}(2020)\citenamefont
  {Gillman}, \citenamefont {Carollo},\ and\ \citenamefont
  {Lesanovsky}}]{Gillman2020}%
  \BibitemOpen
  \bibfield  {author} {\bibinfo {author} {\bibfnamefont {E.}~\bibnamefont
  {Gillman}}, \bibinfo {author} {\bibfnamefont {F.}~\bibnamefont {Carollo}},\
  and\ \bibinfo {author} {\bibfnamefont {I.}~\bibnamefont {Lesanovsky}},\
  }\bibfield  {title} {\bibinfo {title} {Nonequilibrium phase transitions in
  ($1+1$)-dimensional quantum cellular automata with controllable quantum
  correlations},\ }\href {https://doi.org/10.1103/PhysRevLett.125.100403}
  {\bibfield  {journal} {\bibinfo  {journal} {Phys. Rev. Lett.}\ }\textbf
  {\bibinfo {volume} {125}},\ \bibinfo {pages} {100403} (\bibinfo {year}
  {2020})}\BibitemShut {NoStop}%
\bibitem [{\citenamefont {Gillman}\ \emph
  {et~al.}(2021{\natexlab{a}})\citenamefont {Gillman}, \citenamefont
  {Carollo},\ and\ \citenamefont {Lesanovsky}}]{Gillman2021a}%
  \BibitemOpen
  \bibfield  {author} {\bibinfo {author} {\bibfnamefont {E.}~\bibnamefont
  {Gillman}}, \bibinfo {author} {\bibfnamefont {F.}~\bibnamefont {Carollo}},\
  and\ \bibinfo {author} {\bibfnamefont {I.}~\bibnamefont {Lesanovsky}},\
  }\bibfield  {title} {\bibinfo {title} {Numerical simulation of quantum
  nonequilibrium phase transitions without finite-size effects},\ }\href
  {https://doi.org/10.1103/PhysRevA.103.L040201} {\bibfield  {journal}
  {\bibinfo  {journal} {Phys. Rev. A}\ }\textbf {\bibinfo {volume} {103}},\
  \bibinfo {pages} {L040201} (\bibinfo {year}
  {2021}{\natexlab{a}})}\BibitemShut {NoStop}%
\bibitem [{\citenamefont {Gillman}\ \emph
  {et~al.}(2021{\natexlab{b}})\citenamefont {Gillman}, \citenamefont
  {Carollo},\ and\ \citenamefont {Lesanovsky}}]{Gillman2021b}%
  \BibitemOpen
  \bibfield  {author} {\bibinfo {author} {\bibfnamefont {E.}~\bibnamefont
  {Gillman}}, \bibinfo {author} {\bibfnamefont {F.}~\bibnamefont {Carollo}},\
  and\ \bibinfo {author} {\bibfnamefont {I.}~\bibnamefont {Lesanovsky}},\
  }\bibfield  {title} {\bibinfo {title} {Quantum and classical temporal
  correlations in $(1+1)\mathrm{D}$ quantum cellular automata},\ }\href
  {https://doi.org/10.1103/PhysRevLett.127.230502} {\bibfield  {journal}
  {\bibinfo  {journal} {Phys. Rev. Lett.}\ }\textbf {\bibinfo {volume} {127}},\
  \bibinfo {pages} {230502} (\bibinfo {year} {2021}{\natexlab{b}})}\BibitemShut
  {NoStop}%
\bibitem [{\citenamefont {Gillman}\ \emph {et~al.}(2022)\citenamefont
  {Gillman}, \citenamefont {Carollo},\ and\ \citenamefont
  {Lesanovsky}}]{Gillman2022}%
  \BibitemOpen
  \bibfield  {author} {\bibinfo {author} {\bibfnamefont {E.}~\bibnamefont
  {Gillman}}, \bibinfo {author} {\bibfnamefont {F.}~\bibnamefont {Carollo}},\
  and\ \bibinfo {author} {\bibfnamefont {I.}~\bibnamefont {Lesanovsky}},\
  }\bibfield  {title} {\bibinfo {title} {Asynchronism and nonequilibrium phase
  transitions in $(1+1)$-dimensional quantum cellular automata},\ }\href
  {https://doi.org/10.1103/PhysRevE.106.L032103} {\bibfield  {journal}
  {\bibinfo  {journal} {Phys. Rev. E}\ }\textbf {\bibinfo {volume} {106}},\
  \bibinfo {pages} {L032103} (\bibinfo {year} {2022})}\BibitemShut {NoStop}%
\bibitem [{\citenamefont {Marro}\ and\ \citenamefont
  {Dickman}(1999)}]{Marro1999}%
  \BibitemOpen
  \bibfield  {author} {\bibinfo {author} {\bibfnamefont {J.}~\bibnamefont
  {Marro}}\ and\ \bibinfo {author} {\bibfnamefont {R.}~\bibnamefont
  {Dickman}},\ }\href {https://doi.org/10.1017/CBO9780511524288} {\emph
  {\bibinfo {title} {Nonequilibrium Phase Transitions in Lattice Models}}},\
  Collection Alea-Saclay: Monographs and Texts in Statistical Physics\
  (\bibinfo  {publisher} {Cambridge University Press},\ \bibinfo {year}
  {1999})\BibitemShut {NoStop}%
\bibitem [{\citenamefont {\'Odor}(2004)}]{Odor2004}%
  \BibitemOpen
  \bibfield  {author} {\bibinfo {author} {\bibfnamefont {G.}~\bibnamefont
  {\'Odor}},\ }\bibfield  {title} {\bibinfo {title} {Universality classes in
  nonequilibrium lattice systems},\ }\href
  {https://doi.org/10.1103/RevModPhys.76.663} {\bibfield  {journal} {\bibinfo
  {journal} {Rev. Mod. Phys.}\ }\textbf {\bibinfo {volume} {76}},\ \bibinfo
  {pages} {663} (\bibinfo {year} {2004})}\BibitemShut {NoStop}%
\bibitem [{\citenamefont {Jensen}(1999)}]{Jensen1999}%
  \BibitemOpen
  \bibfield  {author} {\bibinfo {author} {\bibfnamefont {I.}~\bibnamefont
  {Jensen}},\ }\bibfield  {title} {\bibinfo {title} {Low-density series
  expansions for directed percolation: I. a new efficient algorithm with
  applications to the square lattice},\ }\href
  {https://doi.org/10.1088/0305-4470/32/28/304} {\bibfield  {journal} {\bibinfo
   {journal} {J. Phys. A}\ }\textbf {\bibinfo {volume} {32}},\ \bibinfo {pages}
  {5233} (\bibinfo {year} {1999})}\BibitemShut {NoStop}%
\bibitem [{\citenamefont {Hinrichsen}(2006)}]{Hinrichsen2006}%
  \BibitemOpen
  \bibfield  {author} {\bibinfo {author} {\bibfnamefont {H.}~\bibnamefont
  {Hinrichsen}},\ }\bibfield  {title} {\bibinfo {title} {Non-equilibrium phase
  transitions},\ }\href
  {https://doi.org/https://doi.org/10.1016/j.physa.2006.04.007} {\bibfield
  {journal} {\bibinfo  {journal} {Physica A}\ }\textbf {\bibinfo {volume}
  {369}},\ \bibinfo {pages} {1} (\bibinfo {year} {2006})}\BibitemShut {NoStop}%
\bibitem [{\citenamefont {Ryan-Anderson}\ \emph {et~al.}(2022)\citenamefont
  {Ryan-Anderson}, \citenamefont {Brown}, \citenamefont {Allman}, \citenamefont
  {Arkin}, \citenamefont {Asa-Attuah}, \citenamefont {Baldwin}, \citenamefont
  {Berg}, \citenamefont {Bohnet}, \citenamefont {Braxton}, \citenamefont
  {Burdick}, \citenamefont {Campora}, \citenamefont {Chernoguzov},
  \citenamefont {Esposito}, \citenamefont {Evans}, \citenamefont {Francois},
  \citenamefont {Gaebler}, \citenamefont {Gatterman}, \citenamefont {Gerber},
  \citenamefont {Gilmore}, \citenamefont {Gresh}, \citenamefont {Hall},
  \citenamefont {Hankin}, \citenamefont {Hostetter}, \citenamefont {Lucchetti},
  \citenamefont {Mayer}, \citenamefont {Myers}, \citenamefont {Neyenhuis},
  \citenamefont {Santiago}, \citenamefont {Sedlacek}, \citenamefont {Skripka},
  \citenamefont {Slattery}, \citenamefont {Stutz}, \citenamefont {Tait},
  \citenamefont {Tobey}, \citenamefont {Vittorini}, \citenamefont {Walker},\
  and\ \citenamefont {Hayes}}]{RyanAnderson2022}%
  \BibitemOpen
  \bibfield  {author} {\bibinfo {author} {\bibfnamefont {C.}~\bibnamefont
  {Ryan-Anderson}}, \bibinfo {author} {\bibfnamefont {N.~C.}\ \bibnamefont
  {Brown}}, \bibinfo {author} {\bibfnamefont {M.~S.}\ \bibnamefont {Allman}},
  \bibinfo {author} {\bibfnamefont {B.}~\bibnamefont {Arkin}}, \bibinfo
  {author} {\bibfnamefont {G.}~\bibnamefont {Asa-Attuah}}, \bibinfo {author}
  {\bibfnamefont {C.}~\bibnamefont {Baldwin}}, \bibinfo {author} {\bibfnamefont
  {J.}~\bibnamefont {Berg}}, \bibinfo {author} {\bibfnamefont {J.~G.}\
  \bibnamefont {Bohnet}}, \bibinfo {author} {\bibfnamefont {S.}~\bibnamefont
  {Braxton}}, \bibinfo {author} {\bibfnamefont {N.}~\bibnamefont {Burdick}},
  \bibinfo {author} {\bibfnamefont {J.~P.}\ \bibnamefont {Campora}}, \bibinfo
  {author} {\bibfnamefont {A.}~\bibnamefont {Chernoguzov}}, \bibinfo {author}
  {\bibfnamefont {J.}~\bibnamefont {Esposito}}, \bibinfo {author}
  {\bibfnamefont {B.}~\bibnamefont {Evans}}, \bibinfo {author} {\bibfnamefont
  {D.}~\bibnamefont {Francois}}, \bibinfo {author} {\bibfnamefont {J.~P.}\
  \bibnamefont {Gaebler}}, \bibinfo {author} {\bibfnamefont {T.~M.}\
  \bibnamefont {Gatterman}}, \bibinfo {author} {\bibfnamefont {J.}~\bibnamefont
  {Gerber}}, \bibinfo {author} {\bibfnamefont {K.}~\bibnamefont {Gilmore}},
  \bibinfo {author} {\bibfnamefont {D.}~\bibnamefont {Gresh}}, \bibinfo
  {author} {\bibfnamefont {A.}~\bibnamefont {Hall}}, \bibinfo {author}
  {\bibfnamefont {A.}~\bibnamefont {Hankin}}, \bibinfo {author} {\bibfnamefont
  {J.}~\bibnamefont {Hostetter}}, \bibinfo {author} {\bibfnamefont
  {D.}~\bibnamefont {Lucchetti}}, \bibinfo {author} {\bibfnamefont
  {K.}~\bibnamefont {Mayer}}, \bibinfo {author} {\bibfnamefont
  {J.}~\bibnamefont {Myers}}, \bibinfo {author} {\bibfnamefont
  {B.}~\bibnamefont {Neyenhuis}}, \bibinfo {author} {\bibfnamefont
  {J.}~\bibnamefont {Santiago}}, \bibinfo {author} {\bibfnamefont
  {J.}~\bibnamefont {Sedlacek}}, \bibinfo {author} {\bibfnamefont
  {T.}~\bibnamefont {Skripka}}, \bibinfo {author} {\bibfnamefont
  {A.}~\bibnamefont {Slattery}}, \bibinfo {author} {\bibfnamefont {R.~P.}\
  \bibnamefont {Stutz}}, \bibinfo {author} {\bibfnamefont {J.}~\bibnamefont
  {Tait}}, \bibinfo {author} {\bibfnamefont {R.}~\bibnamefont {Tobey}},
  \bibinfo {author} {\bibfnamefont {G.}~\bibnamefont {Vittorini}}, \bibinfo
  {author} {\bibfnamefont {J.}~\bibnamefont {Walker}},\ and\ \bibinfo {author}
  {\bibfnamefont {D.}~\bibnamefont {Hayes}},\ }\bibfield  {title} {\bibinfo
  {title} {Implementing fault-tolerant entangling gates on the five-qubit code
  and the color code},\ }\href {https://arxiv.org/abs/2208.01863} {\bibfield
  {journal} {\bibinfo  {journal} {arXiv:2208.01863}\ } (\bibinfo {year}
  {2022})}\BibitemShut {NoStop}%
\bibitem [{\citenamefont {Pino}\ \emph {et~al.}(2021)\citenamefont {Pino},
  \citenamefont {Dreiling}, \citenamefont {Figgatt}, \citenamefont {Gaebler},
  \citenamefont {Moses}, \citenamefont {Allman}, \citenamefont {Baldwin},
  \citenamefont {Foss-Feig}, \citenamefont {Hayes}, \citenamefont {Mayer},
  \citenamefont {Ryan-Anderson},\ and\ \citenamefont {Neyenhuis}}]{pino2020}%
  \BibitemOpen
  \bibfield  {author} {\bibinfo {author} {\bibfnamefont {J.~M.}\ \bibnamefont
  {Pino}}, \bibinfo {author} {\bibfnamefont {J.~M.}\ \bibnamefont {Dreiling}},
  \bibinfo {author} {\bibfnamefont {C.}~\bibnamefont {Figgatt}}, \bibinfo
  {author} {\bibfnamefont {J.~P.}\ \bibnamefont {Gaebler}}, \bibinfo {author}
  {\bibfnamefont {S.~A.}\ \bibnamefont {Moses}}, \bibinfo {author}
  {\bibfnamefont {M.~S.}\ \bibnamefont {Allman}}, \bibinfo {author}
  {\bibfnamefont {C.~H.}\ \bibnamefont {Baldwin}}, \bibinfo {author}
  {\bibfnamefont {M.}~\bibnamefont {Foss-Feig}}, \bibinfo {author}
  {\bibfnamefont {D.}~\bibnamefont {Hayes}}, \bibinfo {author} {\bibfnamefont
  {K.}~\bibnamefont {Mayer}}, \bibinfo {author} {\bibfnamefont
  {C.}~\bibnamefont {Ryan-Anderson}},\ and\ \bibinfo {author} {\bibfnamefont
  {B.}~\bibnamefont {Neyenhuis}},\ }\bibfield  {title} {\bibinfo {title}
  {{Demonstration of the trapped-ion quantum CCD computer architecture}},\
  }\href {https://doi.org/10.1038/s41586-021-03318-4} {\bibfield  {journal}
  {\bibinfo  {journal} {Nature}\ }\textbf {\bibinfo {volume} {592}},\ \bibinfo
  {pages} {209} (\bibinfo {year} {2021})}\BibitemShut {NoStop}%
\bibitem [{\citenamefont {Kim}(2017)}]{kim2017}%
  \BibitemOpen
  \bibfield  {author} {\bibinfo {author} {\bibfnamefont {I.~H.}\ \bibnamefont
  {Kim}},\ }\bibfield  {title} {\bibinfo {title} {Holographic quantum
  simulation},\ }\href {https://arxiv.org/abs/1702.02093} {\bibfield  {journal}
  {\bibinfo  {journal} {arXiv:1702.02093}\ } (\bibinfo {year}
  {2017})}\BibitemShut {NoStop}%
\bibitem [{\citenamefont {Foss-Feig}\ \emph {et~al.}(2021)\citenamefont
  {Foss-Feig}, \citenamefont {Hayes}, \citenamefont {Dreiling}, \citenamefont
  {Figgatt}, \citenamefont {Gaebler}, \citenamefont {Moses}, \citenamefont
  {Pino},\ and\ \citenamefont {Potter}}]{fossfeig2020}%
  \BibitemOpen
  \bibfield  {author} {\bibinfo {author} {\bibfnamefont {M.}~\bibnamefont
  {Foss-Feig}}, \bibinfo {author} {\bibfnamefont {D.}~\bibnamefont {Hayes}},
  \bibinfo {author} {\bibfnamefont {J.~M.}\ \bibnamefont {Dreiling}}, \bibinfo
  {author} {\bibfnamefont {C.}~\bibnamefont {Figgatt}}, \bibinfo {author}
  {\bibfnamefont {J.~P.}\ \bibnamefont {Gaebler}}, \bibinfo {author}
  {\bibfnamefont {S.~A.}\ \bibnamefont {Moses}}, \bibinfo {author}
  {\bibfnamefont {J.~M.}\ \bibnamefont {Pino}},\ and\ \bibinfo {author}
  {\bibfnamefont {A.~C.}\ \bibnamefont {Potter}},\ }\bibfield  {title}
  {\bibinfo {title} {{Holographic quantum algorithms for simulating correlated
  spin systems}},\ }\href
  {https://link.aps.org/doi/10.1103/PhysRevResearch.3.033002} {\bibfield
  {journal} {\bibinfo  {journal} {Phys. Rev. Research}\ }\textbf {\bibinfo
  {volume} {3}},\ \bibinfo {pages} {033002} (\bibinfo {year}
  {2021})}\BibitemShut {NoStop}%
\bibitem [{\citenamefont {Barratt}\ \emph {et~al.}(2021)\citenamefont
  {Barratt}, \citenamefont {Dborin}, \citenamefont {Bal}, \citenamefont
  {Stojevic}, \citenamefont {Pollmann},\ and\ \citenamefont
  {Green}}]{Barratt2021}%
  \BibitemOpen
  \bibfield  {author} {\bibinfo {author} {\bibfnamefont {F.}~\bibnamefont
  {Barratt}}, \bibinfo {author} {\bibfnamefont {J.}~\bibnamefont {Dborin}},
  \bibinfo {author} {\bibfnamefont {M.}~\bibnamefont {Bal}}, \bibinfo {author}
  {\bibfnamefont {V.}~\bibnamefont {Stojevic}}, \bibinfo {author}
  {\bibfnamefont {F.}~\bibnamefont {Pollmann}},\ and\ \bibinfo {author}
  {\bibfnamefont {A.~G.}\ \bibnamefont {Green}},\ }\bibfield  {title} {\bibinfo
  {title} {Parallel quantum simulation of large systems on small {NISQ}
  computers},\ }\href {https://doi.org/10.1038%2Fs41534-021-00420-3} {\bibfield
   {journal} {\bibinfo  {journal} {{Npj Quantum Inf.}}\ }\textbf {\bibinfo
  {volume} {7}} (\bibinfo {year} {2021})}\BibitemShut {NoStop}%
\bibitem [{\citenamefont {Chertkov}\ \emph {et~al.}(2022)\citenamefont
  {Chertkov}, \citenamefont {Bohnet}, \citenamefont {Francois}, \citenamefont
  {Gaebler}, \citenamefont {Gresh}, \citenamefont {Hankin}, \citenamefont
  {Lee}, \citenamefont {Hayes}, \citenamefont {Neyenhuis}, \citenamefont
  {Stutz}, \citenamefont {Potter},\ and\ \citenamefont
  {Foss-Feig}}]{Chertkov2022}%
  \BibitemOpen
  \bibfield  {author} {\bibinfo {author} {\bibfnamefont {E.}~\bibnamefont
  {Chertkov}}, \bibinfo {author} {\bibfnamefont {J.}~\bibnamefont {Bohnet}},
  \bibinfo {author} {\bibfnamefont {D.}~\bibnamefont {Francois}}, \bibinfo
  {author} {\bibfnamefont {J.}~\bibnamefont {Gaebler}}, \bibinfo {author}
  {\bibfnamefont {D.}~\bibnamefont {Gresh}}, \bibinfo {author} {\bibfnamefont
  {A.}~\bibnamefont {Hankin}}, \bibinfo {author} {\bibfnamefont
  {K.}~\bibnamefont {Lee}}, \bibinfo {author} {\bibfnamefont {D.}~\bibnamefont
  {Hayes}}, \bibinfo {author} {\bibfnamefont {B.}~\bibnamefont {Neyenhuis}},
  \bibinfo {author} {\bibfnamefont {R.}~\bibnamefont {Stutz}}, \bibinfo
  {author} {\bibfnamefont {A.~C.}\ \bibnamefont {Potter}},\ and\ \bibinfo
  {author} {\bibfnamefont {M.}~\bibnamefont {Foss-Feig}},\ }\bibfield  {title}
  {\bibinfo {title} {Holographic dynamics simulations with a trapped-ion
  quantum computer},\ }\href {{https://doi.org/10.1038%2Fs41567-022-01689-7}}
  {\bibfield  {journal} {\bibinfo  {journal} {Nat. Phys.}\ }\textbf {\bibinfo
  {volume} {18}},\ \bibinfo {pages} {1074} (\bibinfo {year}
  {2022})}\BibitemShut {NoStop}%
\bibitem [{\citenamefont {Niu}\ \emph {et~al.}(2021)\citenamefont {Niu},
  \citenamefont {Haghshenas}, \citenamefont {Zhang}, \citenamefont {Foss-Feig},
  \citenamefont {Chan},\ and\ \citenamefont {Potter}}]{Niu2021}%
  \BibitemOpen
  \bibfield  {author} {\bibinfo {author} {\bibfnamefont {D.}~\bibnamefont
  {Niu}}, \bibinfo {author} {\bibfnamefont {R.}~\bibnamefont {Haghshenas}},
  \bibinfo {author} {\bibfnamefont {Y.}~\bibnamefont {Zhang}}, \bibinfo
  {author} {\bibfnamefont {M.}~\bibnamefont {Foss-Feig}}, \bibinfo {author}
  {\bibfnamefont {G.~K.-L.}\ \bibnamefont {Chan}},\ and\ \bibinfo {author}
  {\bibfnamefont {A.~C.}\ \bibnamefont {Potter}},\ }\bibfield  {title}
  {\bibinfo {title} {Holographic simulation of correlated electrons on a
  trapped ion quantum processor},\ }\href
  {https://doi.org/10.48550/ARXIV.2112.10810} {\bibfield  {journal} {\bibinfo
  {journal} {arXiv:2112.10810}\ } (\bibinfo {year} {2021})}\BibitemShut
  {NoStop}%
\bibitem [{\citenamefont {Lin}\ \emph {et~al.}(2021)\citenamefont {Lin},
  \citenamefont {Dilip}, \citenamefont {Green}, \citenamefont {Smith},\ and\
  \citenamefont {Pollmann}}]{Lin2021}%
  \BibitemOpen
  \bibfield  {author} {\bibinfo {author} {\bibfnamefont {S.-H.}\ \bibnamefont
  {Lin}}, \bibinfo {author} {\bibfnamefont {R.}~\bibnamefont {Dilip}}, \bibinfo
  {author} {\bibfnamefont {A.~G.}\ \bibnamefont {Green}}, \bibinfo {author}
  {\bibfnamefont {A.}~\bibnamefont {Smith}},\ and\ \bibinfo {author}
  {\bibfnamefont {F.}~\bibnamefont {Pollmann}},\ }\bibfield  {title} {\bibinfo
  {title} {Real- and imaginary-time evolution with compressed quantum
  circuits},\ }\href {https://doi.org/10.1103/PRXQuantum.2.010342} {\bibfield
  {journal} {\bibinfo  {journal} {PRX Quantum}\ }\textbf {\bibinfo {volume}
  {2}},\ \bibinfo {pages} {010342} (\bibinfo {year} {2021})}\BibitemShut
  {NoStop}%
\bibitem [{\citenamefont {Zhang}\ \emph {et~al.}(2022)\citenamefont {Zhang},
  \citenamefont {Jahanbani}, \citenamefont {Niu}, \citenamefont {Haghshenas},\
  and\ \citenamefont {Potter}}]{Zhang2022}%
  \BibitemOpen
  \bibfield  {author} {\bibinfo {author} {\bibfnamefont {Y.}~\bibnamefont
  {Zhang}}, \bibinfo {author} {\bibfnamefont {S.}~\bibnamefont {Jahanbani}},
  \bibinfo {author} {\bibfnamefont {D.}~\bibnamefont {Niu}}, \bibinfo {author}
  {\bibfnamefont {R.}~\bibnamefont {Haghshenas}},\ and\ \bibinfo {author}
  {\bibfnamefont {A.~C.}\ \bibnamefont {Potter}},\ }\bibfield  {title}
  {\bibinfo {title} {Qubit-efficient simulation of thermal states with quantum
  tensor networks},\ }\href {https://arxiv.org/abs/2205.06299} {\bibfield
  {journal} {\bibinfo  {journal} {arXiv:2205.06299}\ } (\bibinfo {year}
  {2022})}\BibitemShut {NoStop}%
\bibitem [{\citenamefont {Dborin}\ \emph {et~al.}(2022)\citenamefont {Dborin},
  \citenamefont {Wimalaweera}, \citenamefont {Barratt}, \citenamefont {Ostby},
  \citenamefont {O'Brien},\ and\ \citenamefont {Green}}]{Dborin2022}%
  \BibitemOpen
  \bibfield  {author} {\bibinfo {author} {\bibfnamefont {J.}~\bibnamefont
  {Dborin}}, \bibinfo {author} {\bibfnamefont {V.}~\bibnamefont {Wimalaweera}},
  \bibinfo {author} {\bibfnamefont {F.}~\bibnamefont {Barratt}}, \bibinfo
  {author} {\bibfnamefont {E.}~\bibnamefont {Ostby}}, \bibinfo {author}
  {\bibfnamefont {T.~E.}\ \bibnamefont {O'Brien}},\ and\ \bibinfo {author}
  {\bibfnamefont {A.~G.}\ \bibnamefont {Green}},\ }\bibfield  {title} {\bibinfo
  {title} {Simulating groundstate and dynamical quantum phase transitions on a
  superconducting quantum computer},\ }\href {https://arxiv.org/abs/2205.12996}
  {\bibfield  {journal} {\bibinfo  {journal} {arXiv:2205.12996}\ } (\bibinfo
  {year} {2022})}\BibitemShut {NoStop}%
\bibitem [{\citenamefont {DeCross}\ \emph {et~al.}(2022)\citenamefont
  {DeCross}, \citenamefont {Chertkov}, \citenamefont {Kohagen},\ and\
  \citenamefont {Foss-Feig}}]{DeCross2022}%
  \BibitemOpen
  \bibfield  {author} {\bibinfo {author} {\bibfnamefont {M.}~\bibnamefont
  {DeCross}}, \bibinfo {author} {\bibfnamefont {E.}~\bibnamefont {Chertkov}},
  \bibinfo {author} {\bibfnamefont {M.}~\bibnamefont {Kohagen}},\ and\ \bibinfo
  {author} {\bibfnamefont {M.}~\bibnamefont {Foss-Feig}},\ }\bibfield  {title}
  {\bibinfo {title} {Qubit-reuse compilation with mid-circuit measurement and
  reset},\ }\href {https://arxiv.org/abs/2210.08039} {\bibfield  {journal}
  {\bibinfo  {journal} {arXiv:2210.08039}\ } (\bibinfo {year}
  {2022})}\BibitemShut {NoStop}%
\bibitem [{\citenamefont {{Quantinuum System Model H1 Product Data
  Sheet}}()}]{H11specsheet}%
  \BibitemOpen
  \bibfield  {author} {\bibinfo {author} {\bibnamefont {{Quantinuum System
  Model H1 Product Data Sheet}}},\ }\href@noop {} {}\bibinfo {howpublished}
  {https://www.quantinuum.com/products/h1},\ \bibinfo {note} {{Version 5.00.
  June 14, 2022}}\BibitemShut {NoStop}%
\bibitem [{\citenamefont {Bonnes}\ and\ \citenamefont
  {Läuchli}(2014)}]{Bonnes2014}%
  \BibitemOpen
  \bibfield  {author} {\bibinfo {author} {\bibfnamefont {L.}~\bibnamefont
  {Bonnes}}\ and\ \bibinfo {author} {\bibfnamefont {A.~M.}\ \bibnamefont
  {Läuchli}},\ }\bibfield  {title} {\bibinfo {title} {Superoperators vs.
  trajectories for matrix product state simulations of open quantum system: A
  case study},\ }\href {https://arxiv.org/abs/1411.4831} {\bibfield  {journal}
  {\bibinfo  {journal} {arXiv:1411.4831}\ } (\bibinfo {year}
  {2014})}\BibitemShut {NoStop}%
\bibitem [{\citenamefont {Temme}\ \emph {et~al.}(2017)\citenamefont {Temme},
  \citenamefont {Bravyi},\ and\ \citenamefont {Gambetta}}]{Temme2017}%
  \BibitemOpen
  \bibfield  {author} {\bibinfo {author} {\bibfnamefont {K.}~\bibnamefont
  {Temme}}, \bibinfo {author} {\bibfnamefont {S.}~\bibnamefont {Bravyi}},\ and\
  \bibinfo {author} {\bibfnamefont {J.~M.}\ \bibnamefont {Gambetta}},\
  }\bibfield  {title} {\bibinfo {title} {Error mitigation for short-depth
  quantum circuits},\ }\href {https://doi.org/10.1103/PhysRevLett.119.180509}
  {\bibfield  {journal} {\bibinfo  {journal} {Phys. Rev. Lett.}\ }\textbf
  {\bibinfo {volume} {119}},\ \bibinfo {pages} {180509} (\bibinfo {year}
  {2017})}\BibitemShut {NoStop}%
\bibitem [{\citenamefont {Li}\ and\ \citenamefont {Benjamin}(2017)}]{Li2017}%
  \BibitemOpen
  \bibfield  {author} {\bibinfo {author} {\bibfnamefont {Y.}~\bibnamefont
  {Li}}\ and\ \bibinfo {author} {\bibfnamefont {S.~C.}\ \bibnamefont
  {Benjamin}},\ }\bibfield  {title} {\bibinfo {title} {Efficient variational
  quantum simulator incorporating active error minimization},\ }\href
  {https://doi.org/10.1103/PhysRevX.7.021050} {\bibfield  {journal} {\bibinfo
  {journal} {Phys. Rev. X}\ }\textbf {\bibinfo {volume} {7}},\ \bibinfo {pages}
  {021050} (\bibinfo {year} {2017})}\BibitemShut {NoStop}%
\bibitem [{\citenamefont {Verstraete}\ \emph {et~al.}(2004)\citenamefont
  {Verstraete}, \citenamefont {Garc\'{\i}a-Ripoll},\ and\ \citenamefont
  {Cirac}}]{Verstraete2004}%
  \BibitemOpen
  \bibfield  {author} {\bibinfo {author} {\bibfnamefont {F.}~\bibnamefont
  {Verstraete}}, \bibinfo {author} {\bibfnamefont {J.~J.}\ \bibnamefont
  {Garc\'{\i}a-Ripoll}},\ and\ \bibinfo {author} {\bibfnamefont {J.~I.}\
  \bibnamefont {Cirac}},\ }\bibfield  {title} {\bibinfo {title} {Matrix product
  density operators: Simulation of finite-temperature and dissipative
  systems},\ }\href {https://doi.org/10.1103/PhysRevLett.93.207204} {\bibfield
  {journal} {\bibinfo  {journal} {Phys. Rev. Lett.}\ }\textbf {\bibinfo
  {volume} {93}},\ \bibinfo {pages} {207204} (\bibinfo {year}
  {2004})}\BibitemShut {NoStop}%
\bibitem [{\citenamefont {Cui}\ \emph {et~al.}(2015)\citenamefont {Cui},
  \citenamefont {Cirac},\ and\ \citenamefont {Ba\~nuls}}]{Cui2015}%
  \BibitemOpen
  \bibfield  {author} {\bibinfo {author} {\bibfnamefont {J.}~\bibnamefont
  {Cui}}, \bibinfo {author} {\bibfnamefont {J.~I.}\ \bibnamefont {Cirac}},\
  and\ \bibinfo {author} {\bibfnamefont {M.~C.}\ \bibnamefont {Ba\~nuls}},\
  }\bibfield  {title} {\bibinfo {title} {Variational matrix product operators
  for the steady state of dissipative quantum systems},\ }\href
  {https://doi.org/10.1103/PhysRevLett.114.220601} {\bibfield  {journal}
  {\bibinfo  {journal} {Phys. Rev. Lett.}\ }\textbf {\bibinfo {volume} {114}},\
  \bibinfo {pages} {220601} (\bibinfo {year} {2015})}\BibitemShut {NoStop}%
\bibitem [{\citenamefont {Mascarenhas}\ \emph {et~al.}(2015)\citenamefont
  {Mascarenhas}, \citenamefont {Flayac},\ and\ \citenamefont
  {Savona}}]{Mascarenhas2015}%
  \BibitemOpen
  \bibfield  {author} {\bibinfo {author} {\bibfnamefont {E.}~\bibnamefont
  {Mascarenhas}}, \bibinfo {author} {\bibfnamefont {H.}~\bibnamefont
  {Flayac}},\ and\ \bibinfo {author} {\bibfnamefont {V.}~\bibnamefont
  {Savona}},\ }\bibfield  {title} {\bibinfo {title} {Matrix-product-operator
  approach to the nonequilibrium steady state of driven-dissipative quantum
  arrays},\ }\href {https://doi.org/10.1103/PhysRevA.92.022116} {\bibfield
  {journal} {\bibinfo  {journal} {Phys. Rev. A}\ }\textbf {\bibinfo {volume}
  {92}},\ \bibinfo {pages} {022116} (\bibinfo {year} {2015})}\BibitemShut
  {NoStop}%
\bibitem [{\citenamefont {Werner}\ \emph {et~al.}(2016)\citenamefont {Werner},
  \citenamefont {Jaschke}, \citenamefont {Silvi}, \citenamefont {Kliesch},
  \citenamefont {Calarco}, \citenamefont {Eisert},\ and\ \citenamefont
  {Montangero}}]{Werner2016}%
  \BibitemOpen
  \bibfield  {author} {\bibinfo {author} {\bibfnamefont {A.~H.}\ \bibnamefont
  {Werner}}, \bibinfo {author} {\bibfnamefont {D.}~\bibnamefont {Jaschke}},
  \bibinfo {author} {\bibfnamefont {P.}~\bibnamefont {Silvi}}, \bibinfo
  {author} {\bibfnamefont {M.}~\bibnamefont {Kliesch}}, \bibinfo {author}
  {\bibfnamefont {T.}~\bibnamefont {Calarco}}, \bibinfo {author} {\bibfnamefont
  {J.}~\bibnamefont {Eisert}},\ and\ \bibinfo {author} {\bibfnamefont
  {S.}~\bibnamefont {Montangero}},\ }\bibfield  {title} {\bibinfo {title}
  {Positive tensor network approach for simulating open quantum many-body
  systems},\ }\href {https://doi.org/10.1103/PhysRevLett.116.237201} {\bibfield
   {journal} {\bibinfo  {journal} {Phys. Rev. Lett.}\ }\textbf {\bibinfo
  {volume} {116}},\ \bibinfo {pages} {237201} (\bibinfo {year}
  {2016})}\BibitemShut {NoStop}%
\bibitem [{\citenamefont {White}\ \emph {et~al.}(2018)\citenamefont {White},
  \citenamefont {Zaletel}, \citenamefont {Mong},\ and\ \citenamefont
  {Refael}}]{White2018}%
  \BibitemOpen
  \bibfield  {author} {\bibinfo {author} {\bibfnamefont {C.~D.}\ \bibnamefont
  {White}}, \bibinfo {author} {\bibfnamefont {M.}~\bibnamefont {Zaletel}},
  \bibinfo {author} {\bibfnamefont {R.~S.~K.}\ \bibnamefont {Mong}},\ and\
  \bibinfo {author} {\bibfnamefont {G.}~\bibnamefont {Refael}},\ }\bibfield
  {title} {\bibinfo {title} {Quantum dynamics of thermalizing systems},\ }\href
  {https://doi.org/10.1103/PhysRevB.97.035127} {\bibfield  {journal} {\bibinfo
  {journal} {Phys. Rev. B}\ }\textbf {\bibinfo {volume} {97}},\ \bibinfo
  {pages} {035127} (\bibinfo {year} {2018})}\BibitemShut {NoStop}%
\bibitem [{\citenamefont {Jaschke}\ \emph {et~al.}(2018)\citenamefont
  {Jaschke}, \citenamefont {Montangero},\ and\ \citenamefont
  {Carr}}]{Jaschke2018}%
  \BibitemOpen
  \bibfield  {author} {\bibinfo {author} {\bibfnamefont {D.}~\bibnamefont
  {Jaschke}}, \bibinfo {author} {\bibfnamefont {S.}~\bibnamefont
  {Montangero}},\ and\ \bibinfo {author} {\bibfnamefont {L.~D.}\ \bibnamefont
  {Carr}},\ }\bibfield  {title} {\bibinfo {title} {One-dimensional many-body
  entangled open quantum systems with tensor network methods},\ }\href
  {https://doi.org/10.1088/2058-9565/aae724} {\bibfield  {journal} {\bibinfo
  {journal} {Quantum Sci. Technol.}\ }\textbf {\bibinfo {volume} {4}},\
  \bibinfo {pages} {013001} (\bibinfo {year} {2018})}\BibitemShut {NoStop}%
\bibitem [{\citenamefont {Cheng}\ \emph {et~al.}(2021)\citenamefont {Cheng},
  \citenamefont {Cao}, \citenamefont {Zhang}, \citenamefont {Liu},
  \citenamefont {Hou}, \citenamefont {Xu},\ and\ \citenamefont
  {Zeng}}]{Cheng2021}%
  \BibitemOpen
  \bibfield  {author} {\bibinfo {author} {\bibfnamefont {S.}~\bibnamefont
  {Cheng}}, \bibinfo {author} {\bibfnamefont {C.}~\bibnamefont {Cao}}, \bibinfo
  {author} {\bibfnamefont {C.}~\bibnamefont {Zhang}}, \bibinfo {author}
  {\bibfnamefont {Y.}~\bibnamefont {Liu}}, \bibinfo {author} {\bibfnamefont
  {S.-Y.}\ \bibnamefont {Hou}}, \bibinfo {author} {\bibfnamefont
  {P.}~\bibnamefont {Xu}},\ and\ \bibinfo {author} {\bibfnamefont
  {B.}~\bibnamefont {Zeng}},\ }\bibfield  {title} {\bibinfo {title} {Simulating
  noisy quantum circuits with matrix product density operators},\ }\href
  {https://doi.org/10.1103/PhysRevResearch.3.023005} {\bibfield  {journal}
  {\bibinfo  {journal} {Phys. Rev. Research}\ }\textbf {\bibinfo {volume}
  {3}},\ \bibinfo {pages} {023005} (\bibinfo {year} {2021})}\BibitemShut
  {NoStop}%
\bibitem [{\citenamefont {Cheng}\ and\ \citenamefont
  {Potter}(2022)}]{Cheng2022}%
  \BibitemOpen
  \bibfield  {author} {\bibinfo {author} {\bibfnamefont {Z.}~\bibnamefont
  {Cheng}}\ and\ \bibinfo {author} {\bibfnamefont {A.~C.}\ \bibnamefont
  {Potter}},\ }\bibfield  {title} {\bibinfo {title} {A matrix product operator
  approach to non-equilibrium floquet steady states},\ }\href
  {https://arxiv.org/abs/2206.07740} {\bibfield  {journal} {\bibinfo  {journal}
  {arXiv:2206.07740}\ } (\bibinfo {year} {2022})}\BibitemShut {NoStop}%
\bibitem [{\citenamefont {Buchhold}\ \emph {et~al.}(2022)\citenamefont
  {Buchhold}, \citenamefont {Müller},\ and\ \citenamefont
  {Diehl}}]{Buchhold2022}%
  \BibitemOpen
  \bibfield  {author} {\bibinfo {author} {\bibfnamefont {M.}~\bibnamefont
  {Buchhold}}, \bibinfo {author} {\bibfnamefont {T.}~\bibnamefont {Müller}},\
  and\ \bibinfo {author} {\bibfnamefont {S.}~\bibnamefont {Diehl}},\ }\bibfield
   {title} {\bibinfo {title} {Revealing measurement-induced phase transitions
  by pre-selection},\ }\href {https://doi.org/10.48550/ARXIV.2208.10506}
  {\bibfield  {journal} {\bibinfo  {journal} {arXiv:2208.10506}\ } (\bibinfo
  {year} {2022})}\BibitemShut {NoStop}%
\bibitem [{\citenamefont {Iadecola}\ \emph {et~al.}(2022)\citenamefont
  {Iadecola}, \citenamefont {Ganeshan}, \citenamefont {Pixley},\ and\
  \citenamefont {Wilson}}]{Iadecola2022}%
  \BibitemOpen
  \bibfield  {author} {\bibinfo {author} {\bibfnamefont {T.}~\bibnamefont
  {Iadecola}}, \bibinfo {author} {\bibfnamefont {S.}~\bibnamefont {Ganeshan}},
  \bibinfo {author} {\bibfnamefont {J.~H.}\ \bibnamefont {Pixley}},\ and\
  \bibinfo {author} {\bibfnamefont {J.~H.}\ \bibnamefont {Wilson}},\ }\bibfield
   {title} {\bibinfo {title} {Dynamical entanglement transition in the
  probabilistic control of chaos},\ }\href
  {https://doi.org/10.48550/ARXIV.2207.12415} {\bibfield  {journal} {\bibinfo
  {journal} {arXiv:2207.12415}\ } (\bibinfo {year} {2022})}\BibitemShut
  {NoStop}%
\bibitem [{\citenamefont {{Quantinuum H1-1}}()}]{H11}%
  \BibitemOpen
  \bibfield  {author} {\bibinfo {author} {\bibnamefont {{Quantinuum H1-1}}},\
  }\href@noop {} {}\bibinfo {howpublished} {https://www.quantinuum.com/},\
  \bibinfo {note} {{May 25-August 22, 2022}}\BibitemShut {NoStop}%
\bibitem [{\citenamefont {Fishman}\ \emph {et~al.}(2022)\citenamefont
  {Fishman}, \citenamefont {White},\ and\ \citenamefont
  {Stoudenmire}}]{itensor}%
  \BibitemOpen
  \bibfield  {author} {\bibinfo {author} {\bibfnamefont {M.}~\bibnamefont
  {Fishman}}, \bibinfo {author} {\bibfnamefont {S.~R.}\ \bibnamefont {White}},\
  and\ \bibinfo {author} {\bibfnamefont {E.~M.}\ \bibnamefont {Stoudenmire}},\
  }\bibfield  {title} {\bibinfo {title} {{The ITensor Software Library for
  Tensor Network Calculations}},\ }\href
  {https://doi.org/10.21468/SciPostPhysCodeb.4} {\bibfield  {journal} {\bibinfo
   {journal} {SciPost Phys. Codebases}\ ,\ \bibinfo {pages} {4}} (\bibinfo
  {year} {2022})}\BibitemShut {NoStop}%
\end{thebibliography}%


\begin{thebibliography}{32}%
\makeatletter
\providecommand \@ifxundefined [1]{%
 \@ifx{#1\undefined}
}%
\providecommand \@ifnum [1]{%
 \ifnum #1\expandafter \@firstoftwo
 \else \expandafter \@secondoftwo
 \fi
}%
\providecommand \@ifx [1]{%
 \ifx #1\expandafter \@firstoftwo
 \else \expandafter \@secondoftwo
 \fi
}%
\providecommand \natexlab [1]{#1}%
\providecommand \enquote  [1]{``#1''}%
\providecommand \bibnamefont  [1]{#1}%
\providecommand \bibfnamefont [1]{#1}%
\providecommand \citenamefont [1]{#1}%
\providecommand \href@noop [0]{\@secondoftwo}%
\providecommand \href [0]{\begingroup \@sanitize@url \@href}%
\providecommand \@href[1]{\@@startlink{#1}\@@href}%
\providecommand \@@href[1]{\endgroup#1\@@endlink}%
\providecommand \@sanitize@url [0]{\catcode `\\12\catcode `\$12\catcode
  `\&12\catcode `\#12\catcode `\^12\catcode `\_12\catcode `\%12\relax}%
\providecommand \@@startlink[1]{}%
\providecommand \@@endlink[0]{}%
\providecommand \url  [0]{\begingroup\@sanitize@url \@url }%
\providecommand \@url [1]{\endgroup\@href {#1}{\urlprefix }}%
\providecommand \urlprefix  [0]{URL }%
\providecommand \Eprint [0]{\href }%
\providecommand \doibase [0]{https://doi.org/}%
\providecommand \selectlanguage [0]{\@gobble}%
\providecommand \bibinfo  [0]{\@secondoftwo}%
\providecommand \bibfield  [0]{\@secondoftwo}%
\providecommand \translation [1]{[#1]}%
\providecommand \BibitemOpen [0]{}%
\providecommand \bibitemStop [0]{}%
\providecommand \bibitemNoStop [0]{.\EOS\space}%
\providecommand \EOS [0]{\spacefactor3000\relax}%
\providecommand \BibitemShut  [1]{\csname bibitem#1\endcsname}%
\let\auto@bib@innerbib\@empty
\bibitem [{\citenamefont {Deutsch}(1991)}]{Deutsch1991}%
  \BibitemOpen
  \bibfield  {author} {\bibinfo {author} {\bibfnamefont {J.~M.}\ \bibnamefont
  {Deutsch}},\ }\bibfield  {title} {\bibinfo {title} {Quantum statistical
  mechanics in a closed system},\ }\href
  {https://doi.org/10.1103/PhysRevA.43.2046} {\bibfield  {journal} {\bibinfo
  {journal} {Phys. Rev. A}\ }\textbf {\bibinfo {volume} {43}},\ \bibinfo
  {pages} {2046} (\bibinfo {year} {1991})}\BibitemShut {NoStop}%
\bibitem [{\citenamefont {Srednicki}(1994)}]{Srednicki1994}%
  \BibitemOpen
  \bibfield  {author} {\bibinfo {author} {\bibfnamefont {M.}~\bibnamefont
  {Srednicki}},\ }\bibfield  {title} {\bibinfo {title} {Chaos and quantum
  thermalization},\ }\href {https://doi.org/10.1103/PhysRevE.50.888} {\bibfield
   {journal} {\bibinfo  {journal} {Phys. Rev. E}\ }\textbf {\bibinfo {volume}
  {50}},\ \bibinfo {pages} {888} (\bibinfo {year} {1994})}\BibitemShut
  {NoStop}%
\bibitem [{\citenamefont {Rigol}\ \emph {et~al.}(2008)\citenamefont {Rigol},
  \citenamefont {Dunjko},\ and\ \citenamefont {Olshanii}}]{Rigol2008}%
  \BibitemOpen
  \bibfield  {author} {\bibinfo {author} {\bibfnamefont {M.}~\bibnamefont
  {Rigol}}, \bibinfo {author} {\bibfnamefont {V.}~\bibnamefont {Dunjko}},\ and\
  \bibinfo {author} {\bibfnamefont {M.}~\bibnamefont {Olshanii}},\ }\bibfield
  {title} {\bibinfo {title} {Thermalization and its mechanism for generic
  isolated quantum systems},\ }\href {https://doi.org/10.1038/nature06838}
  {\bibfield  {journal} {\bibinfo  {journal} {Nature}\ }\textbf {\bibinfo
  {volume} {452}},\ \bibinfo {pages} {854} (\bibinfo {year}
  {2008})}\BibitemShut {NoStop}%
\bibitem [{\citenamefont {Marcuzzi}\ \emph {et~al.}(2016)\citenamefont
  {Marcuzzi}, \citenamefont {Buchhold}, \citenamefont {Diehl},\ and\
  \citenamefont {Lesanovsky}}]{Marcuzzi2016}%
  \BibitemOpen
  \bibfield  {author} {\bibinfo {author} {\bibfnamefont {M.}~\bibnamefont
  {Marcuzzi}}, \bibinfo {author} {\bibfnamefont {M.}~\bibnamefont {Buchhold}},
  \bibinfo {author} {\bibfnamefont {S.}~\bibnamefont {Diehl}},\ and\ \bibinfo
  {author} {\bibfnamefont {I.}~\bibnamefont {Lesanovsky}},\ }\bibfield  {title}
  {\bibinfo {title} {Absorbing state phase transition with competing quantum
  and classical fluctuations},\ }\href
  {https://doi.org/10.1103/PhysRevLett.116.245701} {\bibfield  {journal}
  {\bibinfo  {journal} {Phys. Rev. Lett.}\ }\textbf {\bibinfo {volume} {116}},\
  \bibinfo {pages} {245701} (\bibinfo {year} {2016})}\BibitemShut {NoStop}%
\bibitem [{\citenamefont {Carollo}\ \emph {et~al.}(2019)\citenamefont
  {Carollo}, \citenamefont {Gillman}, \citenamefont {Weimer},\ and\
  \citenamefont {Lesanovsky}}]{Carollo2019}%
  \BibitemOpen
  \bibfield  {author} {\bibinfo {author} {\bibfnamefont {F.}~\bibnamefont
  {Carollo}}, \bibinfo {author} {\bibfnamefont {E.}~\bibnamefont {Gillman}},
  \bibinfo {author} {\bibfnamefont {H.}~\bibnamefont {Weimer}},\ and\ \bibinfo
  {author} {\bibfnamefont {I.}~\bibnamefont {Lesanovsky}},\ }\bibfield  {title}
  {\bibinfo {title} {Critical behavior of the quantum contact process in one
  dimension},\ }\href {https://doi.org/10.1103/PhysRevLett.123.100604}
  {\bibfield  {journal} {\bibinfo  {journal} {Phys. Rev. Lett.}\ }\textbf
  {\bibinfo {volume} {123}},\ \bibinfo {pages} {100604} (\bibinfo {year}
  {2019})}\BibitemShut {NoStop}%
\bibitem [{\citenamefont {Gillman}\ \emph {et~al.}(2019)\citenamefont
  {Gillman}, \citenamefont {Carollo},\ and\ \citenamefont
  {Lesanovsky}}]{Gillman2019}%
  \BibitemOpen
  \bibfield  {author} {\bibinfo {author} {\bibfnamefont {E.}~\bibnamefont
  {Gillman}}, \bibinfo {author} {\bibfnamefont {F.}~\bibnamefont {Carollo}},\
  and\ \bibinfo {author} {\bibfnamefont {I.}~\bibnamefont {Lesanovsky}},\
  }\bibfield  {title} {\bibinfo {title} {Numerical simulation of critical
  dissipative non-equilibrium quantum systems with an absorbing state},\ }\href
  {https://doi.org/10.1088/1367-2630/ab43b0} {\bibfield  {journal} {\bibinfo
  {journal} {New J. Phys.}\ }\textbf {\bibinfo {volume} {21}},\ \bibinfo
  {pages} {093064} (\bibinfo {year} {2019})}\BibitemShut {NoStop}%
\bibitem [{\citenamefont {Jo}\ \emph {et~al.}(2021)\citenamefont {Jo},
  \citenamefont {Lee}, \citenamefont {Choi},\ and\ \citenamefont
  {Kahng}}]{Minjae2021}%
  \BibitemOpen
  \bibfield  {author} {\bibinfo {author} {\bibfnamefont {M.}~\bibnamefont
  {Jo}}, \bibinfo {author} {\bibfnamefont {J.}~\bibnamefont {Lee}}, \bibinfo
  {author} {\bibfnamefont {K.}~\bibnamefont {Choi}},\ and\ \bibinfo {author}
  {\bibfnamefont {B.}~\bibnamefont {Kahng}},\ }\bibfield  {title} {\bibinfo
  {title} {Absorbing phase transition with a continuously varying exponent in a
  quantum contact process: A neural network approach},\ }\href
  {https://doi.org/10.1103/PhysRevResearch.3.013238} {\bibfield  {journal}
  {\bibinfo  {journal} {Phys. Rev. Research}\ }\textbf {\bibinfo {volume}
  {3}},\ \bibinfo {pages} {013238} (\bibinfo {year} {2021})}\BibitemShut
  {NoStop}%
\bibitem [{\citenamefont {Atas}\ \emph {et~al.}(2013)\citenamefont {Atas},
  \citenamefont {Bogomolny}, \citenamefont {Giraud},\ and\ \citenamefont
  {Roux}}]{Atas2013}%
  \BibitemOpen
  \bibfield  {author} {\bibinfo {author} {\bibfnamefont {Y.~Y.}\ \bibnamefont
  {Atas}}, \bibinfo {author} {\bibfnamefont {E.}~\bibnamefont {Bogomolny}},
  \bibinfo {author} {\bibfnamefont {O.}~\bibnamefont {Giraud}},\ and\ \bibinfo
  {author} {\bibfnamefont {G.}~\bibnamefont {Roux}},\ }\bibfield  {title}
  {\bibinfo {title} {Distribution of the ratio of consecutive level spacings in
  random matrix ensembles},\ }\href
  {https://doi.org/10.1103/PhysRevLett.110.084101} {\bibfield  {journal}
  {\bibinfo  {journal} {Phys. Rev. Lett.}\ }\textbf {\bibinfo {volume} {110}},\
  \bibinfo {pages} {084101} (\bibinfo {year} {2013})}\BibitemShut {NoStop}%
\bibitem [{\citenamefont {D'Alessio}\ and\ \citenamefont
  {Rigol}(2014)}]{DAlessio2014}%
  \BibitemOpen
  \bibfield  {author} {\bibinfo {author} {\bibfnamefont {L.}~\bibnamefont
  {D'Alessio}}\ and\ \bibinfo {author} {\bibfnamefont {M.}~\bibnamefont
  {Rigol}},\ }\bibfield  {title} {\bibinfo {title} {Long-time behavior of
  isolated periodically driven interacting lattice systems},\ }\href
  {https://doi.org/10.1103/PhysRevX.4.041048} {\bibfield  {journal} {\bibinfo
  {journal} {Phys. Rev. X}\ }\textbf {\bibinfo {volume} {4}},\ \bibinfo {pages}
  {041048} (\bibinfo {year} {2014})}\BibitemShut {NoStop}%
\bibitem [{\citenamefont {Mizuta}\ \emph {et~al.}(2020)\citenamefont {Mizuta},
  \citenamefont {Takasan},\ and\ \citenamefont {Kawakami}}]{Mizuta2020}%
  \BibitemOpen
  \bibfield  {author} {\bibinfo {author} {\bibfnamefont {K.}~\bibnamefont
  {Mizuta}}, \bibinfo {author} {\bibfnamefont {K.}~\bibnamefont {Takasan}},\
  and\ \bibinfo {author} {\bibfnamefont {N.}~\bibnamefont {Kawakami}},\
  }\bibfield  {title} {\bibinfo {title} {Exact floquet quantum many-body scars
  under rydberg blockade},\ }\href
  {https://doi.org/10.1103/PhysRevResearch.2.033284} {\bibfield  {journal}
  {\bibinfo  {journal} {Phys. Rev. Research}\ }\textbf {\bibinfo {volume}
  {2}},\ \bibinfo {pages} {033284} (\bibinfo {year} {2020})}\BibitemShut
  {NoStop}%
\bibitem [{\citenamefont {Sugiura}\ \emph {et~al.}(2021)\citenamefont
  {Sugiura}, \citenamefont {Kuwahara},\ and\ \citenamefont
  {Saito}}]{Sugiura2021}%
  \BibitemOpen
  \bibfield  {author} {\bibinfo {author} {\bibfnamefont {S.}~\bibnamefont
  {Sugiura}}, \bibinfo {author} {\bibfnamefont {T.}~\bibnamefont {Kuwahara}},\
  and\ \bibinfo {author} {\bibfnamefont {K.}~\bibnamefont {Saito}},\ }\bibfield
   {title} {\bibinfo {title} {Many-body scar state intrinsic to periodically
  driven system},\ }\href {https://doi.org/10.1103/PhysRevResearch.3.L012010}
  {\bibfield  {journal} {\bibinfo  {journal} {Phys. Rev. Research}\ }\textbf
  {\bibinfo {volume} {3}},\ \bibinfo {pages} {L012010} (\bibinfo {year}
  {2021})}\BibitemShut {NoStop}%
\bibitem [{\citenamefont {Skinner}\ \emph {et~al.}(2019)\citenamefont
  {Skinner}, \citenamefont {Ruhman},\ and\ \citenamefont
  {Nahum}}]{Skinner2019}%
  \BibitemOpen
  \bibfield  {author} {\bibinfo {author} {\bibfnamefont {B.}~\bibnamefont
  {Skinner}}, \bibinfo {author} {\bibfnamefont {J.}~\bibnamefont {Ruhman}},\
  and\ \bibinfo {author} {\bibfnamefont {A.}~\bibnamefont {Nahum}},\ }\bibfield
   {title} {\bibinfo {title} {Measurement-induced phase transitions in the
  dynamics of entanglement},\ }\href
  {https://doi.org/10.1103/PhysRevX.9.031009} {\bibfield  {journal} {\bibinfo
  {journal} {Phys. Rev. X}\ }\textbf {\bibinfo {volume} {9}},\ \bibinfo {pages}
  {031009} (\bibinfo {year} {2019})}\BibitemShut {NoStop}%
\bibitem [{\citenamefont {Zabalo}\ \emph {et~al.}(2020)\citenamefont {Zabalo},
  \citenamefont {Gullans}, \citenamefont {Wilson}, \citenamefont
  {Gopalakrishnan}, \citenamefont {Huse},\ and\ \citenamefont
  {Pixley}}]{Zabalo2020}%
  \BibitemOpen
  \bibfield  {author} {\bibinfo {author} {\bibfnamefont {A.}~\bibnamefont
  {Zabalo}}, \bibinfo {author} {\bibfnamefont {M.~J.}\ \bibnamefont {Gullans}},
  \bibinfo {author} {\bibfnamefont {J.~H.}\ \bibnamefont {Wilson}}, \bibinfo
  {author} {\bibfnamefont {S.}~\bibnamefont {Gopalakrishnan}}, \bibinfo
  {author} {\bibfnamefont {D.~A.}\ \bibnamefont {Huse}},\ and\ \bibinfo
  {author} {\bibfnamefont {J.~H.}\ \bibnamefont {Pixley}},\ }\bibfield  {title}
  {\bibinfo {title} {Critical properties of the measurement-induced transition
  in random quantum circuits},\ }\href
  {https://doi.org/10.1103/PhysRevB.101.060301} {\bibfield  {journal} {\bibinfo
   {journal} {Phys. Rev. B}\ }\textbf {\bibinfo {volume} {101}},\ \bibinfo
  {pages} {060301} (\bibinfo {year} {2020})}\BibitemShut {NoStop}%
\bibitem [{\citenamefont {Potter}\ and\ \citenamefont
  {Vasseur}(2021)}]{Potter2021}%
  \BibitemOpen
  \bibfield  {author} {\bibinfo {author} {\bibfnamefont {A.~C.}\ \bibnamefont
  {Potter}}\ and\ \bibinfo {author} {\bibfnamefont {R.}~\bibnamefont
  {Vasseur}},\ }\bibfield  {title} {\bibinfo {title} {Entanglement dynamics in
  hybrid quantum circuits},\ }\href {https://arxiv.org/abs/2111.08018}
  {\bibfield  {journal} {\bibinfo  {journal} {arXiv:2111.08018}\ } (\bibinfo
  {year} {2021})}\BibitemShut {NoStop}%
\bibitem [{\citenamefont {Temme}\ \emph {et~al.}(2017)\citenamefont {Temme},
  \citenamefont {Bravyi},\ and\ \citenamefont {Gambetta}}]{Temme2017}%
  \BibitemOpen
  \bibfield  {author} {\bibinfo {author} {\bibfnamefont {K.}~\bibnamefont
  {Temme}}, \bibinfo {author} {\bibfnamefont {S.}~\bibnamefont {Bravyi}},\ and\
  \bibinfo {author} {\bibfnamefont {J.~M.}\ \bibnamefont {Gambetta}},\
  }\bibfield  {title} {\bibinfo {title} {Error mitigation for short-depth
  quantum circuits},\ }\href {https://doi.org/10.1103/PhysRevLett.119.180509}
  {\bibfield  {journal} {\bibinfo  {journal} {Phys. Rev. Lett.}\ }\textbf
  {\bibinfo {volume} {119}},\ \bibinfo {pages} {180509} (\bibinfo {year}
  {2017})}\BibitemShut {NoStop}%
\bibitem [{\citenamefont {Li}\ and\ \citenamefont {Benjamin}(2017)}]{Li2017}%
  \BibitemOpen
  \bibfield  {author} {\bibinfo {author} {\bibfnamefont {Y.}~\bibnamefont
  {Li}}\ and\ \bibinfo {author} {\bibfnamefont {S.~C.}\ \bibnamefont
  {Benjamin}},\ }\bibfield  {title} {\bibinfo {title} {Efficient variational
  quantum simulator incorporating active error minimization},\ }\href
  {https://doi.org/10.1103/PhysRevX.7.021050} {\bibfield  {journal} {\bibinfo
  {journal} {Phys. Rev. X}\ }\textbf {\bibinfo {volume} {7}},\ \bibinfo {pages}
  {021050} (\bibinfo {year} {2017})}\BibitemShut {NoStop}%
\bibitem [{\citenamefont {Ryan-Anderson}\ \emph {et~al.}(2022)\citenamefont
  {Ryan-Anderson}, \citenamefont {Brown}, \citenamefont {Allman}, \citenamefont
  {Arkin}, \citenamefont {Asa-Attuah}, \citenamefont {Baldwin}, \citenamefont
  {Berg}, \citenamefont {Bohnet}, \citenamefont {Braxton}, \citenamefont
  {Burdick}, \citenamefont {Campora}, \citenamefont {Chernoguzov},
  \citenamefont {Esposito}, \citenamefont {Evans}, \citenamefont {Francois},
  \citenamefont {Gaebler}, \citenamefont {Gatterman}, \citenamefont {Gerber},
  \citenamefont {Gilmore}, \citenamefont {Gresh}, \citenamefont {Hall},
  \citenamefont {Hankin}, \citenamefont {Hostetter}, \citenamefont {Lucchetti},
  \citenamefont {Mayer}, \citenamefont {Myers}, \citenamefont {Neyenhuis},
  \citenamefont {Santiago}, \citenamefont {Sedlacek}, \citenamefont {Skripka},
  \citenamefont {Slattery}, \citenamefont {Stutz}, \citenamefont {Tait},
  \citenamefont {Tobey}, \citenamefont {Vittorini}, \citenamefont {Walker},\
  and\ \citenamefont {Hayes}}]{RyanAnderson2022}%
  \BibitemOpen
  \bibfield  {author} {\bibinfo {author} {\bibfnamefont {C.}~\bibnamefont
  {Ryan-Anderson}}, \bibinfo {author} {\bibfnamefont {N.~C.}\ \bibnamefont
  {Brown}}, \bibinfo {author} {\bibfnamefont {M.~S.}\ \bibnamefont {Allman}},
  \bibinfo {author} {\bibfnamefont {B.}~\bibnamefont {Arkin}}, \bibinfo
  {author} {\bibfnamefont {G.}~\bibnamefont {Asa-Attuah}}, \bibinfo {author}
  {\bibfnamefont {C.}~\bibnamefont {Baldwin}}, \bibinfo {author} {\bibfnamefont
  {J.}~\bibnamefont {Berg}}, \bibinfo {author} {\bibfnamefont {J.~G.}\
  \bibnamefont {Bohnet}}, \bibinfo {author} {\bibfnamefont {S.}~\bibnamefont
  {Braxton}}, \bibinfo {author} {\bibfnamefont {N.}~\bibnamefont {Burdick}},
  \bibinfo {author} {\bibfnamefont {J.~P.}\ \bibnamefont {Campora}}, \bibinfo
  {author} {\bibfnamefont {A.}~\bibnamefont {Chernoguzov}}, \bibinfo {author}
  {\bibfnamefont {J.}~\bibnamefont {Esposito}}, \bibinfo {author}
  {\bibfnamefont {B.}~\bibnamefont {Evans}}, \bibinfo {author} {\bibfnamefont
  {D.}~\bibnamefont {Francois}}, \bibinfo {author} {\bibfnamefont {J.~P.}\
  \bibnamefont {Gaebler}}, \bibinfo {author} {\bibfnamefont {T.~M.}\
  \bibnamefont {Gatterman}}, \bibinfo {author} {\bibfnamefont {J.}~\bibnamefont
  {Gerber}}, \bibinfo {author} {\bibfnamefont {K.}~\bibnamefont {Gilmore}},
  \bibinfo {author} {\bibfnamefont {D.}~\bibnamefont {Gresh}}, \bibinfo
  {author} {\bibfnamefont {A.}~\bibnamefont {Hall}}, \bibinfo {author}
  {\bibfnamefont {A.}~\bibnamefont {Hankin}}, \bibinfo {author} {\bibfnamefont
  {J.}~\bibnamefont {Hostetter}}, \bibinfo {author} {\bibfnamefont
  {D.}~\bibnamefont {Lucchetti}}, \bibinfo {author} {\bibfnamefont
  {K.}~\bibnamefont {Mayer}}, \bibinfo {author} {\bibfnamefont
  {J.}~\bibnamefont {Myers}}, \bibinfo {author} {\bibfnamefont
  {B.}~\bibnamefont {Neyenhuis}}, \bibinfo {author} {\bibfnamefont
  {J.}~\bibnamefont {Santiago}}, \bibinfo {author} {\bibfnamefont
  {J.}~\bibnamefont {Sedlacek}}, \bibinfo {author} {\bibfnamefont
  {T.}~\bibnamefont {Skripka}}, \bibinfo {author} {\bibfnamefont
  {A.}~\bibnamefont {Slattery}}, \bibinfo {author} {\bibfnamefont {R.~P.}\
  \bibnamefont {Stutz}}, \bibinfo {author} {\bibfnamefont {J.}~\bibnamefont
  {Tait}}, \bibinfo {author} {\bibfnamefont {R.}~\bibnamefont {Tobey}},
  \bibinfo {author} {\bibfnamefont {G.}~\bibnamefont {Vittorini}}, \bibinfo
  {author} {\bibfnamefont {J.}~\bibnamefont {Walker}},\ and\ \bibinfo {author}
  {\bibfnamefont {D.}~\bibnamefont {Hayes}},\ }\bibfield  {title} {\bibinfo
  {title} {Implementing fault-tolerant entangling gates on the five-qubit code
  and the color code},\ }\href {https://arxiv.org/abs/2208.01863} {\bibfield
  {journal} {\bibinfo  {journal} {arXiv:2208.01863}\ } (\bibinfo {year}
  {2022})}\BibitemShut {NoStop}%
\bibitem [{\citenamefont {Jensen}(1999)}]{Jensen1999}%
  \BibitemOpen
  \bibfield  {author} {\bibinfo {author} {\bibfnamefont {I.}~\bibnamefont
  {Jensen}},\ }\bibfield  {title} {\bibinfo {title} {Low-density series
  expansions for directed percolation: I. a new efficient algorithm with
  applications to the square lattice},\ }\href
  {https://doi.org/10.1088/0305-4470/32/28/304} {\bibfield  {journal} {\bibinfo
   {journal} {J. Phys. A}\ }\textbf {\bibinfo {volume} {32}},\ \bibinfo {pages}
  {5233} (\bibinfo {year} {1999})}\BibitemShut {NoStop}%
\bibitem [{\citenamefont {Hinrichsen}(2000)}]{Hinrichsen2000}%
  \BibitemOpen
  \bibfield  {author} {\bibinfo {author} {\bibfnamefont {H.}~\bibnamefont
  {Hinrichsen}},\ }\bibfield  {title} {\bibinfo {title} {Non-equilibrium
  critical phenomena and phase transitions into absorbing states},\ }\href
  {https://doi.org/10.1080/00018730050198152} {\bibfield  {journal} {\bibinfo
  {journal} {Adv. Phys.}\ }\textbf {\bibinfo {volume} {49}},\ \bibinfo {pages}
  {815} (\bibinfo {year} {2000})}\BibitemShut {NoStop}%
\bibitem [{\citenamefont {Bezanson}\ \emph {et~al.}(2017)\citenamefont
  {Bezanson}, \citenamefont {Edelman}, \citenamefont {Karpinski},\ and\
  \citenamefont {Shah}}]{julia}%
  \BibitemOpen
  \bibfield  {author} {\bibinfo {author} {\bibfnamefont {J.}~\bibnamefont
  {Bezanson}}, \bibinfo {author} {\bibfnamefont {A.}~\bibnamefont {Edelman}},
  \bibinfo {author} {\bibfnamefont {S.}~\bibnamefont {Karpinski}},\ and\
  \bibinfo {author} {\bibfnamefont {V.~B.}\ \bibnamefont {Shah}},\ }\bibfield
  {title} {\bibinfo {title} {{Julia: A fresh approach to numerical
  computing}},\ }\href {https://doi.org/10.1137/141000671} {\bibfield
  {journal} {\bibinfo  {journal} {SIAM {R}eview}\ }\textbf {\bibinfo {volume}
  {59}},\ \bibinfo {pages} {65} (\bibinfo {year} {2017})}\BibitemShut {NoStop}%
\bibitem [{\citenamefont {Kliesch}\ \emph {et~al.}(2014)\citenamefont
  {Kliesch}, \citenamefont {Gross},\ and\ \citenamefont
  {Eisert}}]{Kliesch2014}%
  \BibitemOpen
  \bibfield  {author} {\bibinfo {author} {\bibfnamefont {M.}~\bibnamefont
  {Kliesch}}, \bibinfo {author} {\bibfnamefont {D.}~\bibnamefont {Gross}},\
  and\ \bibinfo {author} {\bibfnamefont {J.}~\bibnamefont {Eisert}},\
  }\bibfield  {title} {\bibinfo {title} {Matrix-product operators and states:
  Np-hardness and undecidability},\ }\href
  {https://doi.org/10.1103/PhysRevLett.113.160503} {\bibfield  {journal}
  {\bibinfo  {journal} {Phys. Rev. Lett.}\ }\textbf {\bibinfo {volume} {113}},\
  \bibinfo {pages} {160503} (\bibinfo {year} {2014})}\BibitemShut {NoStop}%
\bibitem [{\citenamefont {Werner}\ \emph {et~al.}(2016)\citenamefont {Werner},
  \citenamefont {Jaschke}, \citenamefont {Silvi}, \citenamefont {Kliesch},
  \citenamefont {Calarco}, \citenamefont {Eisert},\ and\ \citenamefont
  {Montangero}}]{Werner2016}%
  \BibitemOpen
  \bibfield  {author} {\bibinfo {author} {\bibfnamefont {A.~H.}\ \bibnamefont
  {Werner}}, \bibinfo {author} {\bibfnamefont {D.}~\bibnamefont {Jaschke}},
  \bibinfo {author} {\bibfnamefont {P.}~\bibnamefont {Silvi}}, \bibinfo
  {author} {\bibfnamefont {M.}~\bibnamefont {Kliesch}}, \bibinfo {author}
  {\bibfnamefont {T.}~\bibnamefont {Calarco}}, \bibinfo {author} {\bibfnamefont
  {J.}~\bibnamefont {Eisert}},\ and\ \bibinfo {author} {\bibfnamefont
  {S.}~\bibnamefont {Montangero}},\ }\bibfield  {title} {\bibinfo {title}
  {Positive tensor network approach for simulating open quantum many-body
  systems},\ }\href {https://doi.org/10.1103/PhysRevLett.116.237201} {\bibfield
   {journal} {\bibinfo  {journal} {Phys. Rev. Lett.}\ }\textbf {\bibinfo
  {volume} {116}},\ \bibinfo {pages} {237201} (\bibinfo {year}
  {2016})}\BibitemShut {NoStop}%
\bibitem [{\citenamefont {White}\ \emph {et~al.}(2018)\citenamefont {White},
  \citenamefont {Zaletel}, \citenamefont {Mong},\ and\ \citenamefont
  {Refael}}]{White2018}%
  \BibitemOpen
  \bibfield  {author} {\bibinfo {author} {\bibfnamefont {C.~D.}\ \bibnamefont
  {White}}, \bibinfo {author} {\bibfnamefont {M.}~\bibnamefont {Zaletel}},
  \bibinfo {author} {\bibfnamefont {R.~S.~K.}\ \bibnamefont {Mong}},\ and\
  \bibinfo {author} {\bibfnamefont {G.}~\bibnamefont {Refael}},\ }\bibfield
  {title} {\bibinfo {title} {Quantum dynamics of thermalizing systems},\ }\href
  {https://doi.org/10.1103/PhysRevB.97.035127} {\bibfield  {journal} {\bibinfo
  {journal} {Phys. Rev. B}\ }\textbf {\bibinfo {volume} {97}},\ \bibinfo
  {pages} {035127} (\bibinfo {year} {2018})}\BibitemShut {NoStop}%
\bibitem [{\citenamefont {Fishman}\ \emph {et~al.}(2022)\citenamefont
  {Fishman}, \citenamefont {White},\ and\ \citenamefont
  {Stoudenmire}}]{itensor}%
  \BibitemOpen
  \bibfield  {author} {\bibinfo {author} {\bibfnamefont {M.}~\bibnamefont
  {Fishman}}, \bibinfo {author} {\bibfnamefont {S.~R.}\ \bibnamefont {White}},\
  and\ \bibinfo {author} {\bibfnamefont {E.~M.}\ \bibnamefont {Stoudenmire}},\
  }\bibfield  {title} {\bibinfo {title} {{The ITensor Software Library for
  Tensor Network Calculations}},\ }\href
  {https://doi.org/10.21468/SciPostPhysCodeb.4} {\bibfield  {journal} {\bibinfo
   {journal} {SciPost Phys. Codebases}\ ,\ \bibinfo {pages} {4}} (\bibinfo
  {year} {2022})}\BibitemShut {NoStop}%
\bibitem [{\citenamefont {Dickman}(2008)}]{Dickman2008}%
  \BibitemOpen
  \bibfield  {author} {\bibinfo {author} {\bibfnamefont {R.}~\bibnamefont
  {Dickman}},\ }\bibfield  {title} {\bibinfo {title} {Absorbing-state phase
  transitions: Exact solutions of small systems},\ }\href
  {https://doi.org/10.1103/PhysRevE.77.030102} {\bibfield  {journal} {\bibinfo
  {journal} {Phys. Rev. E}\ }\textbf {\bibinfo {volume} {77}},\ \bibinfo
  {pages} {030102} (\bibinfo {year} {2008})}\BibitemShut {NoStop}%
\bibitem [{\citenamefont {Filho}\ and\ \citenamefont
  {Dickman}(2011)}]{Filho2011}%
  \BibitemOpen
  \bibfield  {author} {\bibinfo {author} {\bibfnamefont {J.~C.~M.}\
  \bibnamefont {Filho}}\ and\ \bibinfo {author} {\bibfnamefont
  {R.}~\bibnamefont {Dickman}},\ }\bibfield  {title} {\bibinfo {title}
  {Conserved directed percolation: exact quasistationary distribution of small
  systems and monte carlo simulations},\ }\href
  {https://doi.org/10.1088/1742-5468/2011/05/p05029} {\bibfield  {journal}
  {\bibinfo  {journal} {J. Stat. Mech.: Theory Exp.}\ }\textbf {\bibinfo
  {volume} {2011}}\bibinfo  {number} { (05)},\ \bibinfo {pages}
  {P05029}}\BibitemShut {NoStop}%
\bibitem [{\citenamefont {{Carlon}}\ \emph {et~al.}(1999)\citenamefont
  {{Carlon}}, \citenamefont {{Henkel}},\ and\ \citenamefont
  {{Schollw{\"o}ck}}}]{Carlon1999}%
  \BibitemOpen
\bibfield  {number} {  }\bibfield  {author} {\bibinfo {author} {\bibfnamefont
  {E.}~\bibnamefont {{Carlon}}}, \bibinfo {author} {\bibfnamefont
  {M.}~\bibnamefont {{Henkel}}},\ and\ \bibinfo {author} {\bibfnamefont
  {U.}~\bibnamefont {{Schollw{\"o}ck}}},\ }\bibfield  {title} {\bibinfo {title}
  {{Density matrix renormalization group and reaction-diffusion processes}},\
  }\href {https://doi.org/10.1007/s100510050983} {\bibfield  {journal}
  {\bibinfo  {journal} {Eur. Phys. J. B}\ }\textbf {\bibinfo {volume} {12}},\
  \bibinfo {pages} {99} (\bibinfo {year} {1999})}\BibitemShut {NoStop}%
\bibitem [{\citenamefont {Henkel}\ and\ \citenamefont
  {Schutz}(1988)}]{Henkel1988}%
  \BibitemOpen
  \bibfield  {author} {\bibinfo {author} {\bibfnamefont {M.}~\bibnamefont
  {Henkel}}\ and\ \bibinfo {author} {\bibfnamefont {G.}~\bibnamefont
  {Schutz}},\ }\bibfield  {title} {\bibinfo {title} {Finite-lattice
  extrapolation algorithms},\ }\href
  {https://doi.org/10.1088/0305-4470/21/11/019} {\bibfield  {journal} {\bibinfo
   {journal} {J. Phys. A}\ }\textbf {\bibinfo {volume} {21}},\ \bibinfo {pages}
  {2617} (\bibinfo {year} {1988})}\BibitemShut {NoStop}%
\bibitem [{\citenamefont {Noh}\ \emph {et~al.}(2020)\citenamefont {Noh},
  \citenamefont {Jiang},\ and\ \citenamefont {Fefferman}}]{Noh2020}%
  \BibitemOpen
  \bibfield  {author} {\bibinfo {author} {\bibfnamefont {K.}~\bibnamefont
  {Noh}}, \bibinfo {author} {\bibfnamefont {L.}~\bibnamefont {Jiang}},\ and\
  \bibinfo {author} {\bibfnamefont {B.}~\bibnamefont {Fefferman}},\ }\bibfield
  {title} {\bibinfo {title} {Efficient classical simulation of noisy random
  quantum circuits in one dimension},\ }\href
  {https://doi.org/10.22331/q-2020-09-11-318} {\bibfield  {journal} {\bibinfo
  {journal} {{Quantum}}\ }\textbf {\bibinfo {volume} {4}},\ \bibinfo {pages}
  {318} (\bibinfo {year} {2020})}\BibitemShut {NoStop}%
\bibitem [{\citenamefont {Cui}\ \emph {et~al.}(2015)\citenamefont {Cui},
  \citenamefont {Cirac},\ and\ \citenamefont {Ba\~nuls}}]{Cui2015}%
  \BibitemOpen
  \bibfield  {author} {\bibinfo {author} {\bibfnamefont {J.}~\bibnamefont
  {Cui}}, \bibinfo {author} {\bibfnamefont {J.~I.}\ \bibnamefont {Cirac}},\
  and\ \bibinfo {author} {\bibfnamefont {M.~C.}\ \bibnamefont {Ba\~nuls}},\
  }\bibfield  {title} {\bibinfo {title} {Variational matrix product operators
  for the steady state of dissipative quantum systems},\ }\href
  {https://doi.org/10.1103/PhysRevLett.114.220601} {\bibfield  {journal}
  {\bibinfo  {journal} {Phys. Rev. Lett.}\ }\textbf {\bibinfo {volume} {114}},\
  \bibinfo {pages} {220601} (\bibinfo {year} {2015})}\BibitemShut {NoStop}%
\bibitem [{\citenamefont {Mascarenhas}\ \emph {et~al.}(2015)\citenamefont
  {Mascarenhas}, \citenamefont {Flayac},\ and\ \citenamefont
  {Savona}}]{Mascarenhas2015}%
  \BibitemOpen
  \bibfield  {author} {\bibinfo {author} {\bibfnamefont {E.}~\bibnamefont
  {Mascarenhas}}, \bibinfo {author} {\bibfnamefont {H.}~\bibnamefont
  {Flayac}},\ and\ \bibinfo {author} {\bibfnamefont {V.}~\bibnamefont
  {Savona}},\ }\bibfield  {title} {\bibinfo {title} {Matrix-product-operator
  approach to the nonequilibrium steady state of driven-dissipative quantum
  arrays},\ }\href {https://doi.org/10.1103/PhysRevA.92.022116} {\bibfield
  {journal} {\bibinfo  {journal} {Phys. Rev. A}\ }\textbf {\bibinfo {volume}
  {92}},\ \bibinfo {pages} {022116} (\bibinfo {year} {2015})}\BibitemShut
  {NoStop}%
\bibitem [{\citenamefont {Cheng}\ and\ \citenamefont
  {Potter}(2022)}]{Cheng2022}%
  \BibitemOpen
  \bibfield  {author} {\bibinfo {author} {\bibfnamefont {Z.}~\bibnamefont
  {Cheng}}\ and\ \bibinfo {author} {\bibfnamefont {A.~C.}\ \bibnamefont
  {Potter}},\ }\bibfield  {title} {\bibinfo {title} {A matrix product operator
  approach to non-equilibrium floquet steady states},\ }\href
  {https://arxiv.org/abs/2206.07740} {\bibfield  {journal} {\bibinfo  {journal}
  {arXiv:2206.07740}\ } (\bibinfo {year} {2022})}\BibitemShut {NoStop}%
\end{thebibliography}%

\end{document}


\title{Supplementary Information: \\
Characterizing a non-equilibrium phase transition on a quantum computer}

\author{Eli Chertkov}
\email{eli.chertkov@quantinuum.com}
\affiliation{Quantinuum, 303 South Technology Court, Broomfield, Colorado 80021, USA}
\author{Zihan Cheng}
\affiliation{Department of Physics, University of Texas at Austin, Austin, TX 78712, USA}
\author{Andrew C. Potter}
\affiliation{Department of Physics, University of Texas at Austin, Austin, TX 78712, USA}
\affiliation{Department of Physics and Astronomy, and Stewart Blusson Quantum Matter Institute, University of British Columbia, Vancouver, BC, Canada V6T 1Z1}
\author{Sarang Gopalakrishnan}
\affiliation{Department of Electrical and Computer Engineering, Princeton University, Princeton, NJ 08544, USA}
\author{Thomas M. Gatterman}
\affiliation{Quantinuum, 303 South Technology Court, Broomfield, Colorado 80021, USA}
\author{Justin A. Gerber}
\affiliation{Quantinuum, 303 South Technology Court, Broomfield, Colorado 80021, USA}
\author{Kevin Gilmore}
\affiliation{Quantinuum, 303 South Technology Court, Broomfield, Colorado 80021, USA}
\author{Dan Gresh}
\affiliation{Quantinuum, 303 South Technology Court, Broomfield, Colorado 80021, USA}
\author{Alex Hall}
\affiliation{Quantinuum, 303 South Technology Court, Broomfield, Colorado 80021, USA}
\author{Aaron Hankin}
\affiliation{Quantinuum, 303 South Technology Court, Broomfield, Colorado 80021, USA}
\author{Mitchell Matheny}
\affiliation{Quantinuum, 303 South Technology Court, Broomfield, Colorado 80021, USA}
\author{Tanner Mengle}
\affiliation{Quantinuum, 303 South Technology Court, Broomfield, Colorado 80021, USA}
\author{David Hayes}
\affiliation{Quantinuum, 303 South Technology Court, Broomfield, Colorado 80021, USA}
\author{Brian Neyenhuis}
\affiliation{Quantinuum, 303 South Technology Court, Broomfield, Colorado 80021, USA}
\author{Russell Stutz}
\affiliation{Quantinuum, 303 South Technology Court, Broomfield, Colorado 80021, USA}
\author{Michael Foss-Feig}
\affiliation{Quantinuum, 303 South Technology Court, Broomfield, Colorado 80021, USA}

\maketitle

\beginsupplement

\tableofcontents

\section{Properties of the model}

In this work, we numerically and experimentally probe the one-dimensional Floquet quantum contact process (FQCP), a quantum circuit containing elements that drive the system to infinite temperature and dissipative elements that push the system into an absorbing state $\ket{0\cdots 0}$. Below we detail some of the properties of this model. 

\subsection{Space-time symmetries}

The FQCP is periodic in time and executes a full Floquet period in $t=2$ time steps. Within a time step, controlled rotation gates are applied left and right with control qubits on the odd sublattice followed by left and right with control qubits on the even sublattice, resulting in four layers of two-qubit gates between times $t$ and $t+1$. Note that two identical controlled-unitary gates with the same control qubit or the same target qubit commute with one another. This implies that commensurate neighboring layers of left- and right-facing controlled-rotation gates can be commuted past one another (e.g., the first and second layers or third and fourth layers between $t=0$ and $t=1$ in Fig.~1\textbf{a}). Therefore, the model is symmetric under inversion $r \rightarrow -r$ (for open boundary condition chains of odd length or periodic boundary condition chains of even length). While we only study the model with open boundary conditions, the model with periodic boundary conditions is spatially periodic with a two-site unit cell.

In this work, we have focused on the FQCP applied to an initial state $\ket{0\cdots010\cdots0}$ with a single active site at the center site $r=0$ of the chain. This state breaks translational invariance, but does preserve inversion symmetry.

\subsection{Internal symmetries}

In additional to temporal and spatial symmetries, quantum many-body systems can have internal symmetries, such as charge conservation. When such symmetries exist, they can obstruct the ability of the system to thermalize according to the eigenstate thermalization hypothesis \cite{Deutsch1991,Srednicki1994,Rigol2008}. Here we provide numerical evidence that the FQCP model does produce thermalizing behavior and so does not contain hidden internal symmetries.

\begin{figure}
    \centering
    \includegraphics[width=0.5\textwidth]{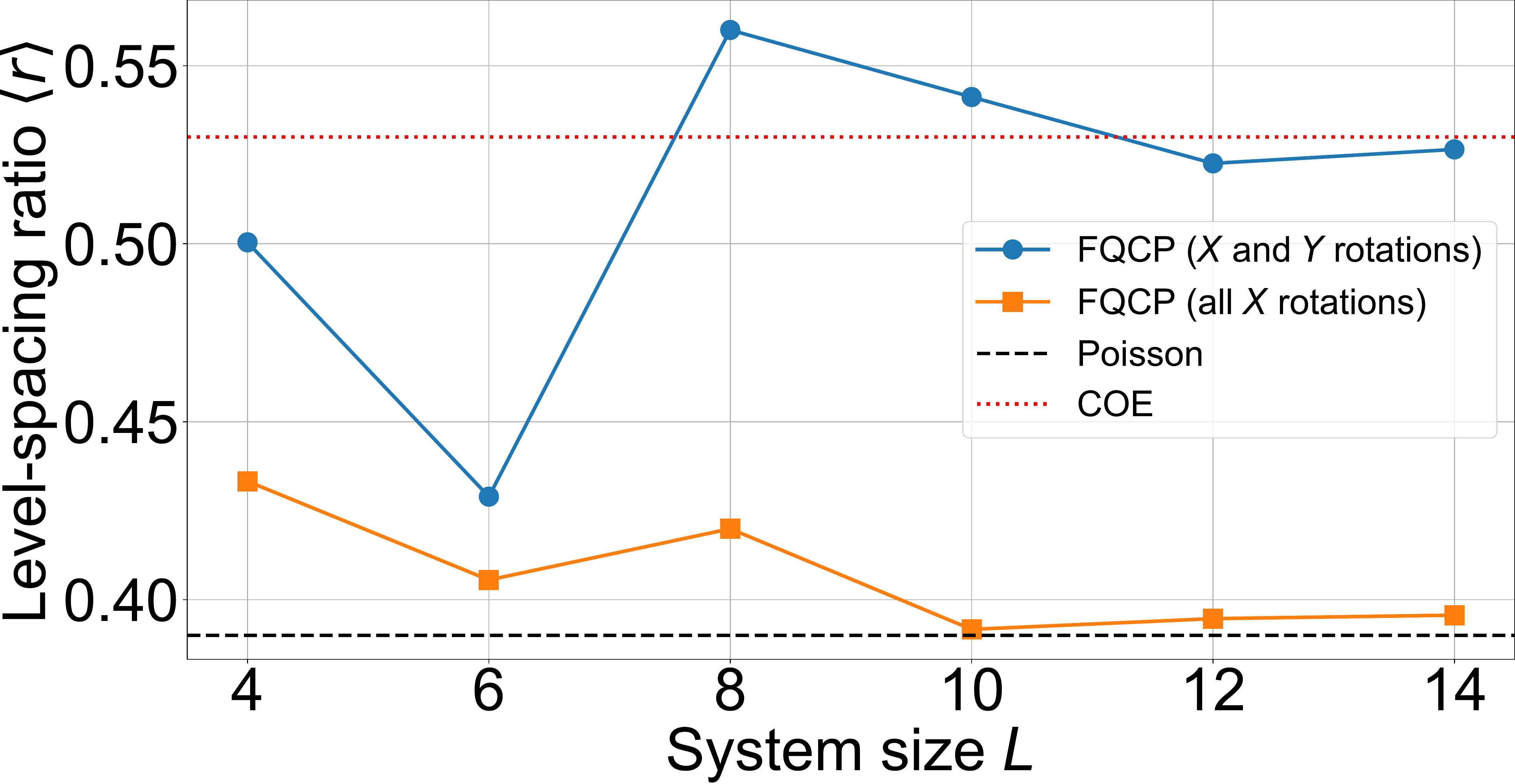}
    \caption{\textbf{Level-spacing statistics.} The average nearest-neighbor level-spacing ratio $r=\min(E_{j+1}-E_{j}, E_{j+2}-E_{j+1})/\max(E_{j+1}-E_{j}, E_{j+2}-E_{j+1})$ of the Floquet unitary $U$ with eigenvalues $e^{-iE_j}$ (sorted so that $E_j \leq E_{j+1}$) for the FQCP with no resets ($p=0$) for different system sizes $L$. The approximate value expected for Poisson (COE) statistics is marked with a dashed black line (dotted red line). The model as described in the main text, which contains $X$ and $Y$ rotations, has level-spacing statistics consistent with COE, while the model with \emph{only} $X$ rotation appears consistent with Poisson level-spacings.}
    \label{fig:levelspacings}
\end{figure}

The FQCP contains both $X$ and $Y$ axis controlled rotations. This differs from the structure of the (Hamiltonian) quantum contact process studied in previous work \cite{Marcuzzi2016,Carollo2019,Gillman2019,Minjae2021}, which only contained $X$ axis rotations. We chose this particular structure to ensure that the model, at $p=0$ when it is a non-dissipative unitary model, is not accidentally at a fine-tuned integrable point, which could cause it to display non-generic (and potentially non-thermalizing) behavior. 

One diagnostic used to assess whether a Hamiltonian (or Floquet unitary) model is integrable or non-integrable is the average level-spacing ratio $\langle r \rangle$ between the spacings of neighboring energy levels (or Floquet eigenvalues) \cite{Atas2013}. Models that possess many conserved quantities, such as integrable models, have eigenvalues that do not repel one another, which causes them to have an average level-spacing ratio of $\langle r \rangle_{Poisson}\approx 0.39$. On the other hand, ergodic models without symmetries generally have average level-spacing ratios corresponding to values obtained from random matrix theory, such as $\langle r \rangle_{COE} \approx 0.53$ for the circular orthogonal ensemble (COE) of random matrices \cite{Atas2013,DAlessio2014}.

As shown in Fig.~\ref{fig:levelspacings}, the all-$X$-rotation version of the FQCP (at $p=0$) appears to possess Poisson level-spacing statistics, while the $X$ and $Y$ rotation model is consistent with COE level-spacing statistics. This suggests that the all-$X$-rotation model has unknown conserved quantities and that the $X$-and-$Y$-rotation FQCP model has no conserved quantities and is ergodic. The average level-spacing ratios were computed using full exact diagonalization on systems with open boundary conditions and an even number of sites (so these models do not possess inversion symmetry, which requires an odd number of sites).

\subsection{Connections to other ergodic phenomena}

As we showed in the previous section, in the $p=0$ dissipation-less limit, the FQCP is an ergodic (thermalizing) unitary Floquet circuit. This means that generic eigenstates of the Floquet unitary appear as volume-law infinite-temperature eigenstates. However, by construction, the completely unentangled absorbing state $|0\ldots0\rangle$ is also an eigenstate of the Floquet unitary. Therefore, the absorbing state is a Floquet quantum many-body scar \cite{Mizuta2020,Sugiura2021}, an athermal eigenstate of the thermalizing Floquet unitary that does not possess extensive entanglement entropy.  

The FQCP is also closely related to recently studied measurement-induced phase transition (MIPT) models \cite{Skinner2019,Zabalo2020,Potter2021}, quantum circuits with unitary gates interspersed with random measurements that demonstrate a phase transition in the entanglement entropies of quantum trajectories. The FQCP can be thought of as essentially an MIPT model with a reset following each measurement and with a specially constrained set of unitary gates (controlled rotation gates instead of Haar random or Clifford gates). For this reason, it is likely that the FQCP undergoes an MIPT transition. It would be interesting future work to validate whether such an MIPT transition exists in the model and whether it coincides exactly with the directed percolation transition or whether it occurs deeper in the active phase.

\section{Methods}

Here we present additional details on the methods utilized in our FQCP experiments performed on the H1-1 quantum computer.

\subsection{Implementation details}

\begin{figure*}
    \centering
    \includegraphics[width=0.75\textwidth]{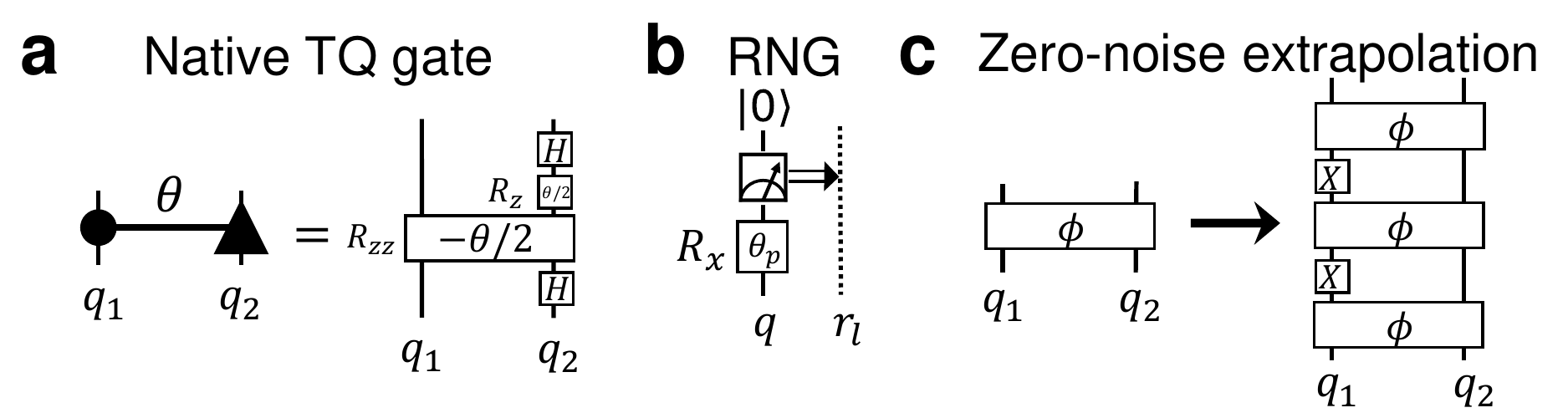}
    \caption{\textbf{Implementations of gates, random number generation, and zero-noise extrapolation.} \textbf{a} The gate decomposition of a controlled-$R_x(\theta)$ gate in terms of the native $R_{zz}(\phi)$ two-qubit gate available in the H1-1 device. \textbf{b} At the beginning of each quantum circuit, we generate random bits $r_l$ that are 1 with probability $p$ using the quantum computer, by performing single qubit $R_x(\theta_p)=\exp(-i\theta_p \hat{\sigma}^x/2)$ rotations followed by measurements, where $\theta_p = 2\arcsin(\sqrt{p})$. We can generate arbitrarily many random bits this way by resetting the qubit and repeating the procedure. \textbf{c} When performing zero-noise extrapolation, we replace each $R_{zz}(\phi)$ gate with three $R_{zz}(\phi)$ gates and two single-qubit $\hat{\sigma}^x$ gates.}
    \label{fig:implementations}
\end{figure*}

Each controlled-rotation gate in the FQCP model is realized using a single arbitrary-angle $R_{zz}(\phi)=\exp(-i\phi \hat{\sigma}^z\otimes \hat{\sigma}^z/2)$ gate that is native to the H1-1 device, as shown in Fig.~\ref{fig:implementations}\textbf{a}. These gates, as well as the random resets in the model, are conditionally applied based on the values of classical bits $z_j, r_l$ (see Fig.~2\textbf{b},\textbf{c}). The $z_j$ bits are used to track in real-time which of the qubits are known to be in the $\ket{0}$ state due to prior mid-circuit resets. Before each conditional reset, we perform a conditional measurement, which we record to bit $m_l$. After each conditional gate and reset, we update the $z_j$ bits to track the locations of the $\ket{0}$ states. The $r_l$ bits are random bits used to trigger the application of the mid-circuit resets. As shown in Fig.~\ref{fig:implementations}\textbf{b}, the $r_l$ bits are generated by the quantum computer itself at the beginning of each circuit, using single-qubit gates, mid-circuit measurements, and mid-circuit resets. 

To save run-time, we do not apply the first layer of random resets at $t=0$. This is because with probability $p$ the random reset at $r=0,t=0$ is applied to the $\ket{1}$ state at $r=0$, causing the dynamics to immediately fall into the absorbing state. After this time step, we know that the state remains unchanged for the rest of the circuit. To avoid simulating this trivial dynamics on the device, we do not apply the first random reset and simply reweight measured observables by $1-p$, the probability of there \emph{not} being an $r=0$ reset at $t=0$. At the final time $t>0$ in each experiment, we measure all $r\geq 0$ qubits; these measurements as well as all mid-circuit measurements are performed in the $\hat{\sigma}^z$-basis.

\subsection{Error mitigation}

To reduce the effects of errors on our measured results, we implement an error mitigation scheme known as zero-noise extrapolation \cite{Temme2017,Li2017}. We apply zero-noise extrapolation by performing two sets of experiments, one for the original model (TQ error $q$) and one for a modified model with each TQ gate replaced by three TQ gates (effective TQ error $3q$; see Fig.~\ref{fig:implementations}\textbf{c}). Using these two data sets, we extrapolate observables to the zero-noise limit by performing a simple linear extrapolation $O(q=0)=O(q)-\frac{1}{2}(O(3q)-O(q))=3O(q)/2-O(3q)/2$. In our experiments at $p=0.3$, for $t=2,4,6,\ldots,14$ we perform zero-noise extrapolation and collect 600 shots of data for the original circuit as well as 200 shots for the magnified-error circuit (except for the $t=12,14$ circuits which had 190, 100 shots for the magnified-error circuit, respectively). For the $p=0.1,0.5$ experiments, we perform 100 shots of the circuit for $t=2,4,\ldots,18$ and do not use zero-noise extrapolation. For the $p=0.3$ density profile shown in Fig.~1\textbf{e}, we performed 100 shots of the circuit for $t=2,4,\ldots,18$ without zero-noise extrapolation.

\subsection{Experimental resource requirements}

As shown in Fig.~2 and discussed in the main text, the FQCP circuits executed on the H1-1 quantum computer utilized qubit-reuse as well as conditional quantum gates, measurements, and resets that are applied depending on the values of classical bits that are updated in real-time during the calculation. As such, the number of times each conditional operation is triggered varies per circuit. Here we describe the number of resources -- qubits, gates, etc. -- utilized on average in these circuits.

\begin{figure*}
    \centering
    \includegraphics[width=\textwidth]{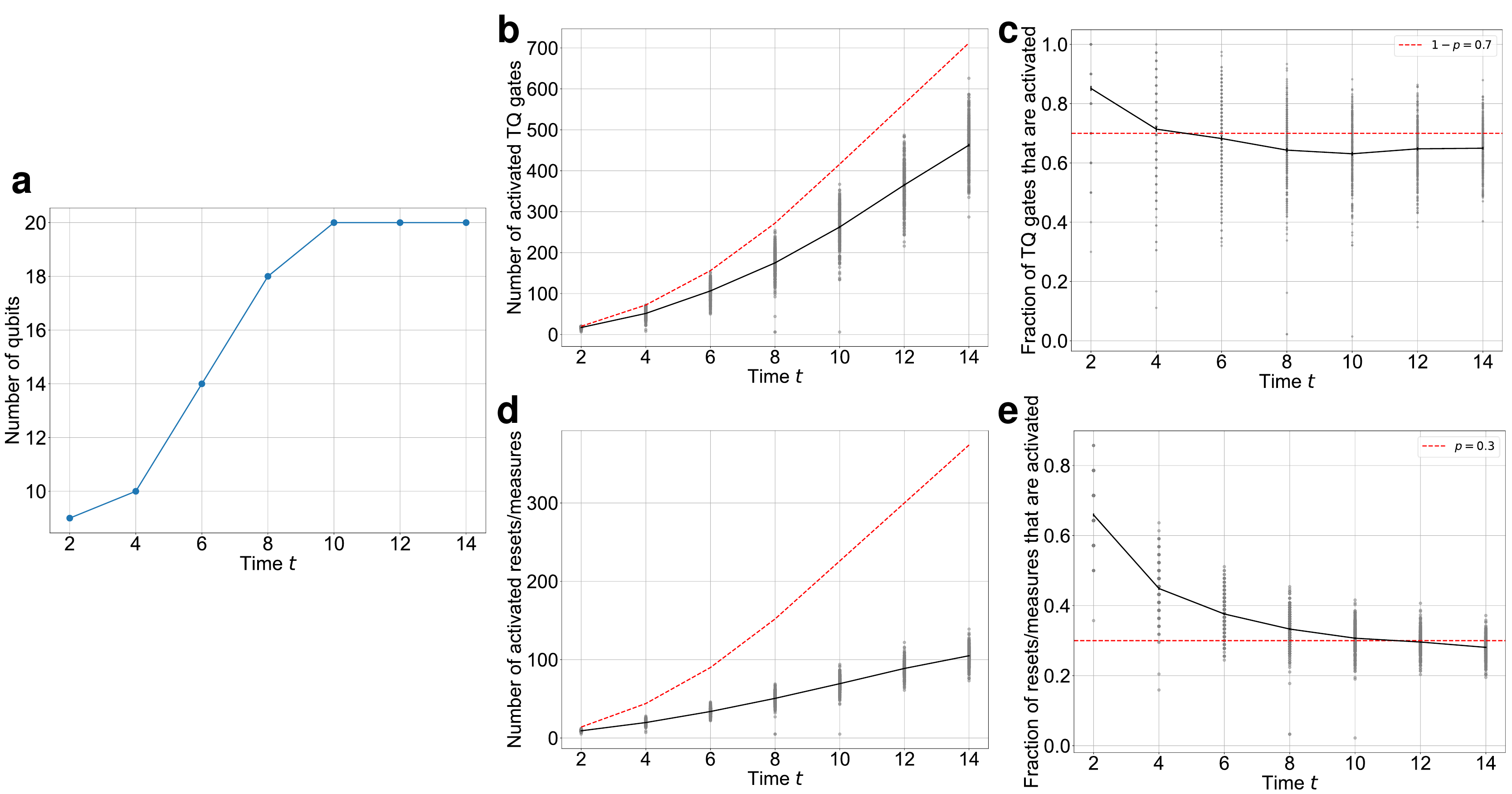}
    \caption{\textbf{Resources used in FQCP circuits.} \textbf{a} The number of qubits used for FQCP circuits with time evolution up to $t$. \textbf{b} The average number of activated two-qubit (TQ) gates versus $t$ for 100 shots of $p=0.3$ circuits (the total number of conditional TQ gates is marked with a red dashed line). \textbf{c} The fraction of TQ gates that are activated. \textbf{d} The average number of activated mid-circuit measures/resets, excluding measures/resets used in random number generation (the total number of conditional measures/resets is marked with a red dashed line). \textbf{e} The fraction of mid-circuit measures/resets that are activated.}
    \label{fig:resources}
\end{figure*}

\begin{figure}
    \centering
    \includegraphics[width=0.45\textwidth]{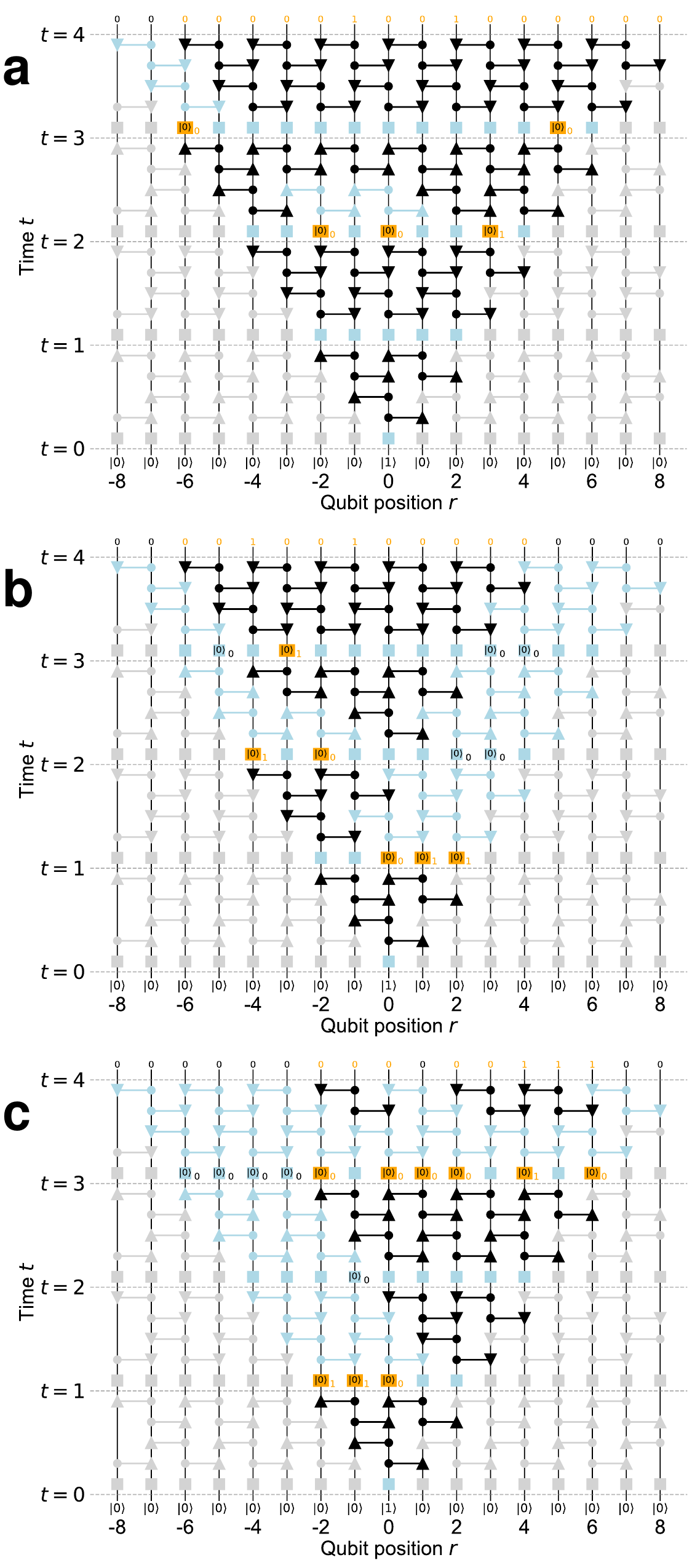}
    \caption{\textbf{Examples of conditionally executed quantum operations.} Examples of the quantum operations executed for different experimentally realized random reset patterns for $t=4$ steps of time evolution at \textbf{a} $p=0.1$, \textbf{b} $p=0.3$, \textbf{c} $p=0.5$. Conditional operations that are activated are colored black and orange, that are not activated are colored light blue, and that are not executed (even conditionally) are colored light gray. The random reset pattern is indicated by the $\ket{0}$ text. The numbers next to the boxes indicate the recorded mid-circuit measurement results obtained in experiment for the given reset pattern. Note that even if a measure is not performed in hardware, a 0 result can be recorded if a qubit is known to be in the $\ket{0}$ state due to a prior reset.}
    \label{fig:conditional_operations}
\end{figure}

The number of qubits used in our calculations varied between 9 and 20, the maximum number of qubits available on H1-1 at the time of these experiments \cite{RyanAnderson2022}. When performing qubit-reused calculations of the form shown in Fig.~2, one is free to choose how many of the highlighted sets of operations ($\varepsilon_1,\varepsilon_2,\ldots$) to perform in parallel before resetting and re-using qubits. For example, one could execute the gates and channels in $\varepsilon_1,\varepsilon_2$, then reset the qubits at $r=-2t,-2t+1$, then re-use them as input qubits at $r=2,3$. Doing more than one set in parallel increases the number of qubits required for the calculation, but speeds up the calculation by increasing the number of TQ gates that can be executed in parallel. We choose to perform the qubit-reuse in such a way so that no qubits are reset and reused until at least 10 qubits participate in the calculation. We do it this way because the H1-1 quantum computer can perform up to 5 parallel TQ gates at once with its 5 gate zones \cite{RyanAnderson2022}. Given this qubit-reuse strategy, the required number of qubits for the circuits simulating $t$ steps of time evolution is as shown in Fig.~\ref{fig:resources}\textbf{a}. 

For each circuit executed on the quantum computer, there is a different pattern of random resets and therefore a different number of gates, measurements, and resets activated by the real-time conditional logic. Fig.~\ref{fig:resources}\textbf{b}--\textbf{e} shows the statistics of the conditional operations. Roughly, as a first-order approximation, we expect that $\approx p$ of the two-qubit (TQ) controlled-rotation gates are not activated due to their control qubit being reset in the time step before the gate is applied. This is consistent with our findings in Fig.~\ref{fig:resources}\textbf{c}, though in reality the fraction of activated TQ gates is less than $1-p$ because of higher-order processes in which many random resets conspire to prevent the execution of TQ gates. Naively, we would expect that exactly $p$ of the of the conditional mid-circuit measures/resets should be applied since these conditional operations are activated by classical bits specifically sampled from a random number generator to be 1 with probability $p$. However, it is possible for mid-circuit measures/resets to \emph{not} be activated if it is known that the qubit is already in the $\ket{0}$ state from prior resets (see Fig.~3\textbf{c}), which makes the actual activated fraction less than $p$. Fig.~\ref{fig:conditional_operations} shows example realizations of the conditional operations for different reset patterns.

\subsection{Effects of hardware errors}

The ideal FQCP model has a perfect absorbing state $\ket{0\cdots 0}$, which cannot be escaped after it is reached. This property of the model ensures the directed percolation (DP) universality of the phase transition. However, in our experiments, the model is subject to errors that destroy the absorbing state, and thereby the signatures of the DP transition, at late times.

The largest error source in the H1-1 quantum computer, and one that directly affects the absorbing state, is the two-qubit (TQ) gate, which has an average gate infidelity of $3\times10^{-3}$ (see Ref.~\onlinecite{RyanAnderson2022} for a recent description of the errors in the device). Ideally, the TQ controlled-rotation gates in the FQCP do not create (or annihilate) active sites if the control qubit is in the $\ket{0}$ state. In particular, each TQ gate should leave the absorbing state unaffected. However, TQ gate errors break this fine-tuned property, leading to the creation of active sites, even if the absorbing state has been reached. We utilize real-time conditional logic to only apply TQ gates as necessary to avoid the damaging effects of unnecessarily applied TQ gates (see Fig.~2). This ``error avoidance'' technique leads to a $\approx 30\%$ reduction in the number of TQ gates used near the critical point (see Fig.~\ref{fig:resources}\textbf{b}). 

Another source of error in H1-1 is a dephasing error that accumulates on all qubits during idling and ion transport. While such dephasing errors will be detrimental to quantum coherences, they will not necessarily destroy the absorbing state property like the TQ gate errors. Since these dephasing errors are essentially $z$-basis rotations, they do not directly affect the $\ket{0\cdots 0}$ absorbing state. They can, however, create active sites that destroy the absorbing state if the dephasing occurs during the execution of a TQ controlled-rotation gate. As shown in Fig.~\ref{fig:implementations}\textbf{a}, the controlled-rotation gates are implemented with a $R_{zz}$ gate surrounded by single-qubit gates. If there is a significant delay between the application of the single-qubit gates before and after the $R_{zz}$ gate, then a significant dephasing error can accumulate within the conditional rotation gate, leading to an effective TQ gate error that creates active sites. To mitigate this effect, we submit our circuits to the quantum device in such a way so that this within-TQ-gate delay time is minimized.

In principle, others errors such as reset, measurement, reset crosstalk, and measurement crosstalk errors can also lead to the creation of active sites, though in the H1-1 device these effects are significantly less important than the effects of TQ gate and dephasing errors. 

\section{Additional experimental data}

From our experimental data, we were able to measure observables not discussed in the main text. These include
\begin{align}
P(t) &\equiv 1-\textrm{tr}(\mathcal{E}_t(\hat{\rho}_0)|0\cdots 0\rangle \langle 0 \cdots 0|) \nonumber \\
&= 1 - \langle 0 \cdots 0|\mathcal{E}_t(\hat{\rho}_0)|0\cdots 0 \rangle, \nonumber \\
P_{r\geq 0}(t) &\equiv 1-\textrm{tr}_{r<0}(\mathcal{E}_t(\hat{\rho}_0)|0_{r=0}\cdots 0_{r=2t}\rangle \langle 0_{r=0} \cdots 0_{r=2t}|),
\end{align}
which are the survival probability and right-side-survival probability, respectively. These observables directly probe the probability of a cluster seeded by a single initial state surviving for a time $t$. At the critical points of directed percolation (DP) phase transitions, the survival probability has been observed to scale asymptotically as $P(t) \sim t^{-\delta}$, where $\delta \approx 0.159464$ in one dimension \cite{Jensen1999,Hinrichsen2000}. We suspect that $P_{r \geq 0}(t)$ should also obey the same scaling at the critical point. Due to reflection symmetry $r \leftrightarrow -r$, if $P_{r\geq 0}(t)=0$ then $P(t)=0$; however, in general $P_{r \geq 0}(t) \leq P(t)$. 

These survival probabilities are computed from the same set of $\hat{\sigma}^z$-basis measurements used to gather the experimental data presented in the main text. Since these quantities involve measuring the states of all $r$ or $r \geq 0$ qubits at a fixed time $t$, most of the mid-circuit measurement data from the experiments cannot be used to compute $P(t)$ or $P_{r \geq 0}(t)$; only final-time measurement data in which many qubits are measured can be used to compute them. In our circuits (depicted in Fig.~2\textbf{a}), we measure all $-2t \leq r \leq 2t$ qubits for times $t \leq 9$ and measure $0 \leq r \leq 2t$ qubits for time $t \leq 18$. Therefore, we collected $P(t)$ data up to $t=9$ and $P_{r \geq 0}(t)$ data up to $t=18$. 

\begin{figure}
    \centering
    \includegraphics[width=0.5\textwidth]{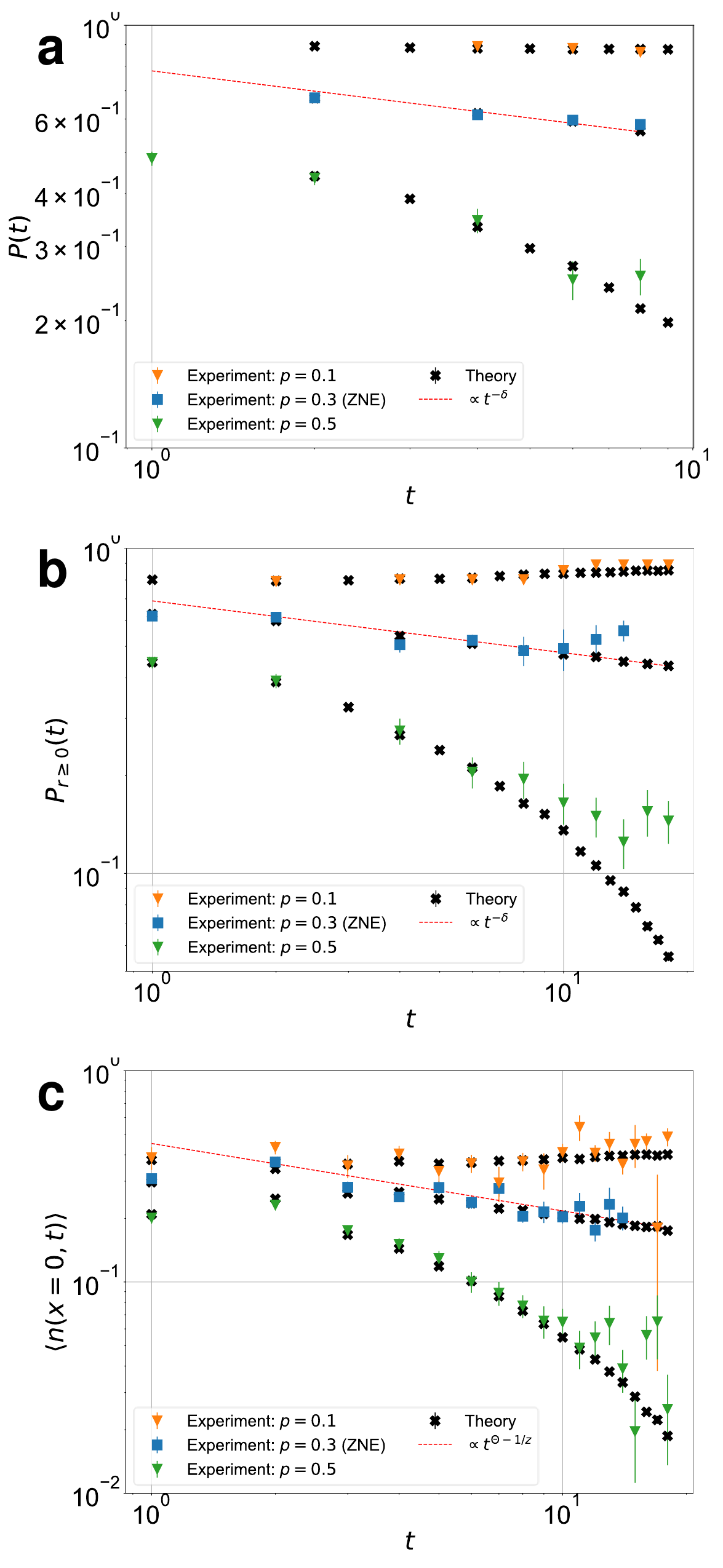}
    \caption{\textbf{Survival probabilities and center-site density.} The experimentally measured \textbf{a} survival probability $P(t)$, \textbf{b} right-side-survival probability $P_{r\geq 0}(t)$, and \textbf{c} center-site density $\langle n(r=0,t)\rangle$ for the Floquet quantum contact process with $p=0.1,0.3,0.5$ and $\theta = 3\pi/4$. The $p=0.3$ data includes zero-noise extrapolation (ZNE), as described in the main text.}
    \label{fig:observables}
\end{figure}

Fig.~\ref{fig:observables}\textbf{a},\textbf{b} show the experimentally measured survival and right-side-survival probabilities, respectively, compared to the theoretical expectation from ideal simulations (with zero-noise extrapolation utilized for $p=0.3$). In Fig.~\ref{fig:observables}\textbf{c}, we also plot the center-site density $\langle n(r=0,t)\rangle$ which at a 1D DP transition is expected to scale as $\langle n(r=0,t)\rangle \sim t^{\Theta - 1/z}$ for $\Theta=0.313686$ and $z=1.580745$ \cite{Jensen1999,Hinrichsen2000}. Though the data are limited to short times and the error bars are large due to limited sampling, the $p=0.3$ data for all of the measured observables are mostly consistent with the expected DP power-law scalings. Notably, the $P_{r \geq 0}(t)$ data for $t \gtrsim 12$ appears to deviate from the expected behavior due to TQ gate errors.

\begin{figure}
    \centering
    \includegraphics[width=0.45\textwidth]{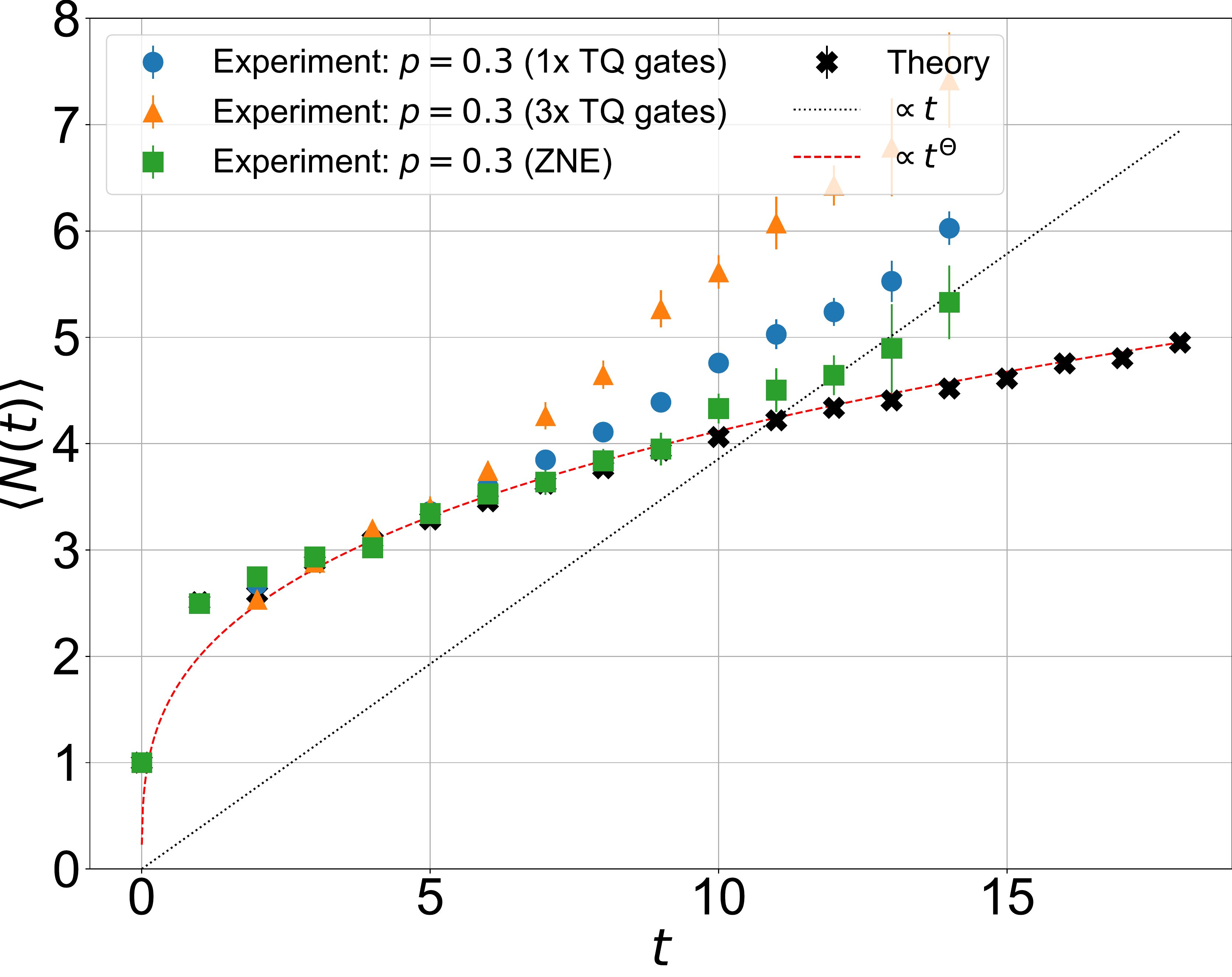}
    \caption{\textbf{Zero-noise extrapolation of observable.} The experimentally measured total number of active sites $\langle N(t) \rangle$ for FQCP circuits with the original number (``1x'') and three-times-as-many (``3x'') two-qubit (TQ) gates, as well as the zero-noise extrapolated (ZNE) values. For comparison, ballistic (black dotted lines) and DP power-law (red dashed lines) scalings are shown, as well as noise-free ideal simulations (``Theory''). Showing data for $p=0.3$ and $\theta = 3\pi/4$.}
    \label{fig:observables_zne}
\end{figure}

Fig.~\ref{fig:observables_zne} shows the effect of zero-noise extrapolation performed for the $\langle N(t) \rangle$ observable. For other observables, any improvements in accuracy due to ZNE were not resolvable within the error bars due to limited sampling.

\section{Classically solvable point numerical simulations}

As discussed in the main text, the dynamics of the FQCP at $\theta=\pi$ can be simulated efficiently classically by evolving a Markov chain. Here we describe our numerical simulations at this ``classically solvable'' point.

\subsubsection{Classical nature of $\theta=\pi$ point}

When $\theta=\pi$, the controlled $x$- and $y$-rotation gates become $-i$ times controlled-$X$ (CX) and controlled-$Y$ (CY) gates, respectivly. When a bitstring is provided as input to one of these gates, a bitstring (times a phase $\pm 1, -i$) is returned as output. The probabilistic reset channel also has this property of mapping bitstrings to bitstrings when considering the behavior of each $z$-basis-measurement quantum trajectory. Therefore, we see that when $\theta=\pi$ and the initial state is a bitstring, each quantum trajectory stays as an unentangled bitstring product state during its entire time evolution (up to an irrelevant phase).

\subsubsection{Details of numerics}

The dynamics of bitstrings due to applications of CX/CY gates and random resets can be simulated highly efficiently classically. We simulate the evolution starting from a single seed initial state $0\cdots 010\cdots 0$ and work in an effectively infinite system-size limit. We do this by storing in memory the locations of the $1$'s in the current state (in a set, or hash table) and updating these locations after each CX/CY gate. When applying a random reset channel at site $j$, we compute a random number $r$ from a pseudo-random number generator and remove the $1$ at site $j$ from the set if $r < p$. This process amounts to sampling a quantum trajectory from the channel. Before each layer of reset channels, we measure observables based on the current state and keep running averages of the observables and their variances. Our implementation of this simulation is written in Julia (version 1.6.2) \cite{julia}.

\subsubsection{Simulation results and determination of critical point properties}

We perform $t=1,\ldots,1000$ time step simulations from the initial seed state for many values of $p \in [0.35,0.45]$, repeating each simulation up to $1.12\times 10^7$ times to obtain small error bars on numerical quantities. We parallelize our calculations over many nodes of a high-performance computing cluster. From these simulations we compute the observables $\langle N(t) \rangle, \langle R^2(t)\rangle, P(t), \langle n(r=0,t)\rangle$, since they are all diagonal in the $\hat{\sigma}^z$ bitstring basis. 

For each observable $O(t)$, we also compute an ``effective exponent''
\begin{equation}
    \delta_O(t) = \frac{\log [O(t+dt)/O(t)]}{\log [(t+dt)/t]} = \frac{\log [O(t+dt)] - \log[O(t)]}{\log [t+dt] - \log[t]}. \label{eq:effective_exponent}
\end{equation}
For an observable scaling as a power-law with exponent $\delta'$ ($O(t)=O_0 t^{\delta'}$), $\delta_O(t)=\delta'$ is a constant. Generally, we expect the observables to scale as power-laws at late times at the critical point $p=p_c$, with $\delta_O$ approaching a constant at late $t$. For the classical point simulations, we use a large value of $dt=500$ to reduce the statistical errors on effective exponents.

\begin{figure*}
    \centering
    \includegraphics[width=\textwidth]{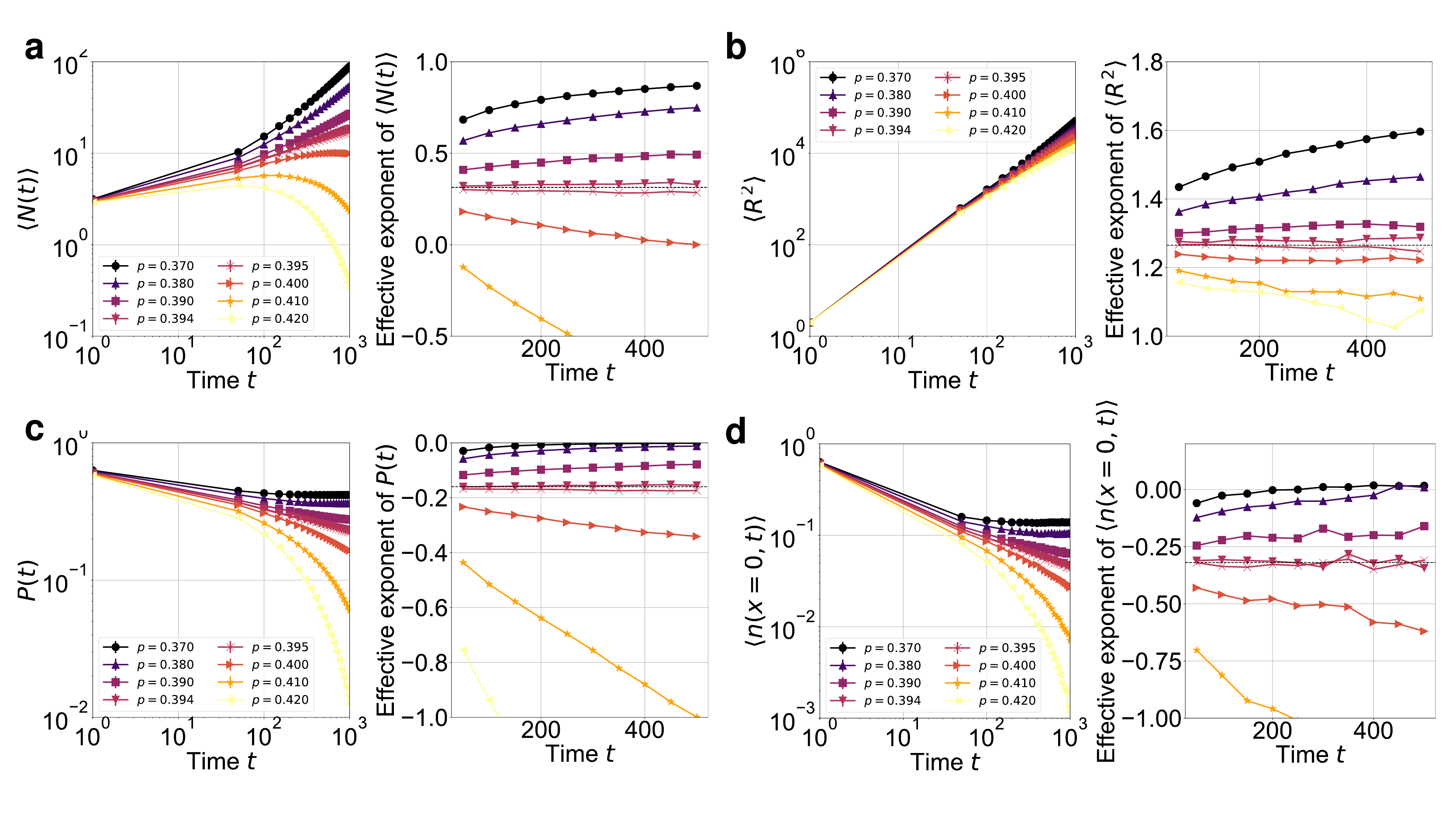}
    \caption{\textbf{Classical point observables and effective exponents.} The time dependence of \textbf{a} the total number of active sites, \textbf{b} the root-mean-squared distance, \textbf{c} the survival probability, and \textbf{d} the center-site density and their effective exponents at the classical point $\theta=\pi$ of the FQCP. The observables were estimated using $10^5$ Markov chain samples. The values of directed percolation exponents are marked with horizontal dashed black lines.}
    \label{fig:observables_and_exponents_classical}
\end{figure*}

\begin{figure*}
    \centering
    \includegraphics[width=\textwidth]{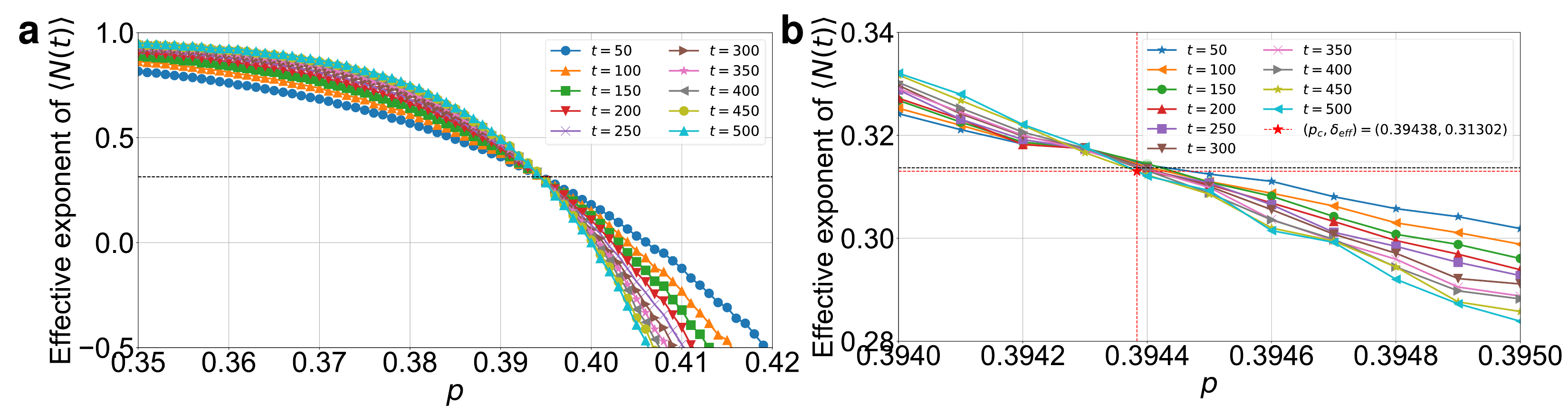}
    \caption{\textbf{Classical point effective exponent crossing.} \textbf{a} The effective exponent of the total number of active sites $\langle N(t) \rangle$ versus $p$ for fixed $t$. The DP value of the critical exponent $\Theta$ is marked with a black dashed line. The values were estimated using $10^5$ Markov chain samples. \textbf{b} A zoomed-in view of the effective exponent crossings, obtained from $1.12\times 10^7$ samples. The crossings of the $t=450$ and $t=500$ curves are marked with a red star and red dashed lines.}
    \label{fig:observables_classical_crossing}
\end{figure*}

Fig.~\ref{fig:observables_and_exponents_classical}\textbf{a} shows the time-dependence of the total number of active sites $\langle N(t) \rangle$ at the classical point for different values of $p$. Fig.~\ref{fig:observables_and_exponents_classical}\textbf{b} shows the time-dependence of the effective exponent of $\langle N(t) \rangle$ for different $p$. From this we can clearly see how the effective exponent asymptotes to a constant close to the DP value of $\Theta=0.313686$ for $p\approx 0.394$, but deviates from this constant at late times when $p$ deviates substantially from this value. One procedure for estimating the critical point $p_c$ is to estimate where the late-time effective exponents \emph{as a function of $p$} cross. That is, where $\delta_{O}(t;p)=\delta_O(t';p)$ for $t \neq t'$, $t,t' \gg 1$. Fig.~\ref{fig:observables_classical_crossing} shows the effective exponents of $\langle N(t) \rangle$ versus $p$ for fixed $t$, showing that they cross near $p=0.3944$ (the inset shows a zoomed-out view). At this crossing point, the effective exponent is $0.21\%$ away from the known DP value. Note that the crossings of the effective exponent curves drift with $t$. For the classical point, we are able to work at large enough $t$ so that this drift is barely detectable. However for the quantum point, where we perform a similar analysis (see below), we are only able to perform simulations at small $t$ and so need to carefully examine the finite-$t$ dependence of the effective exponent crossings.

\section{Quantum point numerical simulations}

For $\theta \neq \pi$, quantum superposition and entanglement affect the physics of the FQCP. In this work, we focus on a single point $\theta = 3\pi/4$, which we call the quantum point. At the quantum point, we use matrix-product operator (MPO) based methods to simulate the dynamics and determine the quasi-steady state of the model. From these simulations, we determine the approximate critical point and critical exponents in the model, which inform the choice of parameters used in our experiments.

\subsection{Dynamics}

\begin{figure}
    \centering
    \includegraphics[width=0.5\textwidth]{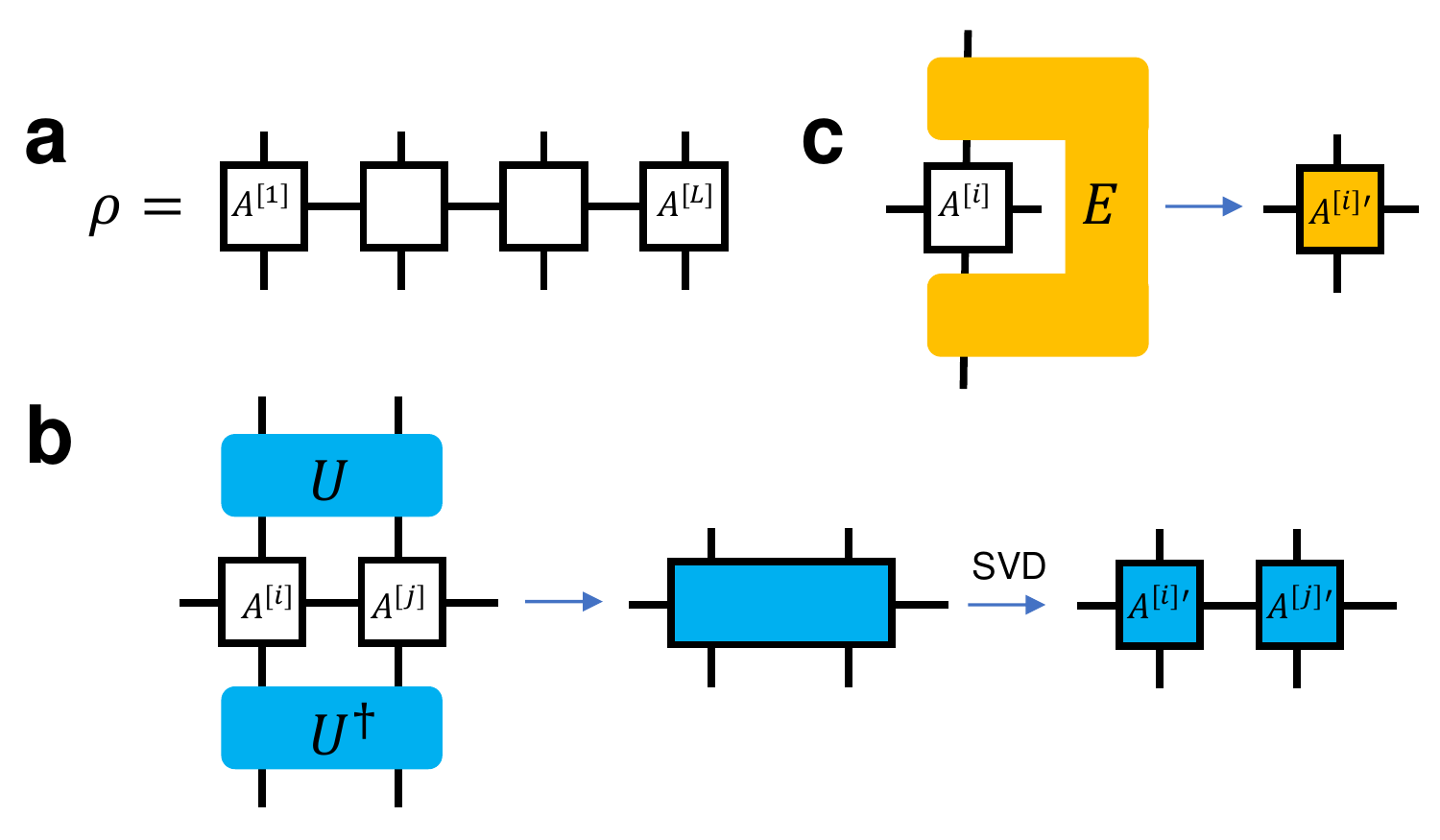}
    \caption{\textbf{MPO dynamics simulation.} \textbf{a} The density matrix $\rho$ as a matrix product operator (MPO). \textbf{b} Update of the MPO upon application of a two-qubit gate $U$, involves a singular value decomposition (SVD). \textbf{c} Update of the MPO upon application of a single-qubit channel $\mathcal{E}$.}
    \label{fig:mpo_dynamics}
\end{figure}

\subsubsection{Details of numerics}

In our dynamics simulations, the time-evolved state is a mixed state density matrix represented as a matrix product operator
\begin{equation}
\rho = \sum_{s_1,s_1',\ldots,s_L,s_L'=0}^{d-1} A^{[1]}_{s_1,s_1'} A^{[2]}_{s_2,s_2'} \cdots A^{[L]}_{s_L,s_L'} \ket{s_1}\bra{s_1'}\otimes \cdots \otimes \ket{s_L}\bra{s_L'}
\end{equation}
where $A^{[j]}_{s_j,s_j'}$ are $D_j \times D_{j+1}$ matrices ($D_1=D_{L+1}=1$), $d=2$ is the physical on-site Hilbert space dimension, and $L$ is the system size. During the evolution, two-qubit unitary gates $U$ (and one-qubit channels $\mathcal{E}$) are applied to the state, updating it according to $\rho \rightarrow U \rho U^\dagger$ ($\rho \rightarrow \mathcal{E}[\rho]$). During the $U$ updates, two neighboring $A$ tensors are contracted together forming a larger two-site tensor (see Fig.~\ref{fig:mpo_dynamics}), which can be converted back into two single-site tensors using a singular value decomposition (SVD). We perform a truncated SVD after each $U$ update, so that only the $D$ largest singular values are kept during each SVD (this results in $D_j \leq D$; $D$ is called the bond dimension). We keep the MPO normalized so that $\textrm{tr}(\rho)=1$ after each SVD. We also keep the MPO in mixed canonical form so that the orthogonality center is centered at one of the two sites where $U$ is applied before each $U$ update. Performing a single layer of gate and channel updates requires $\sim L D^3$ time. For the single seed initial state, performing updates at times $0,1,\ldots,t$ effectively involves performing layer updates on systems of size $L=1,5,\ldots,4t+1$. Therefore, performing a full MPO dynamics simulation up to time $t$ requires $\sim t^2 D^3$ time.

\subsubsection{Analysis of simulation errors}

It is important to note that the approximate density matrix $\rho$ obtained after a truncated SVD is not necessarily Hermitian or positive semi-definite like a true density matrix. Unfortunately, it is in general difficult to even check if a general MPO is positive semi-definite \cite{Kliesch2014,Werner2016,White2018}. Moreover, quantifying the errors for MPO density matrix evolution is not as straight-forward as for matrix product state (MPS) pure-state evolution. In MPS dynamics simulations, the sum of the discarded singular values squared during a truncated SVD bounds the $\ell_2$-norm-error of the pure state, which in turn bounds errors on observables. However, in MPO dynamics simulations, the same quantity only bounds the $\ell_2$-norm-error of the mixed state, but does not tightly bound the errors on observables (they are more tightly bound by the $\ell_1$ or trace norm). In our MPO simulations, we find that errors on observables can be many orders of magnitude larger than the sum of the discarded singular values squared. Due to this observation, we choose to not use an adaptive bond-dimension technique, but instead choose a fixed bond dimension $D$ and examine how observables depend on $D$.

\begin{figure}
    \centering
    \includegraphics[width=0.5\textwidth]{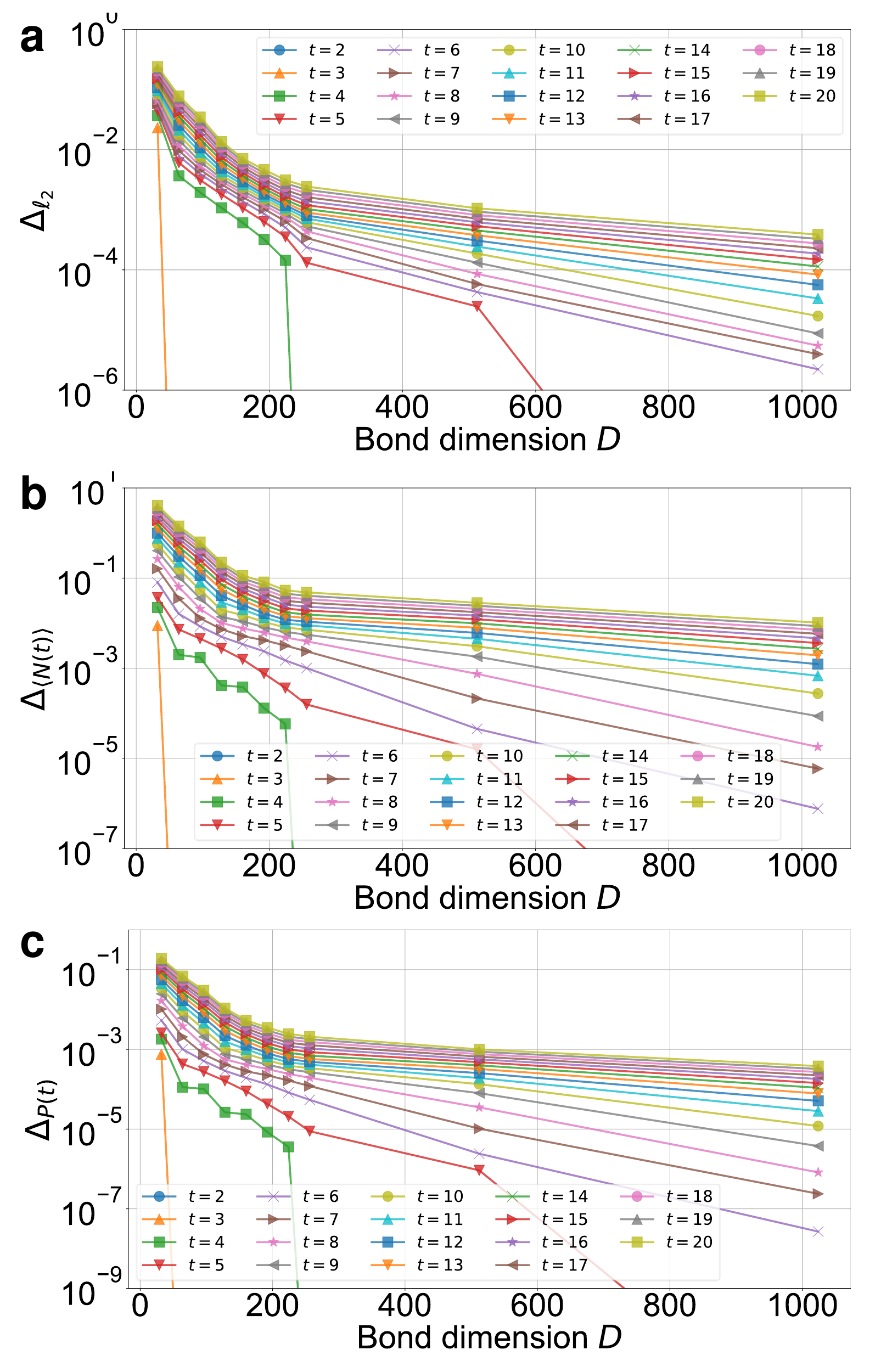}
    \caption{\textbf{MPO dynamics simulation singular value decomposition errors.} The \textbf{a} cumulative sum of the discarded singular values squared $\Delta_{\ell_2}$, \textbf{b} cumulative sum of $\langle N(t) \rangle$ differences from truncated SVD, and \textbf{c} cumulative sum of $P(t)$ differences from truncated SVD versus bond dimension $D$ for many values of $t$ near the critical point at $p=0.3$.}
    \label{fig:mpo_errors}
\end{figure}

We empirically studied the errors in the MPO simulation by looking at a few proxies for simulation inaccuracy. First, we considered the cumulative error due to singular value truncations:
\begin{equation}
\Delta_{\ell_2} = \sum_k \left(\frac{\sum_{j=D+1}^{\chi_k} (s_j^{(k)})^2}{\sum_{l=1}^{\chi_k} (s_l^{(k)})^2}\right)
\end{equation}
where $k$ indexes the SVDs performed during the simulation and $s_1^{(k)}\geq \ldots \geq s_{\chi_k}^{(k)}$ are the ordered singular values obtained during the $k$-th SVD. The term in parentheses is a (normalized) sum of the squares of the singular values discarded during the $k$-th SVD. Additionally, we also recorded the cumulative errors in observables due to SVD truncation:
\begin{equation*}
\Delta_{O} = \sum_k |O_k - O_k'|
\end{equation*}
where $O_k$ is the observable computed before the $k$-th SVD truncation and $O_k'$ is the observable computed after. This quantity does not bound the inaccuracy of the computed finite-$D$ observable relative to the true $D\rightarrow \infty$ answer, but does provide a rough proxy for this quantity. Fig.~\ref{fig:mpo_errors} shows the cumulative SVD errors $\Delta_{\ell_2}$ and cumulative errors in total active sites $\Delta_{\langle N(t) \rangle}$ and survival probability $\Delta_{P(t)}$ for different bond dimensions at $p=0.3$.

\begin{figure*}
    \centering
    \includegraphics[width=\textwidth]{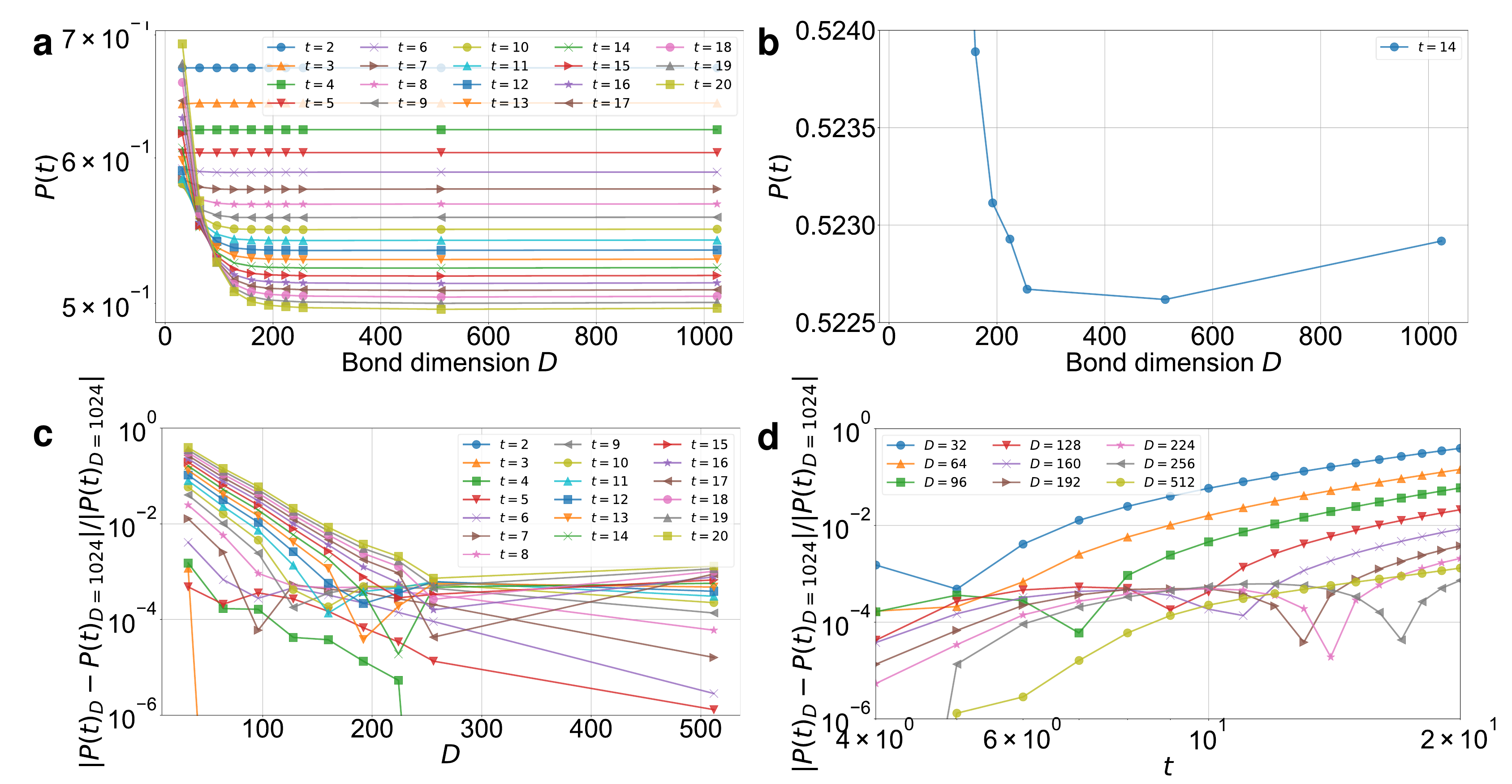}
    \caption{\textbf{MPO dynamics observable errors.} \textbf{a} The survival probability $P(t)$ versus bond dimension $D$ for fixed times $t$ at $p=0.3$. \textbf{b} A zoomed-in view of $P(t=14)$ versus $D$, highlighting the non-monotonic convergence of observables with increasing bond dimension. \textbf{c} The relative error in $P(t)$ versus $D$ compared with the largest bond dimension $D=1024$ result at fixed $t$, on a log-linear scale. \textbf{d} The relative error in $P(t)$ versus time $t$ at fixed $D$, on a log-log scale.}
    \label{fig:mpo_observable_errors}
\end{figure*}

In addition to measuring the accumulated errors from each SVD, we also measured the relative observable error $|O_D - O_{D=D_0}|/|O_{D=D_0}|$, where $O_D$ is the observable obtained from a finite $D$ MPO simulation and $D_0=1024$ is the largest bond dimension simulated. Fig.~\ref{fig:mpo_observable_errors} shows the survival probability $P(t)$ observable and its relative errors versus bond dimension. Interestingly, the relative error does not monotonically decrease with $D$. This could possibly be due to the non-monotonic dependence of observables with bond dimension (see Fig.~\ref{fig:mpo_observable_errors}\textbf{b}) or potentially due to numerical convergence issues in the MPO simulations (though there is not evidence for this in the cumulative SVD errors shown in Fig.~\ref{fig:mpo_errors}).

\subsubsection{Simulation results and determination of critical point properties}

In our simulations, we time-evolve the single seed initial state up to time $t=20$. To avoid any finite system-size effects, we use a finite MPO of length $L=4t+2=82$ sites. We perform simulations at $\theta=3\pi/4$ for many values of $p$ and use bond dimensions up to $D=1024$ (unless specified otherwise, all figures show $\theta=3\pi/4$ and $D=1024$ results). Our MPO code is written in Julia (version 1.6.2) \cite{julia} and uses the ITensor library (version 0.2.9) \cite{itensor}.

\begin{figure*}
    \centering
    \includegraphics[width=\textwidth]{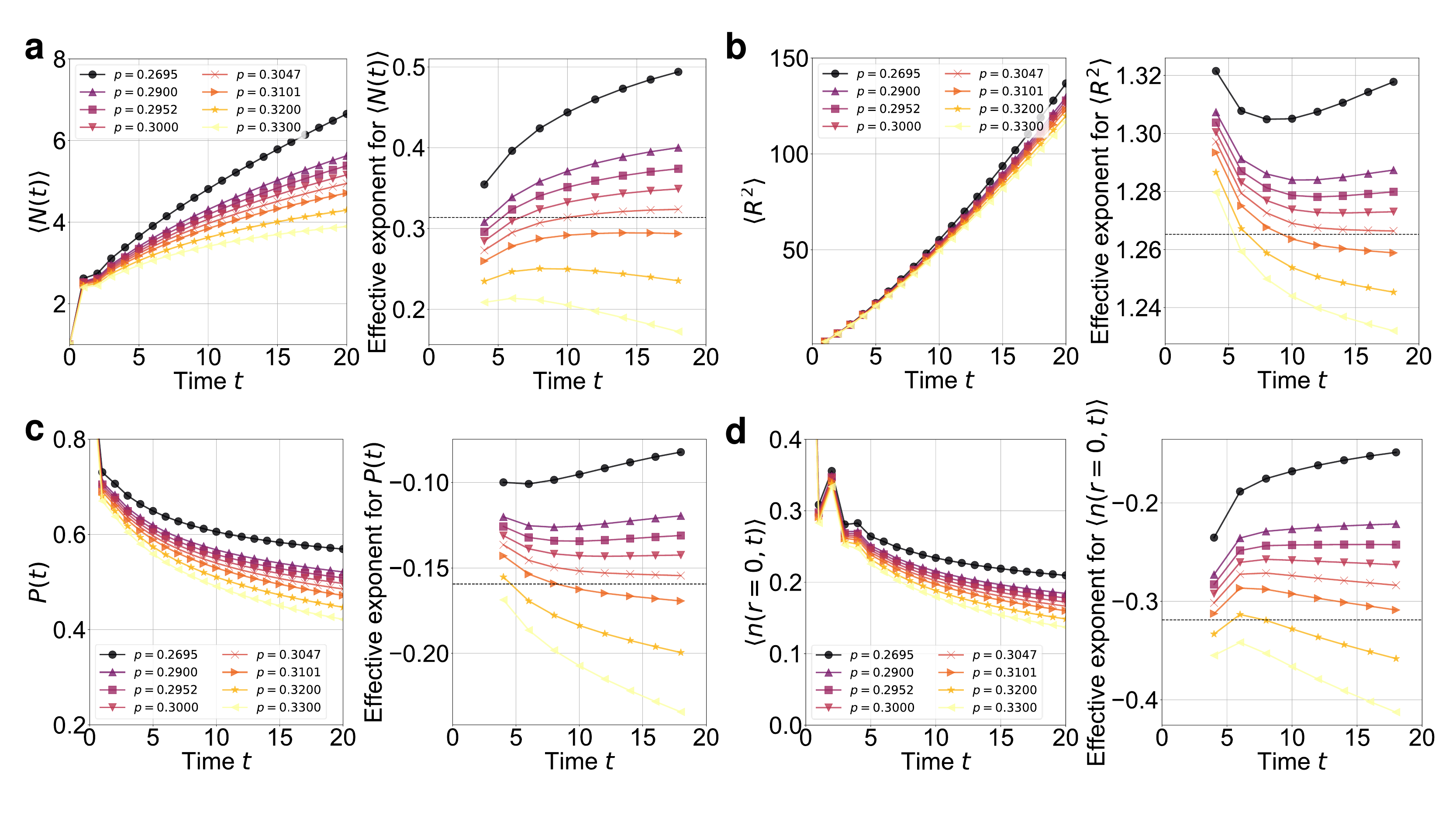}
    \caption{\textbf{Observables and effective exponents.} The time dependence of \textbf{a} the total number of active sites, \textbf{b} the root-mean-squared distance, \textbf{c} the survival probability, and \textbf{d} the center-site density and their effective exponents, for a $D=1024$ bond dimension MPO simulation.}
    \label{fig:observables_and_exponents}
\end{figure*}

Fig.~\ref{fig:observables_and_exponents} shows the values of observables measured from our MPO dynamics simulations versus time $t$ for many fixed $p$ values. For each observable $O(t)$, we also computed an effective exponent (Eq.~(\ref{eq:effective_exponent}) with $dt=2$). 

\begin{figure*}
    \centering
    \includegraphics[width=\textwidth]{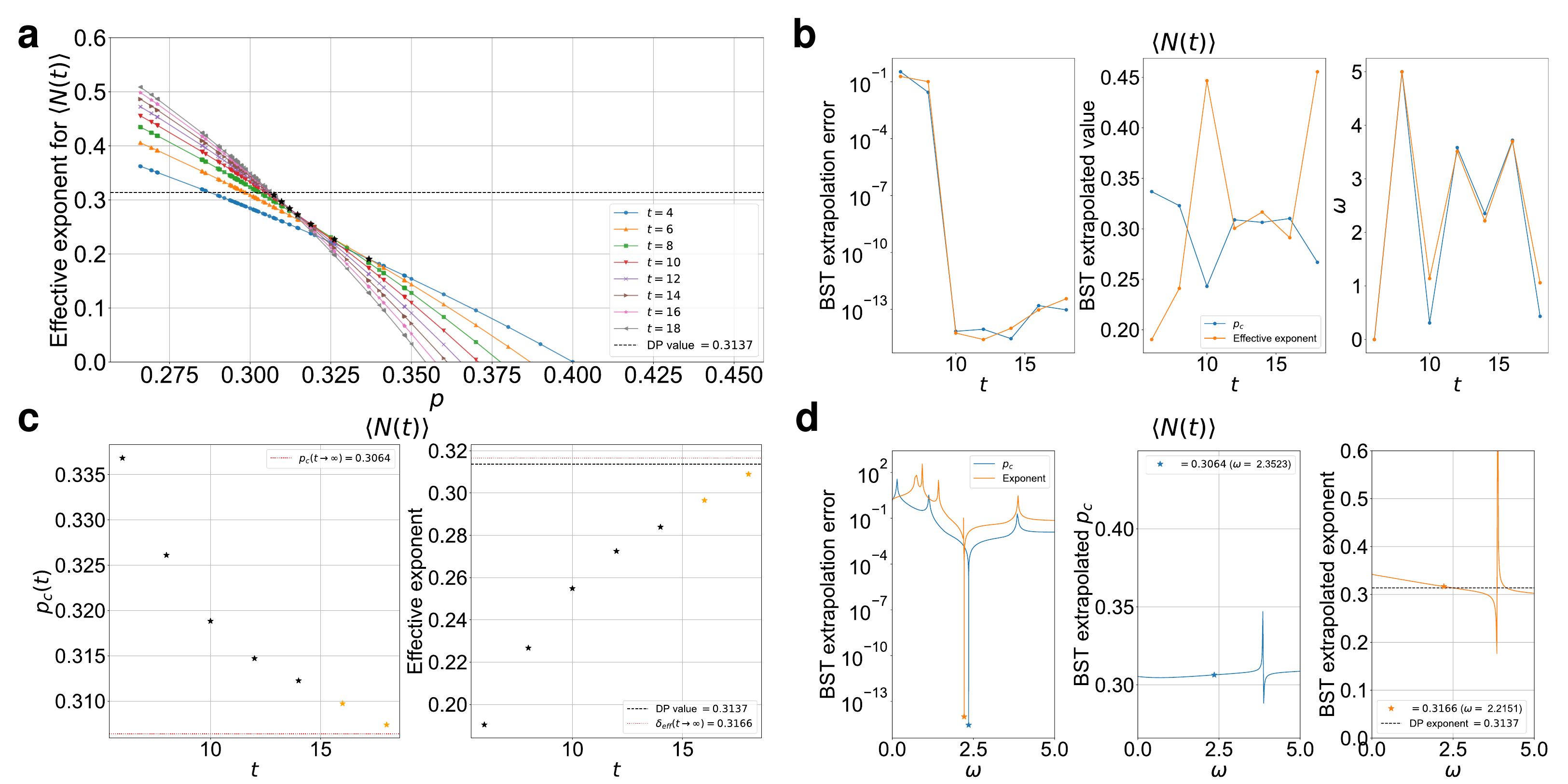}
    \caption{\textbf{BST extrapolation of $\langle N(t) \rangle$.} \textbf{a} The effective exponents of the observable versus $p$ for different times $t$, with crossings between $t$ and $t+2$ curves marked with black stars. \textbf{b} The optimal BST extrapolations using the crossings with times up to $t$. \textbf{c} The $p$ and effective exponent values at the marked crossings in \textbf{a} versus time $t$. The orange stars indicate crossings that are likely inaccurate due to finite bond-dimension effects (in \textbf{b}, we see dramatic changes in the performance of the BST extrapolation when using these crossings). \textbf{d} The optimal BST extrapolation data for $p$ and the effective exponent versus parameter $\omega$ using all of the accurate crossings indicated in \textbf{c}. Black dashed lines indicate the critical exponent values for directed percolation and the red dotted lines indicate our best BST extrapolations.}
    \label{fig:crossings_totalN}
\end{figure*}

To estimate the location of the critical point in our model, we follow a procedure utilized in previous numerical studies of non-equilibrium phase transitions \cite{Dickman2008,Filho2011}. We compute the effective exponents $\delta_O(t;p)$ for fixed $t$ and many different values of $p$. We then identify the $p=p_c(t)$ values at which $\delta_O(t+\tau;p_c(t))=\delta_O(t;p_c(t))$, i.e., where the $\delta_O(t+\tau;p)$ and $\delta_O(t;p)$ curves cross (we use $\tau=2$). Fig.~\ref{fig:crossings_totalN}\textbf{a} shows the effective exponent curves for $\langle N(t) \rangle$, with the $(p_c(t), \delta_{\langle N(t) \rangle}(t;p_c(t)))$ crossing points marked by black stars. These crossing points drift towards the true critical point as $t$ increases so that $p_c(t\rightarrow \infty)=p_c$ approaches the true critical decay probability and $\delta_{\langle N(t) \rangle}(t \rightarrow \infty;p_c(t \rightarrow \infty)) = \Theta$ approaches the true critical exponent. However, for small finite $t$, there can be significant finite-time deviations from the infinite-time limit results. Following prior work \cite{Carlon1999,Filho2011,Dickman2008}, we use a technique known as Bulirsch-Stoer (BST) extrapolation \cite{Henkel1988} to estimate the $t\rightarrow\infty$ limit results for $p_c$ ($\delta_O$) from the finite $p_c(t)$ ($\delta_O(t;p_c(t))$) data. The BST technique involves a free parameter $\omega$, which roughly corresponds to the leading order correction to the infinite-time value (e.g., $p_c(t) \approx p_c + a t^{-\omega} +\cdots$). We determine the optimal $\omega$ value for the extrapolation by performing a fine scan over $\omega$ values from $10^{-3}$ to $5$ and finding the value that minimizes an ``extrapolation error'' objective function. This extrapolation error, used in Ref.~\onlinecite{Filho2011}, is defined as the sum of the differences between BST extrapolated values obtained from the original data set and modified data sets with one value removed. Fig.~\ref{fig:crossings_totalN}\textbf{b} shows the extrapolation error, extrapolated values, and optimal $\omega$ parameters obtained from data sets containing the $\langle N(t) \rangle$ crossings up to a time $t$. Due to the finite bond dimension $D$ used in the MPO numerics, the effective exponents are less accurate at larger $t$ (and smaller $p$), where correlations are the strongest. Because of this, the late-time crossings are likely inaccurate, which can be seen in abrupt changes in the BST extrapolation at late times (e.g., abrupt jumps in extrapolation error, extrapolated value, or optimal $\omega$). Fig.~\ref{fig:crossings_totalN}\textbf{c} shows the $p$ and effective exponent values at the crossings, as well as the optimal BST extrapolation obtained from the accurate early-time crossing data with the DP exponent values marked for comparison. Fig.~\ref{fig:crossings_totalN}\textbf{d} shows the extrapolation error, extrapolated $p_c$, and extrapolated exponent values as a function of $\omega$ for the extrapolation shown in Fig.~\ref{fig:crossings_totalN}\textbf{c}. Note that we estimate the locations of the crossings by performing linear interpolation between the two curves; to ensure accurate crossing values, we perform MPO simulations at enough values of $p$ so that each computed crossing value $p_c(t)$ is between two simulated $p$ values spaced at most $2\times 10^{-4}$ apart.

We apply the same procedure to the the mean-squared displacement $\langle R^2(t) \rangle$, survival probability $P(t)$, and center-site density $\langle n(r=0,t)\rangle$. From these analyses, we estimate the critical point locations $p_c \approx 0.3064, 0.3067, 0.304, 0.29$, which suggests that $0.29 \lesssim p_c \lesssim 0.31$. Most of the effective exponent values obtained this way are close to the DP values, deviating by $0.92\%$, $0.28\%$, $4.4\%$  relative error for $\langle N(t) \rangle$, $\langle R^2(t) \rangle$, and $P(t)$, respectively (see Table~\ref{tab:exponents} for comparison of exponents with DP values). The small spread in $p_c$ estimates and the closeness of the effective exponents to the known DP values provide good evidence that the FQCP has a DP transition at $p_c\approx 0.3$ for $\theta=3\pi/4$. Notably, the center-site density's effective exponent appears to deviate significantly, about $31\%$, from the DP value. Some possible explanations for this deviation are: strong finite-time effects for this observable and strong finite bond-dimension effects (the center site in the 1D MPO is the most highly entangled site and the site most affected by SVD truncation). Evidence that this discrepancy is due to short-time effects is given in Fig.~\ref{fig:mpo_dynamics_classicalpoint}, which shows that the center-site density at the classical point also exhibits a critical exponent far from the DP value at early times. From Fig.~\ref{fig:observables_and_exponents_classical}\textbf{d}, we see that eventually, at late enough times, the classical point center-site density does eventually approach the expected DP power-law scaling.

We also performed MPO dynamics simulations for two modified version of the FQCP model: one with all controlled $X$ rotations (instead of $X$ and $Y$ rotations) and one with single-qubit amplitude damping channels instead of probabilistic reset channels. Using the same effective-exponent-crossing analysis described above, we found similar evidence of a transition at $p_c \approx 0.3$ with effective exponents consistent with directed percolation. These results suggest that the FQCP model is not fine-tuned and that different unitary gates and dissipative channels can also lead to the same directed percolation transition.

\begin{table}
    \centering
    \begin{tabular}{l|c|c|c|c}
        Observable & Exponent & FQCP & DP & Relative error \\
        \hline
        \hline
        $\langle N(t) \rangle$ & $\Theta$ & $0.3166$ & $0.313686$ & $0.92\%$ \\
        $\langle R^2(t)\rangle$ & $2/z$ & $1.2616$ & $1.265226$ & $0.28\%$\\
        $P(t)$ & $-\beta/\nu_\parallel$ & $-0.1525$ & $-0.159464$ & $4.4\%$ \\
        $\langle n(r=0,t)\rangle$ & $\Theta-1/z$ & $-0.2186$ & $-0.318927$ & $31\%$ \\
    \end{tabular}
    \caption{\textbf{Estimated FQCP critical exponents.} For four observables, which at the critical point scale as power-laws in time $\sim t^\delta$ for exponent $\delta$, we present our estimates for the critical exponents for the Floquet quantum contact process (FQCP) at the quantum point $\theta=3\pi/4$. These estimates were obtained from the effective-exponent-crossing analysis described in the text, which used MPO dynamics simulation data. For comparison, we include the directed percolation (DP) values\cite{Jensen1999,Hinrichsen2000} of the exponents as well as relative errors between the FQCP and DP values.}
    \label{tab:exponents}
\end{table}

\subsubsection{Validation of methods on classical point}

\begin{figure*}
    \centering
    \includegraphics[width=\textwidth]{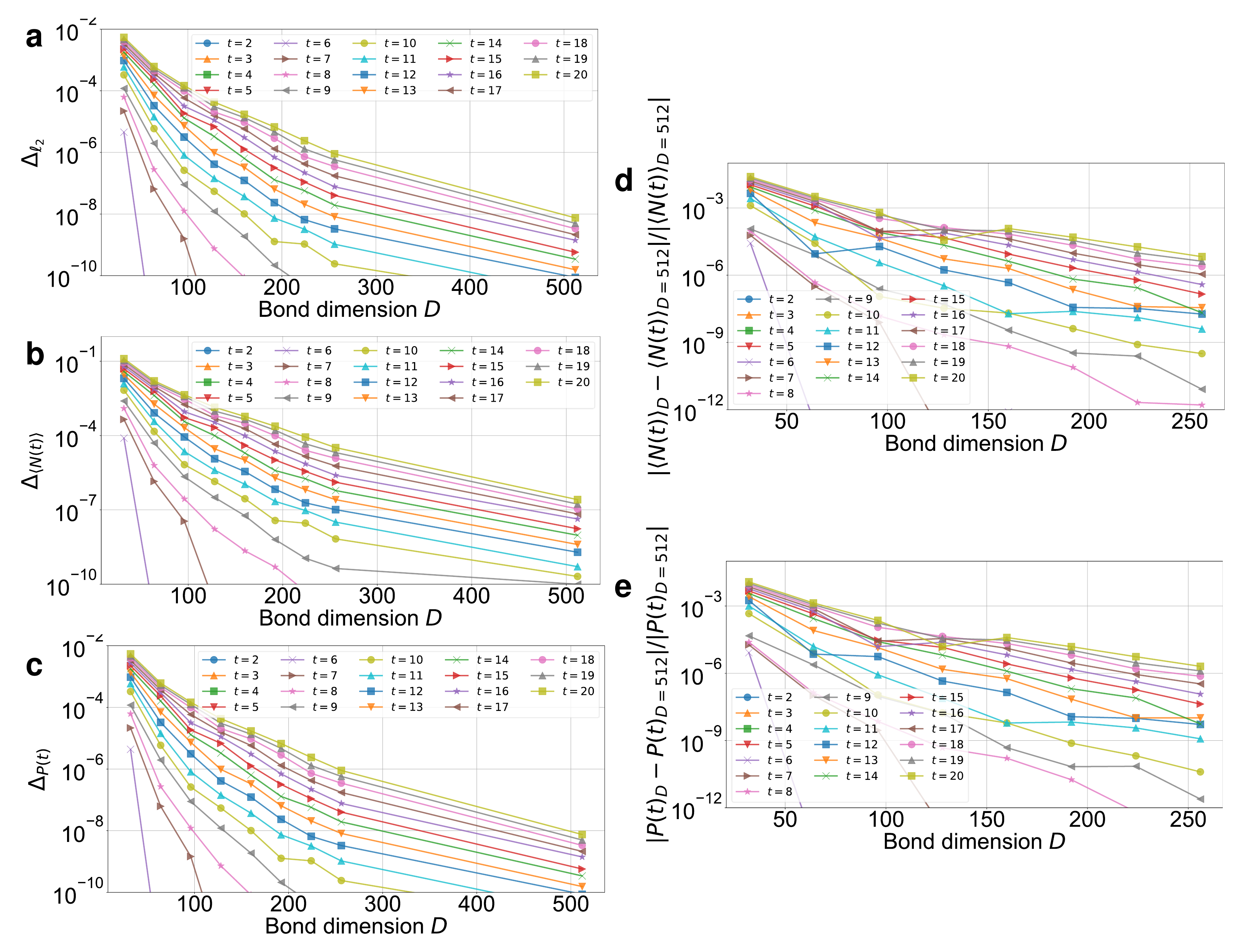}
    \caption{\textbf{MPO dynamics errors at classical point.} The cumulative sum of \textbf{a} the discarded singular values squared $\Delta_{\ell_2}$, \textbf{b} $\langle N(t) \rangle$ differences from truncated SVD, and \textbf{c} $P(t)$ differences from truncated SVD versus bond dimension $D$ for many values of $t$, on a log-linear scale. The relative error in \textbf{d} $\langle N(t) \rangle$ and \textbf{e} $P(t)$ versus $D$ compared with the largest bond dimension $D=512$ result at fixed $t$, on a log-linear scale. The results are at the classical point $\theta=\pi$ at $p=0.394$ near the critical point.}
    \label{fig:mpo_errors_classicalpoint}
\end{figure*}

\begin{figure*}
    \centering
    \includegraphics[width=\textwidth]{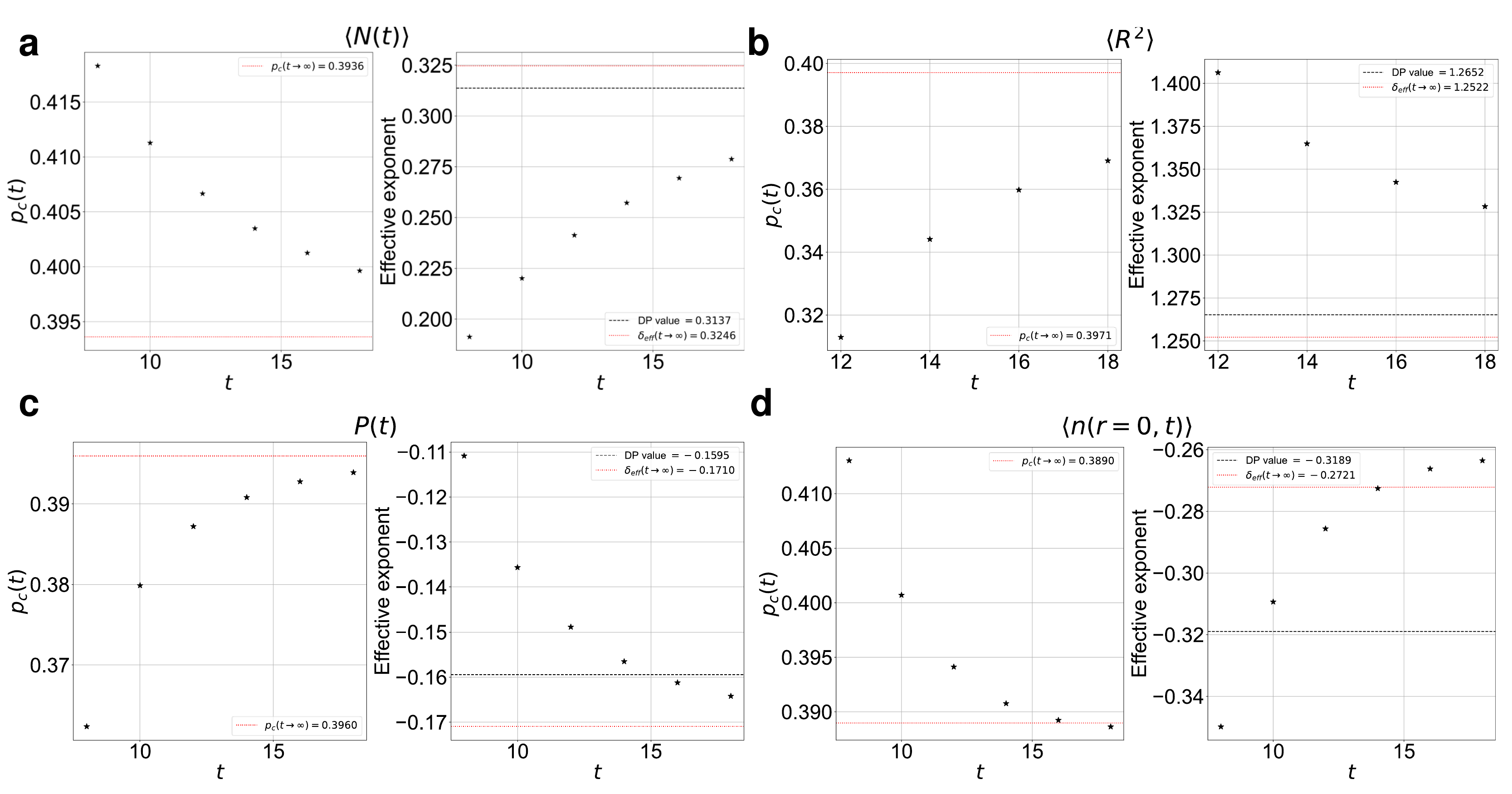}
    \caption{\textbf{Determination of critical point scaling at classical point.}  The crossings of the effective exponents for the \textbf{a} the total number of active sites, \textbf{b} the root-mean-squared distance, \textbf{c} the survival probability, and \textbf{d} the center-site density for a $D=512$ bond dimension MPO simulation at the classical point $\theta=\pi$. The black dashed lines correspond to DP exponent values and the red dashed lines correspond to BST extrapolations based on the crossings.}
    \label{fig:mpo_dynamics_classicalpoint}
\end{figure*}

To validate our approach for locating the transition and determining the critical exponents, we also performed the same MPO dynamics simulations and effective exponent crossing analysis at the classically solvable point $\theta=\pi$. As shown in Fig.~\ref{fig:mpo_errors_classicalpoint}, we found that the errors in the MPO simulation at the classical point were much lower than at the quantum point (similar results were observed in MPO dynamics simulations performed in Ref.~\onlinecite{Gillman2019}). Fig.~\ref{fig:mpo_dynamics_classicalpoint} shows the effective exponent crossings for $\langle N(t) \rangle$, $\langle R^2(t)\rangle$, $P(t)$, and $\langle n(r=0,t)\rangle$ obtained from bond dimension $D=512$ MPO dynamics simulations. The relative errors of the effective exponents from the DP values are $3.5\%,1.0\%, 7.2\%, 15.7\%$, respectively. Like for the quantum point MPO simulations, the $\langle n(r=0,t)\rangle$ effective exponent deviates substantially from the true value due to strong finite $t$ effects. Moreover, the estimates of the critical point of $p_c \approx 0.3936, 0.3971, 0.3960, 0.3890$ obtained from BST extrapolation are within $0.2\%, 0.7\%, 0.4\%, 1.4\%$ of the $p_c=0.3944$ estimate obtained from the high-accuracy classical point numerics discussed above (see Fig.~\ref{fig:observables_classical_crossing}). These results suggest that our procedure for estimating critical points from short-time MPO dynamics simulations can accurately estimate the presence of the phase transition and that the quantum point and classical point both exhibit DP critical scaling.

\subsubsection{Dynamics of entanglement entropies and other quantities}

\begin{figure}
    \centering
    \includegraphics[width=0.45\textwidth]{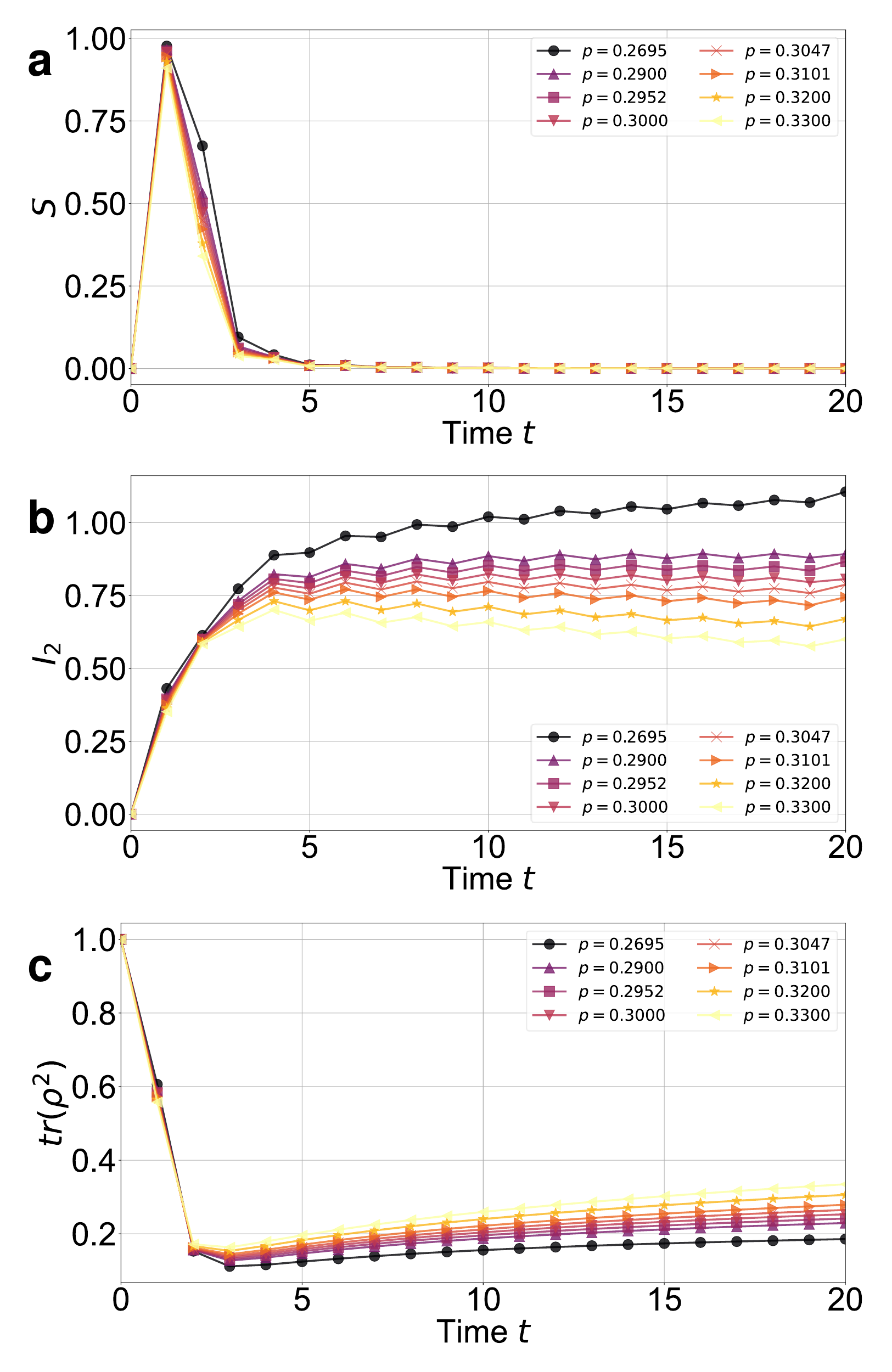}
    \caption{\textbf{MPO dynamics entropies.} \textbf{a} Half-cut MPO entanglement entropy $S$, \textbf{b} half-cut Renyi mutual information $I_2$, and \textbf{c} purity $\textrm{tr}(\rho^2)$ of the full density matrix versus time $t$ for different $p$. Results for $\theta=3\pi/4$.}
    \label{fig:mpo_dynamics_entanglement}
\end{figure}

In our MPO simulations, we also compute quantities, such as entropies, that cannot be measured (or are difficult to measure) experimentally. Fig.~\ref{fig:mpo_dynamics_entanglement}\textbf{a} shows the half-cut MPO entanglement entropy $S = -\sum_m p_m \log p_m$ where $p_m=s_m^2/\sum_k s_k^2$ and $s_m$ are the singular values of the $A^{[j]}$ tensor (shaped as a matrix with the columns corresponding to the bond between sites $j$ and $j+1$). Note that this quantity, despite its name, does not measure entanglement but rather computes the entropy present in the singular value spectra of the tensors connecting sites $j$ and $j+1$. Surprisingly, we find that, after a brief increase in the first time step, the MPO entanglement entropy quickly decays to zero, \emph{for all $p$}. This appears similar to ``entanglement barriers'' observed in prior studies of open quantum systems \cite{Noh2020}. Naively, this suggests that one does not need a large bond dimension MPO to accurately capture the dynamics of the system at late times, even at the critical point. However, by considering the convergence of observables with bond dimension, we find that one does need a significant bond dimension (that increases with $t$) to obtain accurate simulation results for $p \lesssim p_c$. This suggests that the MPO entanglement entropy does not always directly coincide with classical simulation difficulty, as suggested in a recent study \cite{Noh2020}. Fig.~\ref{fig:mpo_dynamics_entanglement}\textbf{b} shows the half-cut Renyi mutual information $I_{2} = S_2(\rho_A) + S_2(\rho_B) - S_2(\rho)$, where $\rho_A$ ($\rho_B$) is the reduced density matrix of the left-half (right-half) of the system and $S_2(\rho)=-\log \textrm{tr}(\rho^2)$. Unlike the MPO entanglement entropy, the Renyi mutual information grows with $t$ for $p \lesssim p_c$; it appears to more accurately capture the presence of (classical and quantum) correlations in the system than the MPO entanglement entropy. Fig.~\ref{fig:mpo_dynamics_entanglement}\textbf{c} shows the purity $\textrm{tr}(\rho^2)$ for the full density matrix $\rho$. Note that for the single seed initial state considered here, the density matrix has $\gtrsim p$ overlap with the absorbing state ($\rho \approx p|0\cdots0\rangle\langle 0\cdots 0| + \cdots$), which means that $\textrm{tr}(\rho^2)\gtrsim p^2$. This prevents the purity from going to zero in the active phase.

\subsubsection{Uniform versus single seed initial state}

Non-equilibrium phase transitions can exhibit critical scaling of observables that depend on initial conditions. In addition to the previously discussed single seed initial state, another initial state commonly studied in absorbing state transitions\cite{Hinrichsen2000} is the uniform active state $\ket{1\cdots 1}$. In numerical studies of the continuous time quantum contact process, it was observed that some observables for the single seed initial state, e.g., $\langle R^2(t) \rangle$ and $\langle N(t)\rangle$, were consistent with DP scaling \cite{Gillman2019} while some observables for the uniform initial state, e.g., $\langle n(t) \rangle = \frac{1}{L}\sum_{r} \langle n(r,t)\rangle$, deviated significantly from DP scaling \cite{Carollo2019}. Here we assess whether this is the case for the FQCP model.

We perform MPO dynamics simulations with a uniform initial state at the quantum point for up to $t=20$ time steps with a length $L=4t+2=82$ site MPO with maximum bond dimension of $D=256$. Empirically, we observe that the uniform initial state simulations converge faster with bond dimension than the single seed state ones. For example, for the $D=256$ uniform state simulations the cumulative discarded singular value squared $\Delta_{\ell^2}$ was always significantly below $10^{-4}$ for all $t\leq 20$, while for the $D=256$ single seed state simulations (shown in Fig.~\ref{fig:mpo_errors}\textbf{a}) the same quantity was always significantly above $10^{-4}$ for all $t$. In our uniform state simulations, we focus on the average active site density observable $\langle n(t) \rangle$. For our numerics, we study systems of size $L = 4t+2$, which are large enough to contain the entire causal cone of the two sites $r=0,1$ in the center of the chain (i.e., those causal cones do not reach the edges of the finite system). This allows us to compute the average density directly in the thermodynamic limit by simply averaging the average densities of the two center sites: $\langle n(t) \rangle_{L=\infty} = (\langle n(r=0,t) \rangle + \langle n(r=1,t) \rangle)/2$.

\begin{figure}
    \centering
    \includegraphics[width=0.5\textwidth]{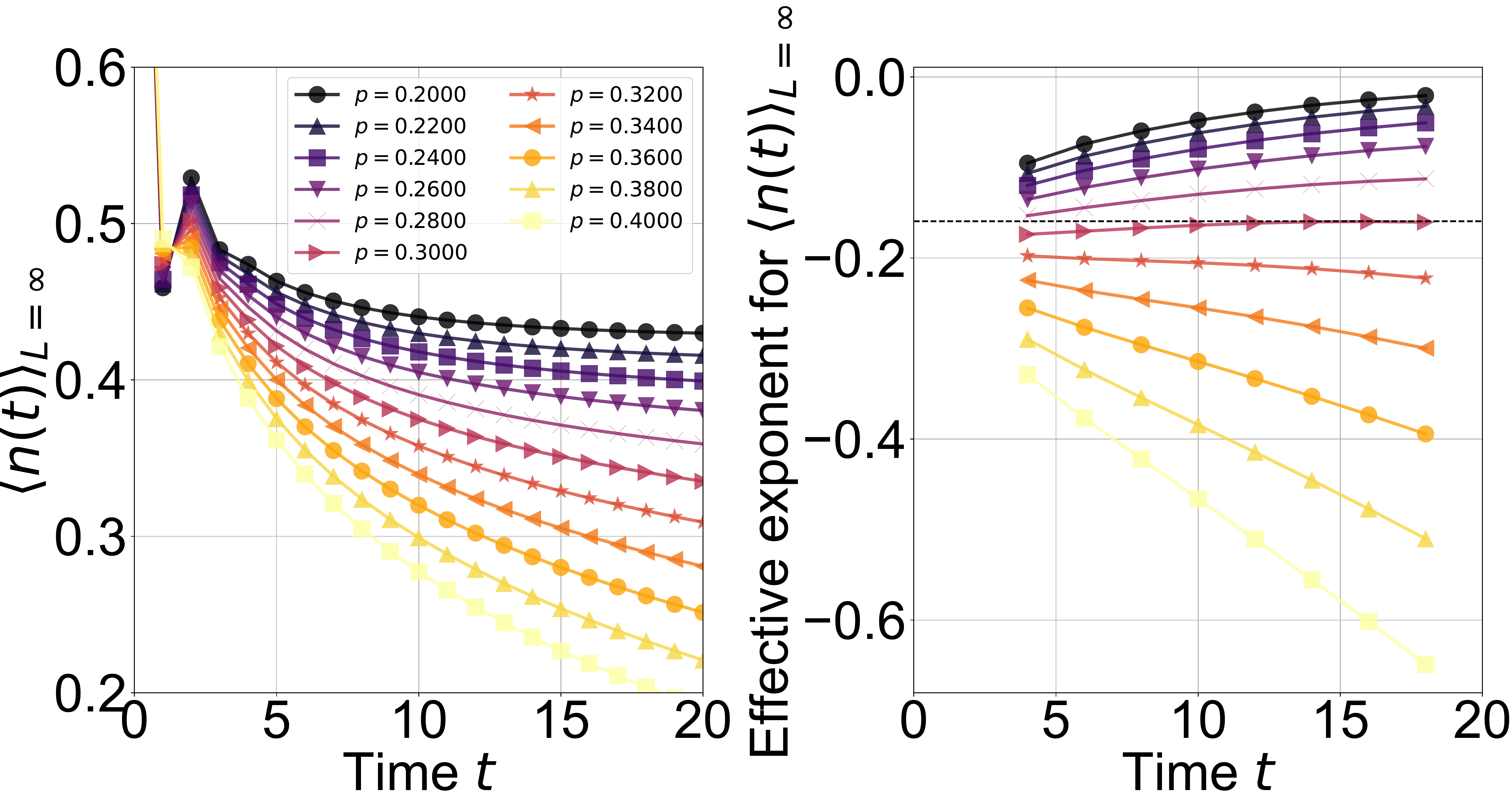}
    \caption{\textbf{Uniform initial state scaling at quantum point.} Left: The time evolution of the thermodynamic limit active site density for the FQCP at the quantum point $\theta=3\pi/4$ starting from a uniform initial state $\ket{\cdots 111\cdots}$, obtained from MPO numerics with bond dimension $D=256$. Right: The effective exponent for the active site density, with the DP exponent value marked with a black dashed line.}
    \label{fig:uniform_state_dynamics}
\end{figure}

Fig.~\ref{fig:uniform_state_dynamics} shows the dynamics of the average density for the uniform initial state obtained from our MPO numerics. We see that the effective exponent at $p=0.3\approx p_c$, closely approaches the DP value marked by a black dashed line. These results suggest that the FQCP model exhibits DP scaling for both the single seed and uniform initial states. 

\subsection{Open DMRG simulation of quasi-steady state}

As a complement to the MPO dynamics simulation, we also implement an open density matrix renormalization group (DMRG) calculation of the low-lying excited states of the Floquet channel. Compared to the dynamical simulation, this gives access to very late times and can be advantageous in cases where there is an operator entanglement barrier to extrapolating from short to long times. The DMRG results are consistent with the DP universality class when appropriately extrapolated to infinite system size.

\begin{figure}
	\centering
	\includegraphics[width=0.5\textwidth]{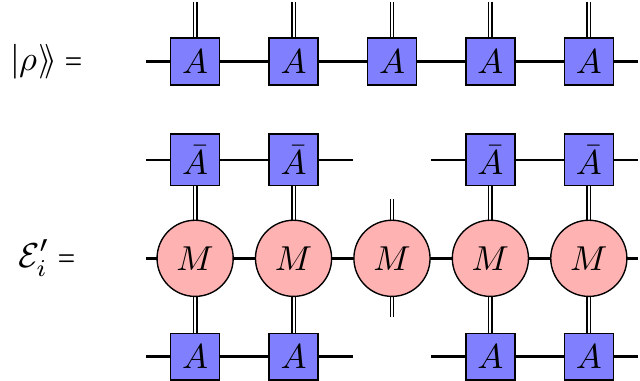}
	\caption{\textbf{DMRG simulation.} (upper) Vectorized density matrix  as a matrix product state (MPS); (lower) Modified local Floquet evolution opertor $\mathcal{E}'_i$.}
	\label{fig:MPS-DMRG}
\end{figure}

\subsubsection{Details of the method}

In general, the Floquet evolution superoperator $\mathcal{E}=\mathcal{E}(\theta, p)$ has eignvalues $e^{\e_\alpha}$ given by $\mathcal{E}[\rho_\alpha]=e^{\e_\alpha}[\rho_\alpha]$. Assuming there is no dark steady state, these eigenvalues can be sorted as $0=\e_0>{\rm Re}\e_1\geqslant{\rm Re}\e_2\geqslant\cdots$.  For the FQCP model, the absorbing state $\rho_0=\vert0\cdots0\rangle\langle0\cdots0\vert$ is the unique steady state in finite size system, thus the key quantity is the relaxation time from the longest-lived active state to the absorbing inactive state, which here is given by $\tau=-\frac{1}{{\rm Re \e_1}}$ corresponding to the lifetime of the slowest decay mode $\rho_1$. Finite-size scaling implies that, at the critical point, the relaxation time grows with systems size as $\tau\sim L^z$, which hence provides an approach to determine $p_c$ and $z$. Since the relaxation time in the active phase is in general exponentially large, it is difficult to calculate the relaxation time directly through time-evolution methods. As an alternative, we use the density-matrix renormalization group (DMRG) scheme to obtain a numerical approximation of the slowest decay mode $\rho_1$ of the Floquet evolution superoperator $\mathcal{E}$\cite{Cui2015,Mascarenhas2015, Cheng2022}. 

We first consider a vectorization of the density matrix $\rho\rightarrow\vert\rho\rrangle$ using the Choi isomorphism $\vert\psi\rangle\langle\phi\vert\rightarrow\vert\psi\otimes\phi\rrangle$. The density matrix can be represented as an MPO or an MPS with squared physical dimension, which is compatible with DMRG (see Fig. \ref{fig:MPS-DMRG} for a graphical representation):
\begin{equation}
	\vert\rho\rrangle = \sum_{\{\mu_i\}}A_{\mu_1}^{[1]}\dots A_{\mu_L}^{[L]} |\mu_1\dots \mu_L\rrangle.
\end{equation}
Here, each $A^{[i]}$ is a $d^2\times D_i\times D_{i+1}$ tensor, $\mu_i\in\{1\dots d^2\}$ labels a basis of physical states for the vectorized density matrix, $i=1\dots L$ label sites of the system, $D_1=D_{L+1}=1$ and $D_i \leq D$ where $D$ is the bond dimension.
In this notation, the Floquet evolution superoperator $\mathcal{E}$, which implements one Floquet period of time evolution, can be written as a matrix-product (super) operator acting on $ \vert\rho\rrangle$:
\begin{align}
	\mathcal{E}=\sum_{\{\mu_i,\nu_i\}}M_{\mu_1\nu_1}^{[1]}\cdots M_{\mu_L\nu_L}^{[L]}\vert \mu_1\cdots\mu_L\rrangle\llangle\nu_1\cdots\nu_L\vert,
\end{align}
where each $M^{[i]}$ is a $d^2\times d^2\times D_{O,i}\times D_{O,i+1}$ tensor, $D_{O,1}=D_{O,L+1}=1$, and $D_{O,i} \leq D_O$ where $D_O$ is the operator bond dimension.

To target the slowest decaying mode, $\rho_1$, we leverage that the trace-preserving nature of $\mathcal{E}$ implies a model-independent left steady state: $\llangle \mathbb{I}\vert\mathcal{E}=\llangle \mathbb{I}\vert$, where $\vert\mathbb{I}\rrangle$ is the maximally mixed state,  i.e., $\llangle \mathbb{I}\vert$ is the left eigenvector corresponding to steady state $\vert\rho_0\rrangle$. We use this fact to construct the modified evolution operator $\mathcal{E}'=\mathcal{E}-w\vert\mathbb{I}\rrangle\llangle\mathbb{I}\vert$, where $w>0$ penalizes the steady (absorbing) state and causes $\rho_1$ to become the ``quasi-steady state'' of $\mathcal{E}'$, or eigenvector with eigenvalue closest to 1. Using the modified Floquet operator $\mathcal{E}'$, we find an MPS approximation of $\rho_1$ by iteratively solving for the eigenvalue with largest real part of the local effective evolution operator $\mathcal{E}'_i$ obtained by contracting all indices of $\llangle\rho\vert\mathcal{E}'\vert\rho\rrangle$ except for the indices on site $i$ (see Fig. \ref{fig:MPS-DMRG} for graphical representation). (In practice, we take $w=1$ and have $D_O=256$ for the FQCP model.)

\begin{figure*}
	\centering
	\includegraphics[width=\textwidth]{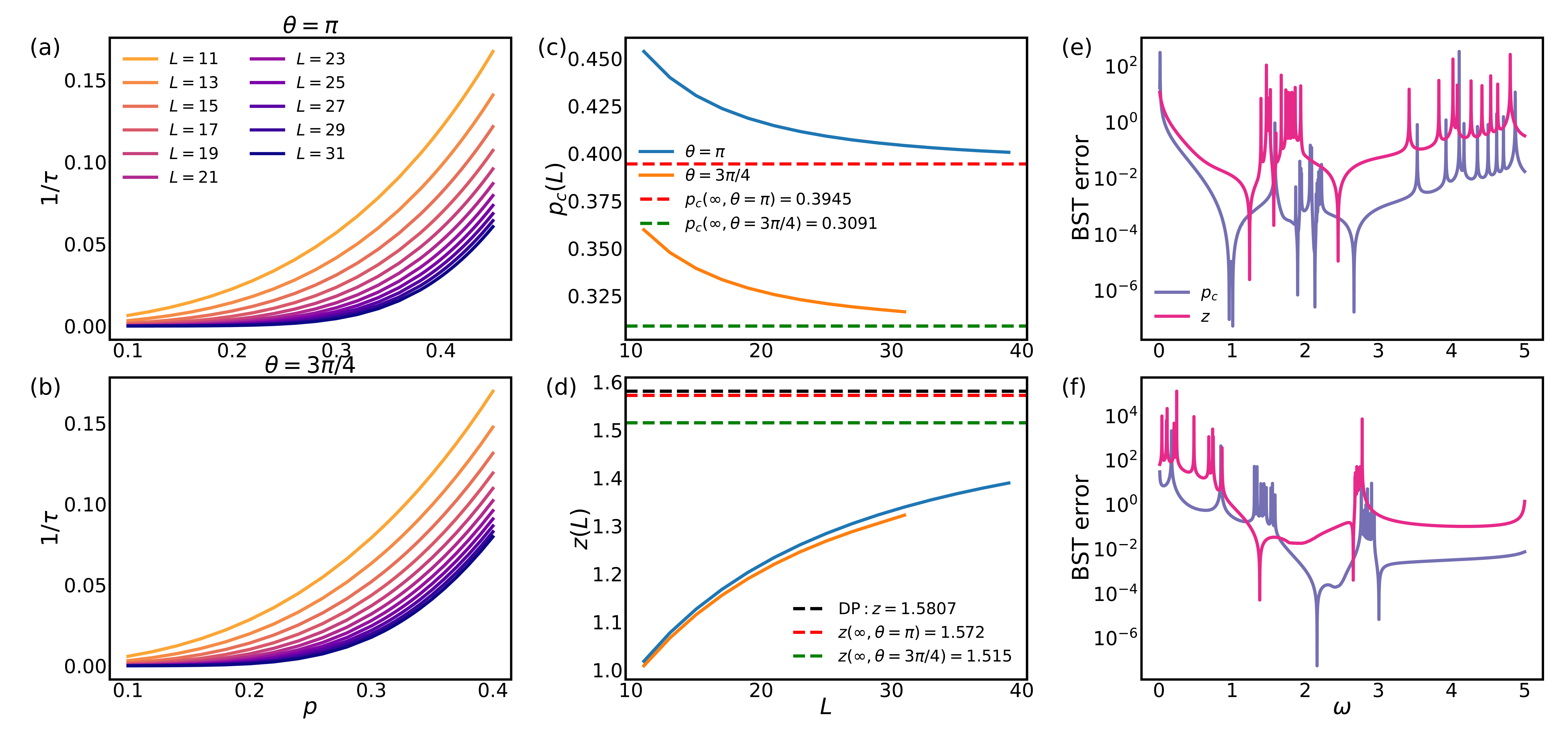}
    \caption{\textbf{Determination of critical point scaling of relaxation time.} Inverse of relaxation time $1/\tau$ versus $p$ at the (a) classical and (b) quantum points. (c) The finite-size critical point $p_c(L)$ versus $L$. (d) The finite-size critical exponent $z(L)$ versus $L$. The BST error from the extrapolation of $p_c(L)$ and $z(L)$ at the (e) classical and (f) quantum points.}
	\label{fig:relaxationtimecritical}
\end{figure*}

\subsubsection{Critical point scaling of relaxation time}

We perform DMRG simulations at both the classical point $\theta = \pi$ and quantum point $\theta = 3\pi/4$. The inverse of the relaxation time, $\tau^{-1}$, which we refer to as the dissipative gap, is shown in Fig.\ref{fig:relaxationtimecritical} (a) and (b). To extract the critical point information, we consider the finite-size scaling ansatz $\tau(p, L)=L^{-z}\mathcal{F}(\vert p-p_c\vert L^{1/\nu_\perp})$, where $\mathcal{F}$ is a scaling function. This scaling form implies that the quantity\cite{Dickman2008,Filho2011}
\begin{align}
	R_L(p)\equiv\frac{\log\[\tau(p, L)/\tau(p,L-2)\]}{\log\[(L-2)/L\]},
\end{align}
converges to $z$ at $p=p_c$ as $L\rightarrow\infty$. By considering three consecutive sizes $L-2, L, L+2$, one can define the finite-size critical point $p_c(L)$ and critical exponent $z(L)$ satisfying $R_L(p_c(L))=R_{L+2}(p_c(L))\equiv z(L)$. Both $p_c(L)$ and $z(L)$ form finite-size sequences, from which we perform BST extrapolation to estimate $p_c=p_c(\infty)$ and $z=z(\infty)$. The simulation used to extract the critical point information is performed from $L=11$ to $L=39$ with maximum bond dimension $D_{\rm max}=324$ for the classical case and from $L=11$ to $L=31$ for the quantum case  with $D_{\rm max}=260$.  As shown in Fig.\ref{fig:relaxationtimecritical}(c,d), for the classical point $\theta=\pi$, we obtain $p_c\approx0.3945$, which is consistent with the high-accuracy result $p_c=0.3944$ mentioned above, and $z\approx1.572$ with $0.5\%$ off the DP value $z=1.5807$. For the quantum case $\theta=\frac{3}{4}\pi$, the estimated critical point is $p_c\approx0.3091$ with $z\approx1.515$, which has $4.15\%$ deviation compared to the DP value. 

We note that that the errors in the DMRG calculation, for example due to bond-dimension truncation or due to the finite tolerance of the non-Hermitian eigensolver used in DMRG, propagate through to errors in the finite-size extrapolation for $p_c$ and hence $z$ in a manner that is difficult to reliably estimate. We expect that such errors account for the small percent-level discrepancies between the extracted $z$ value and the DP values, and view the DMRG results as consistent within error bars of DP scaling.

\bibliography{refs}